\documentclass[11pt]{article}
\usepackage{fullpage}
\usepackage[utf8]{inputenc}
\usepackage[pass]{geometry}
\usepackage{addlines}
\usepackage{amsmath,amsthm,amssymb}
\usepackage{graphics,graphicx,tikz,color,float, hyperref}
\usepackage{mathtools}
\usepackage{amsfonts}
\usepackage{comment}
\usepackage{framed}
\usepackage[section]{placeins}
\usepackage{algorithm}
\usepackage[noend]{algpseudocode}
\usepackage{xcolor}
\newcommand{\mycomment}[1]{\textup{{\color{red}#1}}}

\newcommand{\expect}[2]{\E_{#1}{\left[#2\right]}}
\newtheorem{fact}{Fact}
\newtheorem{definition}{Definition}
\newtheorem{lemma}{Lemma}
\newtheorem{theorem}{Theorem}
\newcommand{\tab}{\hspace*{2em}}
\newcommand{\beq}{\begin{equation}}
	\newcommand{\enq}{\end{equation}}
\newcommand{\bel}{\begin{lemma}}
	\newcommand{\enl}{\end{lemma}}
\newcommand{\bet}{\begin{theorem}}
	\newcommand{\ent}{\end{theorem}}

\newcommand{\tr}{\mathrm{Tr}}

\newcommand{\E}{\mathbb{E}}
\newcommand{\ketbra}[1]{|#1\rangle\langle#1|}

\newcommand{\eps}{\varepsilon}

\newcommand{\generateA}{A= \Ext_1(Y,Z_s) }
\newcommand{\generateT}{C= \Ext_2(Z,A)}
\newcommand{\generateB}{B= \Ext_{1}(Y,C)}
\newcommand{\generateAA}{A^\prime= \Ext_1(Y^\prime,Z^\prime_s) }
\newcommand{\generateTT}{C^\prime= \Ext_2(Z^\prime,A^\prime)}
\newcommand{\generateBB}{B^\prime= \Ext_{1}(Y^\prime,C^\prime)}
\newcommand{\generateAbar}{\overline{A}= \Ext_1(Y,\overline{Z_s}) }
\newcommand{\generateTbar}{\overline{C}= \Ext_2(\overline{Z},\overline{A})}
\newcommand{\generateBbar}{\overline{B}= \Ext_{1}(Y,\overline{C})}
\newcommand{\generateAAbar}{\overline{A}^\prime= \Ext_1(Y^\prime,\overline{Z_s}^\prime) }
\newcommand{\generateTTbar}{\overline{C}^\prime= \Ext_2(\overline{Z}^\prime,\overline{A}^\prime)}
\newcommand{\generateBBbar}{\overline{B}^\prime= \Ext_{1}(Y^\prime,\overline{C}^\prime)}

\newcommand{\generateYSbar}{$\overline{Z}_s =$ Prefix$(\overline{Z},s)$}

\newcommand{\generateYSSbar}{$\overline{Z}_s^\prime =$ Prefix$(\overline{Z}^\prime,s)$}
\newcommand{\sendArl}{$A \longleftarrow A$}

\newcommand{\sendBrl}{$B \longleftarrow B$}

\newcommand{\sendTlr}{$C \longrightarrow C$}
\newcommand{\sendAArl}{$A^\prime \longleftarrow A^\prime$}

\newcommand{\sendBBrl}{$B^\prime \longleftarrow B^\prime$}

\newcommand{\sendTTlr}{$C^\prime \longrightarrow C^\prime$}
\newcommand{\sendAbarrl}{$\overline{A} \longleftarrow \overline{A}$}

\newcommand{\sendBbarrl}{$\overline{B} \longleftarrow \overline{B}$}

\newcommand{\sendTbarlr}{$\overline{C} \longrightarrow \overline{C}$}
\newcommand{\sendAAbarrl}{$\overline{A}^\prime \longleftarrow \overline{A}^\prime$}

\newcommand{\sendBBbarrl}{$\overline{B}^\prime \longleftarrow \overline{B}^\prime$}

\newcommand{\sendTTbarlr}{$\overline{C}^\prime \longrightarrow \overline{C}^\prime$}
\newcommand{\sendYSlr}{$Z_s \longrightarrow Z_s$}

\newcommand{\sendYSbarlr}{$\overline{Z}_s \longrightarrow \overline{Z}_s$}

\newcommand{\sendYSSbarlr}{$\overline{Z}_s^\prime \longrightarrow \overline{Z}_s^\prime$}

\newcommand{\sendYYbarlr}{$\overline{Z}^\prime \longrightarrow \overline{Z}^\prime$}

\newcommand*{\cA}{\mathcal{A}}

\newcommand*{\cH}{\mathcal{H}}

\newcommand*{\cD}{\mathcal{D}}

\newcommand*{\cO}{\mathcal{O}}

\newcommand*{\cX}{\mathcal{X}}

\newcommand*{\cZ}{\mathcal{Z}}
\newcommand*{\cE}{\mathcal{E}}

\newcommand{\cP}{\mathcal{P}}

\newcommand*{\IP}{\mathsf{IP}}
\newcommand{\Ext}{\mathsf{Ext}}
\newcommand{\pre}{\mathsf{Prefix}}
\newcommand{\advc}{\mathsf{AdvGen}}
\newcommand{\advcb}{\mathsf{AdvCB}}
\newcommand{\ff}{\mathsf{FF}}
\newcommand{\ecc}{\mathsf{ECC}}

\newcommand{\supp}{\mathrm{supp}}
\newcommand{\suppress}[1]{}

\newcommand{\defeq}{\ensuremath{ \stackrel{\mathrm{def}}{=} }}
\newcommand{\F}{\mathbb{F}}

\newcommand {\br} [1] {\ensuremath{ \left( #1 \right) }}

\newcommand {\minusspace} {\: \! \!}
\newcommand {\smallspace} {\: \!}
\newcommand {\fn} [2] {\ensuremath{ #1 \minusspace \br{ #2 } }}

\newcommand {\dmax} [2] {\fn{\mathrm{D}_{\max}}{#1 \middle\| #2}}

\newcommand {\mutinf} [2] {\fn{\mathrm{I}}{#1 \smallspace : \smallspace #2}}

\newcommand {\condmutinf} [3] {\mutinf{#1}{#2 \smallspace \middle\vert \smallspace #3}}
\newcommand {\hminone} [1] {\fn{ \mathrm{H }_{\min}}{#1}}
\newcommand {\hminn} [2] {\fn{ \mathrm{ \tilde{H} }_{\min}}{#1 \middle | #2}}
\newcommand {\hmin} [2] {\fn{ \mathrm{H }_{\min}}{#1 \middle | #2}}

\newcommand {\id} {\ensuremath{\mathbb{I}}}

\newcommand {\Hmin}{\mathrm{H}_{\min}}

\newcommand{\samp}{\mathsf{Samp}}
\newcommand{\ind}{\mathcal{S}}
\newcommand{\indbar}{\overline{\mathcal{S}}}
\usepackage{afterpage}
\newcommand*{\qpass}{\mathsf{qpa\mhyphen states}}
\newcommand*{\qpas}{\mathsf{qpa\mhyphen state}}
\newcommand*{\qmas}{\mathsf{qma\mhyphen state}}
\newcommand*{\qmass}{\mathsf{qma\mhyphen states}}
\newcommand*{\nmas}{\mathsf{qnm\mhyphen state}}
\newcommand*{\qma}{\mathsf{qma}}

\newcommand*{\qmra}{\mathsf{qMara}}
\newcommand*{\qia}{\mathsf{qia}}

\newcommand*{\nma}{\mathsf{qnma}}

\newcommand*{\nmext}{\mathsf{nmExt}}

\newcommand{\mac}{\mathsf{MAC}}
\newcommand{\X}{\mathcal{X}}
\newcommand{\Y}{\mathcal{Y}}
\newcommand{\Z}{\mathcal{Z}}

\newcommand*{\cL}{\mathcal{L}}

\newcommand{\bra}[1]{\langle #1|}
\newcommand{\ket}[1]{|#1 \rangle}

\mathchardef\mhyphen="2D

\newcommand{\argmax}{\operatornamewithlimits{arg\ max}}

\makeatletter
\newcommand*{\rom}[1]{\expandafter\@slowromancap\romannumeral #1@}
\makeatother

\mathchardef\mhyphen="2D

\newtheorem{remark}{Remark}
\newtheorem{claim}{Claim}
\newtheorem{corollary}{Corollary}

\newenvironment{changemargin}[2]{%
\begin{list}{}{%
\setlength{\topsep}{0pt}%
\setlength{\leftmargin}{#1}%
\setlength{\rightmargin}{#2}%
\setlength{\listparindent}{\parindent}%
\setlength{\itemindent}{\parindent}%
\setlength{\parsep}{\parskip}%
}%
\item[]}{\end{list}}

\newfloat{Protocol}{tbhp}{lop}
\title{Quantum secure non-malleable extractors}
\author{
	Naresh Goud Boddu\footnote{NTT Research, USA, \texttt{naresh.boddu@ntt-research.com}}
	\and
	Rahul Jain \footnote{Centre for Quantum Technologies and Department of Computer Science, 
  National University of Singapore and MajuLab, UMI 3654, Singapore,  \texttt{rahul@comp.nus.edu.sg}}
  \and
Upendra Kapshikar\footnote{Center for Quantum Technologies, National University of Singapore, \texttt{e0382999@u.nus.edu}}~~\footnote{This paper was presented at 17th Conference on the Theory of Quantum Computation, Communication and Cryptography (TQC 2022) in workshop track.}
}
\begin{document}
\begin{titlepage}
\clearpage
\maketitle
\thispagestyle{empty}
\begin{abstract}

{\em Non-malleable extractors}, introduced by Dodis and Wichs~\cite{DW09}, have found several applications in the study of tamper-resilient cryptography. For example, seeded non-malleable extractors are a key ingredient in the {\em privacy amplification} (PA) protocol with an active classical adversary. Similarly, $2$-source non-malleable extractors provide a way to construct {\em non-malleable codes}, introduced by Dziembowski, Pietrzak, and Wichs~\cite{DPW10}, with further applications to non-malleable secret sharing. Thus, understanding the security of such non-malleable extractors against quantum adversaries is vital.

We construct several efficient quantum secure non-malleable extractors. All our constructions are based on the works of Chattopadhyay, Goyal, and Li~\cite{CGL15}, and Cohen~\cite{Coh15}

    \begin{itemize}
    \item  We construct the first efficient quantum secure non-malleable extractor for source min-entropy $k \geq \textsf{poly}\left(\log \left(\frac{n}{\eps}\right)\right)$ and seed length $d = \textsf{poly}\left(\log \left(\frac{n}{\eps}\right)\right)$, where $n$ is the length of the source and $\eps$ is the error parameter. Previously, Aggarwal, Chung, Lin, and Vidick~\cite{ACLV18} demonstrated that an inner-product based non-malleable extractor proposed by Li~\cite{Li12c} is quantum secure, but it required linear (in $n$) min-entropy and seed length.
       \item By leveraging the connection between non-malleable extractors and PA (first established in the quantum setting by Cohen and Vidick~\cite{CV16}), we obtain a $2$-round PA protocol that is secure against active quantum adversaries with communication $\textsf{poly}\left(\log \left(\frac{n}{\eps}\right)\right)$. This allows for a trade-off between communication and error in a PA protocol, improving on the results of~\cite{ACLV18}, which required communication to be linear in $n$.
        
  \item We construct an efficient quantum secure $2$-source non-malleable extractor for min-entropy $k \geq n - n^{\Omega(1)}$, with an output size of $n/4$ and error $2^{- n^{\Omega(1)}}$.
       \item Additionally, we explore the natural extensions of these extractors when the tampering of the inputs occurs $t$-times. We construct efficient quantum secure $t$-non-malleable extractors for both the seeded case ($t = d^{\Omega(1)}$) and the $2$-source case ($t = n^{\Omega(1)}$).
         We construct efficient quantum secure $t$-non-malleable extractors for both seeded ($t=d^{\Omega(1)}$) as well as $2$-source case ($t=n^{\Omega(1)}$).
    \end{itemize}
\end{abstract}
\end{titlepage}
\section{Introduction}

{\em Extractors} are functions that transform weak sources into uniform randomness. They are crucial because randomized algorithms are designed under the assumption that the randomness used is uniformly distributed. Extractors have numerous applications, including privacy amplification (PA), pseudo-randomness, derandomization, expanders, combinatorics, and cryptography. Some general models of weak sources are the so-called {\em min-entropy} sources and {\em conditional min-entropy} sources. Please refer to Section~\ref{sec:prelims} for definitions of information-theoretic quantities, extractors, and various adversary models.

 Let random variables $X \in \{ 0,1\}^n , Y  \in \{ 0,1\}^n$ and $S \in  \{ 0,1\}^d$ (where $U_d$ is the  uniform distribution on $d$ bits and $X \otimes S$ represents independent random variables $X,S$)
\[\mathcal{C}_1 = \{ X : \Hmin(X)  \geq k \} \quad ; \quad \mathcal{C}_2 = \{ X \otimes S : \Hmin(X)  \geq k \ \textnormal{and} \ S =U_d\} \quad ; \]
\[\mathcal{C}_3 = \{ X \otimes Y : \Hmin(X)  \geq k_1 \ \textnormal{and} \ \Hmin(Y)  \geq k_2 \}. \]

It can be argued that no deterministic function can extract even one uniform bit given an (arbitrary) source $X \in \mathcal{C}_1$, for $k \leq n-1$ \cite{CG85}.
This led to designing extractors using sources from $\mathcal{C}_2$.
They use an additional uniform source (aka {\em seed} $S=U_d$) called {\em seeded extractors}. Subsequent works also considered extraction from class $\mathcal{C}_3$, involving multiple independent weak sources~\cite{CG85,Bou05}. In the classical setting, extractors have been studied extensively both in the seeded and the multi-source settings~\cite{ILM89,GUV09,DW08,Bou05,CG85,KLRZ08,R06,Raz05,KLR09}.

Consider a situation where the input of an extractor is tampered with. 
 For example, for a source $(X,S) \in \mathcal{C}_2$, an adversary may tamper with the seed $S$ to  modify it to some other seed $S'$. In this case, a natural question arises: \emph{`Does the output of tampered input $(X,S')$ have any correlation with the output of the untampered input $(X,S)$?'\ }. To be resilient against such adversarial tampering of the input, it is crucial that the original input produces an output that is (almost) independent of the one generated by the tampered input. Extractors with this property are called non-malleable extractors~\cite{DW09}. A non-malleable extractor, denoted by $\nmext$, produces an output that is nearly uniform and independent of tampering. Formally, this means that the outputs satisfy $\left(\nmext(X,S)\nmext(X,S'\right) \approx U_m \otimes \nmext(X,S')$). 

Applications to cryptography motivate the study of extractors in the presence of an adversary holding some side information $E$ about the source. For simplicity and brevity, we refer to the joint systems, including the adversary's side information, as a source. These sources are considered to be of the form:
\[\mathcal{C}_4 = \{ XE \otimes S : \hmin{X}{E}  \geq k \ \textnormal{and} \ S =U_d\}. \]
Here, we require the output of the extractor $\Ext(X,S)$ to be uniform given the side information $E$, i.e., $\left(\Ext(X,S)E \approx U_m \otimes E\right)$.
Additionally, in the case of a non-malleable extractor, we require that it is (nearly) independent of any potential tampering of the seed: \[\nmext(X,S)\nmext(X,S')E \approx U_m \otimes \nmext(X,S')E.\]
Similarly one can consider $2$-sources with adversary side information (below $k_1,k_2>0$ and $\vert Y\vert=n$): 
$$\mathcal{C} = \{ X-E-Y : \hmin{X}{E}  \geq k_1 \ \textnormal{and} \ \hmin{Y}{E} \geq k_2 \},$$
where $X-E-Y$ represents a Markov-chain (see Definition~\ref{def:markovchain}). $2$-source extractors have been extensively studied as well~\cite{Bou05,CG85,KLRZ08,R06,Raz05,KLR09,CGL15,li15,CZ19}. 
 Similar to seeded extractors, $2$-source extractors also have a stronger variant in terms of non-malleability. 
For a $2$-source non-malleable extractor,  we allow tampering on both $X$ and $Y$. An adversary can modify $(X,Y)$ to some $(X^\prime, Y^\prime)$ such that,  either $\Pr[X\neq X^\prime]=1$ or $\Pr[Y \neq Y^\prime]=1$.  A $2$-source non-malleable extractor is a function  $ 2\nmext : \lbrace 0,1\rbrace^n \times \lbrace 0,1\rbrace^{n} \to \lbrace 0,1 \rbrace^m$ such that:
 \begin{align*}
      2\nmext(X,Y) 2\nmext(X^\prime,Y^\prime) E Y Y^\prime &\approx_\eps U_m \otimes  2\nmext(X^\prime,Y^\prime) E Y Y^\prime.
 \end{align*}
 $2$-source non-malleable extractors have been used in the construction of non-malleable codes in the well studied {\em split-state} model by Chattopadhyay, Goyal and Li~\cite{CGL15}. 
These non-malleable codes are known to have applications in various cryptographic tasks such as non-malleable secret-sharing and non-malleable commitment~\cite{GPR16,GK16,GK18,ADNOP19,AP19}. Seeded non-malleable extractors were used, by Chattopadhyay and Zuckerman~\cite{CZ19}, as a key ingredient in their breakthrough construction of $2$-source extractors for $\mathsf{polylog}(n)$ min-entropy sources. $2$-source extractors also find applications in graph-theory and are related to Ramsey graphs, well studied combinatorial objects.

With the advent of quantum computers, it is natural to investigate the security of extractors against a quantum adversary with quantum side information on weak sources. Such sources are of the form: \[\mathcal{Q}_1 = \{ \sigma_{XES}=\sigma_{XE} \otimes \sigma_S : \hmin{X}{E}_\sigma  \geq k \ \textnormal{and} \ \sigma_S =U_d\}, \]
where side information $E$ is quantum and source as well as seed $(XS)$ are classical. As expected, quantum side information presents many more challenges compared to classical side information. Gavinsky et al.~\cite{GKKRW07} provided an example of a seeded extractor that is secure against a classical adversary but not secure against a quantum adversary, even with very small side information.

Very little is known about the security of non-malleable extractors against quantum side information. The initial challenge lies in defining a non-malleable extractor with quantum side information, as we need to provide security with updated quantum side information when the adversary modifies $(E,S) \to (E',S')$.
Informally, we require (for formal definition see Definition~\ref{nme})  $$\nmext(X,S)\nmext(X,S')E' \approx U_m \otimes \nmext(X,S')E'.$$
In the classical setting it can be argued that, conditioned on $E=e$, $X$ and $S'$ remain independent, since with this conditioning, $S'$ is a deterministic function of $S$. However in the quantum setting, conditioning on quantum side information cannot be done in this manner, so this argument does not hold. 

In this paper, we study seeded non-malleable extractors and extend the definition to $2$-source non-malleable extractors (see Definition~\ref{def:2nme}). We also explore their natural extensions where the tampering is performed 
$t$-times, allowing the adversary to tamper with $(E,S) \to (E',S^1, \ldots, S^t)$.

In this scenario, we require that the output remains nearly independent given the quantum side information and any of the tampered outputs. For example, in the seeded case (see Definition~\ref{tnme}):
$$\nmext(X,S)\nmext(X,S^1) \ldots \nmext(X,S^t) E' \approx U_m \otimes \nmext(X,S^1) \ldots \nmext(X,S^t)E'.$$ Before stating our results, we give a brief overview of some relevant previous works.
\subsection*{Previous works}
Aggarwal et al.~\cite{ACLV18} have shown that an  inner-product based non-malleable extractor proposed by Li~\cite{Li12c} is quantum secure, however it requires linear min-entropy and seed length.\suppress{
\begin{definition}[\cite{Li12c}]\label{ipnme}
	 Let $p \ne 2$ be a prime and 
	 $n$ be an integer. Define $\nmext : \mathbb{F}^n_p \times  \mathbb{F}^{n/2}_p \to  \mathbb{F}_p$ given by $\nmext(X,Y) \defeq \langle X, Y \vert \vert Y^2 \rangle$, where $\vert \vert$ represents concatenation of strings and $Y^2$ is computed via multiplication in   $\mathbb{F}_{p^{n/2}}$. 
\end{definition}}

Recent work by Aggarwal et al.~\cite{ABJO21} has strengthened this result by showing that Li's extractor remains quantum secure even when the seed is not uniform, although it still requires linear min-entropy in the seed. They achieve this by introducing a notion of security against a quantum measurement adversary, to which the eventual quantum security of the inner-product is reduced.

To the best of our knowledge, the inner-product based non-malleable extractor proposed by Li~\cite{Li12c} is the only non-malleable extractor for which quantum security is known.

Earlier works by Cohen and Vidick~\cite{CV16} and Bouman and Fehr~\cite{BF11} attempted to provide quantum security for non-malleable extractors based on the powerful technique of \emph{alternating extraction} introduced by Dziembowski and Pietrzak~\cite{DP07}. Unfortunately, these results were later withdrawn due to subtle issues in the arguments.

\suppress{
In~\cite{BF11}, as per them, a subtle issue concerning conditioning on quantum side information is overlooked. In~\cite{CV16} the authors attempted to deal with it by using the formalism of~\emph{quantum Markov-chains}.
Unfortunately, later it was noticed by them that there is an issue in the argument while using the Markov-chain formalism and the results do not hold. }
\subsection*{Our results} 

Let $\sigma_{XES}$ be a source from $\mathcal{Q}_1$. We have  $\sigma_{XES} =\sigma_{XE} \otimes \sigma_S$, $\hmin{X}{E}_\sigma \geq k$ and $\sigma_S=U_d$. One may consider register $S$ as uniform seed and register $E$ as  adversary quantum side information on source $X$. We consider the pure state extension of $\sigma_{XES}$ denoted by  $\sigma_{X\hat{X}\hat{E}E S\hat{S}} = \sigma_{X\hat{X}\hat{E}E} \otimes \sigma_{S\hat{S}}$, as it helps us in our analysis. Here $\sigma_{X\hat{X}\hat{E}E}$, $\sigma_{S\hat{S}}$ are {\em canonical purifications} of $\sigma_{XE}$ and $\sigma_S$ respectively. For simplicity we call the entire pure state as a source, even though the uniform randomness is extracted from classical registers of a pure state (see Definition~\ref{def:classicalinpurestate}). Note that  $\hat{X}, \hat{S}$ are copies of $X,S$ respectively (see Definition~\ref{def:copyofaclassicalregister}). Consider,
\[\mathcal{Q}_2 = \{ \sigma_{X\hat{X}\hat{E}ES\hat{S}} =\sigma_{X\hat{X}\hat{E}E} \otimes \sigma_{S\hat{S}} : \hmin{X}{E}_\sigma  \geq k \ \textnormal{and} \ \sigma_S =U_d\},\]
where $\sigma_{X\hat{X}\hat{E}ES\hat{S}}$ is a pure state. Note the sources in $\mathcal{Q}_2$ are purifications of sources in $\mathcal{Q}_1$. The conditions in $\mathcal{Q}_2$ are equivalent to, 
\[ \hmin{X}{ES\hat{S}}_\sigma =\hmin{X}{E}_\sigma \geq k \quad ; \quad \hmin{S}{X\hat{X}\hat{E}}_\sigma =\Hmin(S)_\sigma = d. \] More generally, this leads us to consider the following sources (see Definition~\ref{qmadvk1k2}):
\suppress{
Now suppose, adversary chooses an isometry $V: \cH_{E} \otimes \cH_{S} \rightarrow \cH_{E'}\otimes \cH_{S}\otimes \cH_{S'}$ to modify the seed $(S' \ne S)$, and let 
$$\rho_{X\hat{E}E'SS'\hat{S}} = V \sigma_{X\hat{E}ES\hat{S}} V^\dagger.$$ We need, 
$$(\nmext(X,S)\nmext(X,S')E'S)_\rho \approx U_m \otimes (\nmext(X,S')E'S)_\rho$$ to show quantum non-malleable security. 
}
\[\mathcal{Q} = \{ \sigma_{X\hat{X}NMY\hat{Y}} : \sigma_{X\hat{X}NMY\hat{Y}} \ \textnormal{is a} \ (k_1,k_2)\mhyphen\qpas\}.\]
\noindent Here $\mathsf{qpa}$ stands for \emph{quantum purified adversary}. To understand the advantage of considering the sources along with purification registers consider the following example. 
 Consider the Markov-chain $\sigma_{XEY} =\frac{1}{\sqrt{2}} (\ket{00}+\ket{11})\ket{0}$. Note $\condmutinf{X}{Y}{E}_\sigma =0$. Let $\rho_{XEY}$ be the state after applying CNOT gate on qubit $Y$ conditioned on qubit $E$ of $\sigma$.  Note $\rho_{XEY} =  \frac{1}{\sqrt{2}} (\ket{000}+\ket{111})$. We have $\condmutinf{X}{Y}{E}_\rho  \ne 0$. This points us to one of the key difficulty faced by earlier approach of Markov model.  On the other hand, we note that sources in $\mathcal{Q}$ remain in $\mathcal{Q}$ after adversarial tampering. This can be seen as follows: Let 
 $\sigma_{X\hat{X}NMY\hat{Y}} \ \textnormal{be a} \ (k_1,k_2)\mhyphen\qpas$ and adversary tampers $(Y,M) \to (Y,Y',M')$. Let $\rho_{X\hat{X}NM'YY'\hat{Y}}$ be the state after adversary action. It is easy to note (using Fact~\ref{fact102}) that $\hmin{X}{M'YY'\hat{Y}}_\rho \geq k_1$ and $\hmin{Y}{NX \hat{X}}_\rho \geq k_2$.
 
 
 This enables us to analyse the constructions of non-malleable extractors step by step and ensuring the parameters $(k_1,k_2)$ for the $(k_1,k_2)\mhyphen\qpas$ at the end are still in control to extract randomness.  

Also, note the sources $\mathcal{C}_2, \mathcal{C}_3, \mathcal{C}_4, \mathcal{Q}_1, \mathcal{Q}_2$ can all be seen as special cases of $\mathcal{Q}$ (in the purification picture). This provides us a general framework to define extractors and non-malleable extractors, both in the seeded and the $2$-source settings.

We now state our results. Let $n,d,t$ be positive integers and $k,\eps >0$. The following result is about seeded non-malleable extractors.
\begin{theorem}[quantum secure non-malleable extractor]\label{intro:thm:nmext}
Let $d = \cO(\log^{7}(n/\eps))$ and $k = \Omega(d)$. There exists an efficient non-malleable extractor $\nmext :\{0,1 \}^n \times \{0,1 \}^d \to \{0,1 \}^{k/4}$ that is  $(k, \cO(\eps))$-quantum secure (see Definition~\ref{nme}). 
\end{theorem}
The adversary in the result above is the $\nma$ (short for quantum non-malleable adversary). As a corollary,  we get the following result for the standard model of sources considered in the literature (sources in $\mathcal{Q}_1$).
\begin{corollary}\label{intro:corr:add1}
 Let $d = \cO(\log^{7}(n/\eps))$ and $k = \Omega(d)$. Let $\rho_{XEY}$ be a c-q state with registers ($XY$) classical such that  \[ \hmin{X}{E}_\rho \geq k \quad ; \quad \rho_{XEY} =\rho_{XE} \otimes U_d \quad ; \quad \vert X \vert =n.\]Let  $\mathsf{T}: \mathcal{L} (\mathcal{H}_E \otimes \mathcal{H}_{Y}) \rightarrow \mathcal{L} (\mathcal{H}_{E^\prime} \otimes \mathcal{H}_Y \otimes \mathcal{H}_{Y^\prime})$ be a (safe) CPTP map such that for $\sigma_{XE'YY'} =\mathsf{T}(\rho_{XEY})$, we have registers $XYY'$ classical and $\Pr(Y \ne Y')_\sigma=1$. Let the function $\nmext$ be from Theorem~\ref{intro:thm:nmext},  $L= \nmext(X,Y)$ and $L'=\nmext(X,Y')$. Then, 
 $$ \| \sigma_{LL'YY'E'} - U_{k/4} \otimes \sigma_{L'YY'E'} \|_1 \leq \cO(\eps).$$
\end{corollary}

Dodis and Wichs~\cite{DW09} gave a two-round protocol for privacy amplification (PA) against active adversaries with classical side information. The main ingredient in their protocol is a non-malleable extractor, which when combined with an information-theoretically secure message authentication code gives security in PA. As shown in~\cite{CV16}, using quantum secure non-malleable extractor, one can extend the proof of security by Dodis and Wichs to the case of active quantum adversaries. Thus, our quantum secure non-malleable extractor, given by Theorem~\ref{intro:thm:nmext}, enables us to obtain a PA protocol against active quantum adversaries (see Definition~\ref{privamp}). 

\begin{theorem} \label{thm:PA}
Let $d = \cO(\log^{7}(n/\eps))$, $k = \Omega(d)$ and $\delta>0$ be a small enough constant. There exists an efficient two-round PA protocol against active quantum adversaries for min-entropy $k$ sources that can extract $\left(\frac{1}{2} - \delta\right)k$ bits with communication $\cO(d)$ and error $\cO(\eps)$.
\end{theorem}

This result allows for a trade-off in communication and error in a PA protocol, improving on the result of~\cite{ACLV18}, where the communication is required to be linear in $n$. We provide a proof of Theorem~\ref{thm:PA} in Appendix~\ref{sec:PA}. Similar proofs have appeared in~\cite{CV16,ABJO21,ACLV18}. 

We show the following result for $2$-source non-malleable extractors.
\begin{theorem}[quantum secure $2$-source non-malleable extractor]\label{intro:thm:2nmext}
Let $k = \cO(n^{1/4})$ and $\eps =2^{-n^{\Omega(1)}}$. There exists an efficient $2$-source non-malleable extractor $2\nmext :\{0,1 \}^n \times \{0,1 \}^n \to \{0,1 \}^{n/4}$ that is  $(n-k,n-k, \cO(\eps))$-quantum secure (see Definition~\ref{def:2nme}). 
\end{theorem}
The above result is stated for the adversary $\nma$ (quantum non-malleable adversary). As a corollary, we obtain corresponding results for other models of $2$-source adversaries studied in the literature.

Kasher and Kempe~\cite{KK10} introduced the  quantum independent adversary ($\qia$) model, where the adversary obtains independent side-information from both sources.  Informally, $\qia$ gets the registers $\rho_{E_1E_2}$ as quantum side information in $\rho_{XE_1E_2Y} $ such that 
\[\rho_{XE_1E_2Y} =  \left(\rho_{XE_1} \otimes \rho_{YE_2} \right) \quad ;  \quad \hmin{X}{E_1}_\rho \geq k_1 \quad ; \quad \hmin{Y}{E_2}_\rho \geq k_2 .\]
We refer the reader to~\cite{KK10} for complete details. We propose to incorporate non-malleable extractor security against $\qia$ as follows.
\begin{definition}[$2$-source non-malleable extractor against $\qia$]\label{intro2nmextiadvmodel}
Let $\rho_{XE_1E_2Y}$ be a c-q state with registers ($XY$) classical such that $\vert X \vert =\vert Y\vert =n$,  \[\rho_{XE_1E_2Y} =  \left(\rho_{XE_1} \otimes \rho_{YE_2} \right) \quad ;  \quad \hmin{X}{E_1}_\rho \geq k_1 \quad ; \quad \hmin{Y}{E_2}_\rho \geq k_2 .\]Let  $\mathsf{T}_1: \mathcal{L} (\mathcal{H}_{E_2} \otimes \mathcal{H}_{X})  \rightarrow \mathcal{L} ( \mathcal{H}_{E_2^\prime} \otimes \mathcal{H}_X \otimes \mathcal{H}_{X^\prime}) $, $\mathsf{T}_2:\mathcal{L} ( \mathcal{H}_{E_1} \otimes \mathcal{H}_{Y})\rightarrow \mathcal{L} ( \mathcal{H}_{E_1^\prime} \otimes \mathcal{H}_Y \otimes \mathcal{H}_{Y^\prime})$ be (safe) CPTP maps such that for $\sigma_{XX'E_1'E_2'YY'} =(\mathsf{T}_1 \otimes \mathsf{T}_2) (\rho_{XE_1E_2Y})$, we have registers $(XX'YY')$ classical and either  $\Pr(X \ne X')_\sigma=1$ or   $\Pr(Y \ne Y')_\sigma=1$. We say a function $f : \{0,1 \}^n \times \{0,1 \}^n \to \{0,1 \}^m$ is a $(k_1,k_2,\eps)$-quantum secure $2$-source non-malleable extractor against $\qia$ iff for every $\sigma$ as defined above, we have 
\[ \| \sigma_{f(X,Y)f(X',Y')YY' E_1^\prime} - U_m \otimes \sigma_{f(X',Y')YY' E_1^\prime} \|_1 \leq \eps.\]
\suppress{
and\[ \Vert\sigma_{f(X,Y)f(X',Y')XX' E_2^\prime} - U_m \otimes \sigma_{f(X',Y')XX' E_2^\prime} \|_1 \leq \eps.  \] }
\end{definition}

\begin{remark}\label{remark:2nmextiadvmodel}In the Definition~\ref{intro2nmextiadvmodel}, one may ask if we can provide both the registers $E_1'$ and $E_2'$ as side-information to the adversary along with $YY'$. However this may allow adversary to gain complete knowledge of $X,Y$ (since $E_2'$ may contain a copy of $X$ and $E_1'$ may contain a copy of $Y$) making the model uninteresting. Thus we settle on the model as in Definition~\ref{intro2nmextiadvmodel}.
\end{remark}
We have the following corollary of Theorem~\ref{intro:thm:2nmext}.
\begin{corollary}[$2 \nmext$ is a $2$-source non-malleable extractor against $\qia$]\label{intro_corr:add2}
Let the function $2 \nmext$ be from Theorem~\ref{intro:thm:2nmext}. $2 \nmext$ is an $(n-k,n-k,\cO(\eps))$-quantum secure $2$-source non-malleable extractor against $\qia$.
\end{corollary}
Arnon-Friedman, Portmann and Scholz~\cite{APS16} introduced the  quantum Markov adversary ($\qmra$). Informally, $\qmra$ gets the register $\rho_{E}$ as quantum side information in a Markov-chain $\rho_{XEY}$. We propose to incorporate $2 \mhyphen$source non-malleable extractor security against $\qmra$ as follows.
\begin{definition}[$2$-source non-malleable extractor against $\qmra$]\label{intro:def:2nmextmarkov}
Let $\rho_{XEY}$ be a c-q state with registers ($XY$) classical such that  \[\rho_{XEY} = \sum_{t} \Pr(T=t) \ketbra{t} \otimes  \left(\rho^t_{XE_1} \otimes \rho^t_{YE_2} \right)~\footnote{This holds for a Markov-chain $(X-E-Y)_\rho$.} \quad ;  \quad \hmin{X}{E}_\rho \geq k_1 \quad ; \quad \hmin{Y}{E}_\rho \geq k_2,\] where $T$ is classical register over a  finite alphabet. Let  $\mathsf{T}_1: \mathcal{L} (\mathcal{H}_{E_2} \otimes \mathcal{H}_{X} \otimes \mathcal{H}_{T}) \rightarrow \mathcal{L} (\mathcal{H}_{E_2^\prime} \otimes \mathcal{H}_X \otimes \mathcal{H}_{X^\prime} \otimes \mathcal{H}_{T})$, $\mathsf{T}_2: \mathcal{L} (\mathcal{H}_{E_1} \otimes \mathcal{H}_{Y}\otimes \mathcal{H}_{T}) \rightarrow \mathcal{L} (\mathcal{H}_{E_1^\prime} \otimes \mathcal{H}_Y \otimes \mathcal{H}_{Y^\prime}\otimes \mathcal{H}_{T})$ be (safe) CPTP maps such that for $\sigma_{XX'E_1'TE_2'YY'} =(\mathsf{T}_1 \otimes \mathsf{T}_2) (\rho_{XEY})$, we have registers ($XX'TYY'$) classical and either  $\Pr(X \ne X')_\sigma=1$ or   $\Pr(Y \ne Y')_\sigma=1$. We say a function $f : \{0,1 \}^n \times \{0,1 \}^n \to \{0,1 \}^m$ is a $(k_1,k_2,\eps)$-quantum secure $2$-source non-malleable extractor against $\qmra$ iff for every $\sigma$ as defined above, we have  
\[ \| \sigma_{f(X,Y)f(X',Y')YY'E_1^\prime T} - U_m \otimes \sigma_{f(X',Y')YY' E_1^\prime T} \|_1 \leq \eps.\]
\suppress{and 
\[ \| \sigma_{f(X,Y)f(X',Y')XX' E_2^\prime T} - U_m \otimes \sigma_{f(X',Y')XX'  E_2^\prime T} \|_1 \leq \eps. \]}

\end{definition}

\begin{remark}For reasons similar to that of Remark~\ref{remark:2nmextiadvmodel}, in Definition~\ref{intro:def:2nmextmarkov} we do not allow the registers $E_1'$ and $E_2'$ as side-information to the adversary along with $YY'T$. 
\end{remark}

We have the following corollary of Theorem~\ref{intro:thm:2nmext}.

\begin{corollary}[$2 \nmext$ is a $2$-source non-malleable extractor against $\qmra$]\label{intro:corr:add3}
Let the function $2 \nmext$ be from Theorem~\ref{intro:thm:2nmext}. $2 \nmext$ is an $(n-k,n-k,\cO(\eps))$-quantum secure $2$-source non-malleable extractor against $\qmra$.
\end{corollary}
The following are the $t$-tampering extensions of Theorem~\ref{intro:thm:nmext} and Theorem~\ref{intro:thm:2nmext}.
\begin{theorem}[quantum secure $t$-non-malleable extractor]\label{intro:thm:tnmext}
Let $d = \cO(\log^{7}(n/\eps)), t=d^{\Omega(1)}$ and $k = \Omega(d)$. There exists an efficient  non-malleable extractor $t\mhyphen\nmext :\{0,1 \}^n \times \{0,1 \}^d \to \{0,1 \}^{k/4t}$ that is  $(t;k, \cO(\eps))$-quantum secure (see Definition~\ref{tnme}).
\end{theorem}
\begin{theorem}[quantum secure $2$-source $t$-non-malleable extractor]\label{intro:thm:2tnmext}
Let $k = \cO(n^{1/4})$, $\eps =2^{-n^{\Omega(1)}}$ and $t=n^{\Omega(1)}$. There exists an efficient $2$-source non-malleable extractor $t \mhyphen 2\nmext :\{0,1 \}^n \times \{0,1 \}^n \to \{0,1 \}^{n/4t}$ that is  $(t;n-k,n-k, \cO(\eps))$-quantum secure (see Definition~\ref{def:2tsourcenme}).
\end{theorem}

Before going to details of our results, let us sketch of some key elements present in the contsruction of non-malleable extractors.
\subsection*{Some central ingredients}
Similar to classical non-malleable extractors, our theorems regarding quantum-secure  non-malleable extractors are also based on the powerful technique of alternating extraction along with {\em advice generators} and {\em correlation breakers with advice}~\cite{CGL15,Coh16b} that use the clever {\em flip flop}  primitive~\cite{Coh15}. 
{Here, we provide a short overview of these primitives. See \cite{CGL15,Coh16b,Coh15} for more details.}
\begin{itemize}
\item{Alternating extraction:} Alternating extraction is a powerful technique used in the construction of non-malleable extractors, particularly in scenarios involving multiple sources of weak randomness. Consider sources $X,Y$ (with sufficient min-entropy). 
Consider a tampering given by $X \to X'$ and $Y \to Y'$.
The procedure of alternating extraction involves iteratively applying (seeded) extractors to different sources to progressively refine and enhance the uniformity and independence of the extracted randomness. The process can be represented as $\Ext_1 \to \Ext_2 \to \Ext_3 \to \cdots \to \Ext_t $, where each $\Ext_i$ is applied to a part of the source and possibly the output of the previous extractor.
For odd $i$, $\Ext_i$ is applied on source $X$ and the output of the previous extractor, and for even $i$, $\Ext_i$ is applied on source $Y$, (hence the name, alternating extraction) and the output of the previous extractor. By alternating between the sources and carefully managing the interplay between them, this method ensures that the final output is nearly uniform and retains minimal correlation with any tampered inputs. In the quantum setting, alternating extraction becomes even more intricate, as it must account for the complexities introduced by quantum side information and the non-commutative nature of quantum states. 
\item{Flip flop primitive $\ff(X,Y, \text{Advice bit})$:}
The flip-flop primitive uses alternating extraction to break correlations between random variables, leveraging weak sources of randomness and an advice bit.
 The flip-flop function ensures that $\ff(X,Y,0)$ is uniform even given $\ff(X',Y',1)$ (and vice-versa).
\item{Correlation breakers with advice:}
 These are generalizations of the flip-flop primitive, using advice strings instead of advice bits.
 A correlation breaker $ \advcb(X,Y, \alpha)$ ensures uniformity of the output even given correlated inputs with different advice strings $\alpha'$, i.e., $ \advcb(X',Y', \alpha')$.
\item{Advice generators:}
As stated above, correlation breakers need some advice string to work. 
Advice generators produce these strings.
\end{itemize}

 With these main components in place, let us move to the proof overview.
\subsection*{Proof overview} 
As noted before, a key difficulty one faces in analyzing non-malleable extractors in the quantum setting is formulating and manipulating conditional independence relations between random variables since conditioning on quantum side information is tricky. Cohen  and  Vidick~\cite{CV16} attempted to deal with the difficulty by using the formalism of quantum Markov-chains. However, as noted by them, after the adversary tampers with the source, a quantum Markov-chain no longer necessarily remains a quantum Markov-chain. Hence it appeared that a generalization of the quantum Markov-chain model is needed.

We use $\qpass$ instead and consider sources in $\mathcal{Q}$. Additionally, we relate the sources produced by $\qma$ model, $l \mhyphen \qmass$~\cite{ABJO21} (see Definition~\ref{qmadv}) and $(k_1,k_2) \mhyphen\qpass$ (see Lemmas~\ref{lemma:nearby_rho_prime_prime}~and~\ref{lem:qmaqpa}). We note that a source in $\mathcal{Q}$ remains in $\mathcal{Q}$ after adversary tampering, thereby getting over the key difficulty faced by earlier models,
including quantum Markov-chains. Since $\qma$ can simulate other adversary models, we are able to derive as corollaries, the existence of seeded and $2$-source non-malleable extractors in the more standard models of adversaries studied previously in the literature. Similarly, our application for privacy amplification against active quantum adversary is in the standard model of adversary considered in previous works e.g.~\cite{CV16,ACLV18}.

Our proof follows on the lines of~\cite{CGL15,Coh16b}. The key technical lemmas that we use repeatedly in the analysis are, 
\begin{itemize}
    \item a quantum analogue of alternating extraction in a $(k_1,k_2)\mhyphen\qpas$ with approximately uniform seed (Lemma~\ref{lem:2}), and
    \item min-entropy loss under classical interactive communication to ensure enough conditional min-entropy is left for alternating extraction (Lemma~\ref{lem:minentropy}).
\end{itemize}
Lemma~\ref{lem:2} makes use of the powerful Uhlmann's theorem. Lemma~\ref{lem:minentropy} follows using arguments similar to that of a result of Jain and Kundu~\cite{JK21} for quantum communication. In the technique of alternating extraction, we repeatedly extract and generate several approximately uniform random variables. In our analysis, the generation of random variables is viewed as communication protocols (see Protocols~\ref{prot:block1}~to~\ref{prot:Var_GEN(1,1)DiffBefore} for seeded non-malleable extractor analysis). We consider this approach more intuitive and makes the analysis more fine-grained.

As the analysis progresses, several  additional classical random variables need to be generated and considered. We generate them in a manner  such that the requirement of conditional-min-entropy is met for alternating extraction. It is a priori not clear what the sequence of generation of classical random variables should be (for original inputs and tampered inputs) because of the non-commutative nature of quantum side information.
Careful analysis of the classical non-malleable extractor constructions leads us to show that such a sequence of generation of random variables exists. The communication protocols (Protocols~\ref{prot:block1}~to~\ref{prot:Var_GEN(1,1)DiffBefore}) specify the exact sequence in which the additional random variables are generated in various cases.

In the analysis of $2$-source non-malleable extractors, we additionally need a $2$-source extractor
for sources in $\mathcal{Q}$.~\cite{ABJO21} provide security of an inner-product $2$-source extractor for an $l\mhyphen\qmas$ (see Definition~\ref{qmadv}) as long as $l < n$. To use this result we need to relate a $(k_1,k_2)\mhyphen\qpas$ (see Definition~\ref{qmadvk1k2}) with some  $l\mhyphen\qmas$.~\cite{ABJO21} show that a pure state $\sigma_{X\hat{X}NMY\hat{Y}}$ can be generated in the  $l\mhyphen\qmas$ framework if
\[ \hmin{X}{MY\hat{Y}}_\sigma \geq k_1 \quad ; \quad \hminn{Y}{NX\hat{X}}_\sigma \geq k_2.\]
Note that one of the min-entropy bounds is in terms of the modified-conditional-min-entropy $\hminn{\cdot}{\cdot}$. Next we make use of a result that connects $\hmin{\cdot}{\cdot}$ and $\hminn{\cdot}{\cdot}$, i.e for any quantum state $\rho_{XE}$,
\[ \hminn{X}{E}_\rho \leq \hmin{X}{E}_\rho  \leq \hminn{X}{E}_{\rho'} +2 \log (1/\eps) \]
for some $\rho' \approx_\eps \rho$  (see Lemma~\ref{lem:hmin_and_tilde_relation}). 

In the proof, while relating $(k_1,k_2)\mhyphen\qpas$ with some $l\mhyphen\qmas$, we face an additional technical difficulty of correcting a marginal state.  For this we use the {\em substate perturbation lemma} due to~\cite{JK21} and circumvent the issue at the cost of minor loss in parameters (see Lemma~\ref{lemma:nearby_rho_prime_prime}). Thus, we are able to show that any $(k_1,k_2)\mhyphen\qpas$ (say $\sigma$) can be approximated by an $l\mhyphen\qmas$ (say $\sigma^\prime$) for $l \approx 2n-k_1-k_2.$ We also show that $l\mhyphen\qmas$, $\sigma^\prime$  also has the appropriate $\hminn{\cdot}{\cdot}$ bounds, i.e. \[ \hminn{X}{MY\hat{Y}}_{\sigma^\prime} \approx k_1 \quad ; \quad \hminn{Y}{NX\hat{X}}_{\sigma^\prime} \approx k_2.\]Now the security of inner-product in an $l\mhyphen\qmas$ (with appropriate $\hminn{.}{.}$ bounds) from ~\cite{ABJO21}   for $l < n$ implies the security of inner-product in a $(k_1,k_2)\mhyphen\qpas$ for $k_1+k_2 > n$ which is then used in the $2$-source non-malleable extractor construction. The analysis for $2$-source non-malleable extractor then proceeds on similar lines of seeded non-malleable extractor. 

Analysis of the $t$-tampered counterparts proceeds on similar lines by appropriate adjustment of parameters to account for increased communication in communication protocols. 

\subsection*{Comparison with~\cite{ACLV18,ABJO21}}
Both~\cite{ACLV18} and~\cite{ABJO21} have considered the inner-product based non-malleable extractor proposed by Li~\cite{Li12c}. \cite{ACLV18} extends the first step of classical proof, the reduction provided by the non-uniform XOR lemma, to the quantum case. This helps in reducing the task of showing non-malleable extractor property of inner-product to showing security of inner-product in a certain communication game. They then approach the problem of showing security of inner-product in a communication game by using the “reconstruction paradigm” of~\cite{DPVR09} to guess the entire input $X$ from the modified side information.

On the other hand, the work of~\cite{ABJO21} reduces the security of inner-product in a communication game to the security of inner-product against the quantum measurement adversary. In the process, both~\cite{ACLV18} and~\cite{ABJO21} crucially use the combinatorial properties of inner-product. For example,~\cite{ABJO21} heavily uses the pair wise independence property of inner-product.

Note that the communication protocols we use in our analysis are not related to the earlier approach of reduction to a communication game, which is more specific to inner-product.

\subsection*{Comparison with classical constructions}
To the best of our knowledge,~\cite{Li19} provides the construction of seeded non-malleable extractor that works for seed-length $d= \cO \left(\log n + \log^{1+o(1)}(\frac{1}{\eps})\right)$, source min-entropy $k \geq \cO \left( \log \log n + \log(\frac{1}{\eps})\right)$ and output length $m =\Omega(k)$. In the case of $2$-source non-malleable extractor,~\cite{Li19} construction works for sources with min-entropy $k_1,k_2\geq(1-\delta)n$, output length $m= \Omega(n)$ and error $ \eps =2^{-\Omega \left(\frac{n \log \log n}{ \log n}\right)}$.

In the quantum setting,~\cite{ACLV18} provided the first construction of quantum secure seeded non-malleable extractor for seed-length $d= \frac{n}{2}$, source min-entropy $k \geq (\frac{1}{2} + \delta)n$, output length $m =\Omega(n)$ and error $\eps =2^{- \Omega(n)}$. Our work exponentially improves the parameters
for source min-entropy $k \geq  \textsf{poly}\left(\log \left( \frac{n}{\eps} \right)\right)$, seed-length $d=\textsf{poly}\left(\log \left( \frac{n}{\eps} \right)\right)$ and output length $m = \Omega(k)$. In the setting of quantum secure $2$-source non-malleable extractors, we provide the first construction 
for sources with min-entropy $k_1,k_2\geq(1-o(1))n$, output length $m= n/4$  and error $ \eps =2^{-n^{\Omega(1)}}$. We note though we are still far from achieving close to optimal constructions in the  quantum-setting, we hope our techniques find new applications in proving quantum security of other classical non-malleable extractors.

\subsection*{Subsequent works}

\begin{itemize}
    \item ~\cite{ABJ22} (IEEE Trans. Inf. Theory 2024) have extended the connection of~\cite{CG14a} between $2$-source non-malleable extractors and $2$-split-state non-malleable codes (for classical messages) secure against quantum adversaries. They used our quantum secure $2$-source non-malleable extractors to construct the first explicit quantum secure $2$-split-state non-malleable codes (for classical messages) of length $m=n^{\Omega(1)}$, error $\eps=2^{-n^{\Omega(1)}}$ and codeword length $2n$.
    \item Using the techniques introduced in this work, \cite{BBJ23} (QCrypt 2023) constructed a rate 1/2 quantum secure non-malleable randomness encoder. They use this
in a black-box manner, to construct the following:
    \begin{itemize}
        \item rate $1/11$, $3$-split-state non-malleable code for quantum messages
        \item rate $1/3$, $3$-split-state non-malleable code for classical messages against quantum adversaries
        \item rate $1/5$, $2$-split-state non-malleable code for (uniform) classical messages against quantum adversaries.
    \end{itemize}
\item Furthermore, \cite{BGJR23} (QCrypt 2023 and TCC 2024) have constructed 

\begin{itemize}
    \item  rate $1/11$, $2$-split-state non-malleable code for (uniform) quantum messages
    \item $2$-split-state non-malleable code for quantum messages of length $m=n^{\Omega(1)}$, error $\eps=2^{-n^{\Omega(1)}}$ and codeword length $\cO(n)$.
    \item They showed something stronger: the explicit $2$-split-state non-malleable code for quantum messages is, in fact, a $2$-out-of-$2$ non-malleable secret sharing scheme for quantum messages with share size $n$, any message of length at most $n^{\Omega(1)}$, and error $\eps=2^{-n^{\Omega(1)}}$.
\end{itemize} 
\end{itemize}

\subsection*{Organization}
In Section~\ref{sec:prelims} we describe quantum information theoretic and other preliminaries. Section~\ref{sec:claims} contains useful lemmas and claims. We describe the construction and security analysis of seeded non-malleable extractor in Section~\ref{sec:nmext}, of $2$-source non-malleable extractor in Section~\ref{sec:2nm} and of $t$-tampered versions of these in Appendix~\ref{sec:tnmext} and~\ref{sec:2tnm} respectively. The communication protocols used in all the analysis appear in the Appendix~\ref{sec:communication_protocols}. 
We provide a proof of PA against active adversaries in Appendix~\ref{sec:PA}.
\section{Preliminaries}
\label{sec:prelims}
Let $n,m,d,t$  represent positive integers and $l, k, k_1, k_2, \delta, \gamma, \eps \geq 0$ represent reals.
\subsection*{Quantum information theory} All the logarithms are evaluated to the base $2$. Let $\X, \Y, \Z$ be finite sets (we only consider finite sets in this paper). Let $\vert \X \vert$ represent the size of $\X$, that is the number of elements in $\X$. For a {\em random variable} $X \in \X$, we use $X$ to denote both the random variable and its distribution~\footnote{Some works use $P_X$ to denote distribution of $X$, however we use this non-standard notation for brevity.}, whenever it is clear form the context. We use $x \leftarrow X$ to denote $x$ drawn according to $X$. We call random variables $X, Y$, {\em copies} of each other if $\Pr[X=Y]=1$.  For a random variable $X \in \{0,1 \}^n$ and $d\leq n$, let $\pre(X, d)$ represent the first $d$ bits of $X$. Let $U_d$ represent the uniform distribution over $\{0,1 \}^d$. Let $Y^1, Y^2, \ldots, Y^t$ be random variables. We denote the joint random variable  $Y^1 Y^2 \ldots Y^t$ by $Y^{[t]}$.
Similarly for any subset $\mathcal{S} \subseteq [t]$, we use $Y^{\mathcal{S}}$ to denote the joint random variable comprised of all the $Y^s$ such that $s \in \mathcal{S}$.

Consider a finite-dimensional Hilbert space $\cH$ endowed with an inner-product $\langle \cdot, \cdot \rangle$ (we only consider finite-dimensional Hilbert-spaces). A quantum state (or a density matrix or a state) is a positive semi-definite operator on $\cH$ with trace value  equal to $1$. It is called {\em pure} iff its rank is $1$.  Let $\ket{\psi}$ be a unit vector on $\cH$, that is $\langle \psi,\psi \rangle=1$.  With some abuse of notation, we use $\psi$ to represent the state and also the density matrix $\ketbra{\psi}$, associated with $\ket{\psi}$. Given a quantum state $\rho$ on $\cH$, {\em support of $\rho$}, called $\text{supp}(\rho)$ is the subspace of $\cH$ spanned by all eigenvectors of $\rho$ with non-zero eigenvalues.
 
A {\em quantum register} $A$ is associated with some Hilbert space $\cH_A$. Define $\vert A \vert := \log\left(\dim(\cH_A)\right)$. Let $\mathcal{L}(\cH_A)$ represent the set of all linear operators on the Hilbert space $\cH_A$. For operators $O, O'\in \cL(\cH_A)$, the notation $O \leq O'$ represents the L\"{o}wner order, that is, $O'-O$ is a positive semi-definite operator. We denote by $\mathcal{D}(\cH_A)$, the set of all quantum states on the Hilbert space $\cH_A$. State $\rho$ with subscript $A$ indicates $\rho_A \in \mathcal{D}(\cH_A)$. If two registers $A,B$ are associated with the same Hilbert space, we shall represent the relation by $A\equiv B$. For two states $\rho, \sigma$, we let $\rho \equiv \sigma$ represent that they are identical as states (potentially in different registers). Composition of two registers $A$ and $B$, denoted $AB$, is associated with the Hilbert space $\cH_A \otimes \cH_B$.  For two quantum states $\rho\in \mathcal{D}(\cH_A)$ and $\sigma\in \mathcal{D}(\cH_B)$, $\rho\otimes\sigma \in \mathcal{D}(\cH_{AB})$ represents the tensor product ({\em Kronecker} product) of $\rho$ and $\sigma$. The identity operator on $\cH_A$ is denoted $\id_A$. Let $U_A$ denote maximally mixed state in $\cH_A$. Let $\rho_{AB} \in \mathcal{D}(\cH_{AB})$. Define
$$ \rho_{B} \defeq \tr_{A}{\rho_{AB}} \defeq \sum_i (\bra{i} \otimes \id_{B})
\rho_{AB} (\ket{i} \otimes \id_{B}) , $$
where $\{\ket{i}\}_i$ is an orthonormal basis for the Hilbert space $\cH_A$.
The state $\rho_B\in \mathcal{D}(\cH_B)$ is referred to as the marginal state of $\rho_{AB}$ on the register $B$. Unless otherwise stated, a missing register from subscript in a state will represent partial trace over that register. Given $\rho_A\in\mathcal{D}(\cH_A)$, a {\em purification} of $\rho_A$ is a pure state $\rho_{AB}\in \mathcal{D}(\cH_{AB})$ such that $\tr_{B}{\rho_{AB}}=\rho_A$. Purification of a quantum state is not unique.
Suppose $A\equiv B$. Given $\{\ket{i}_A\}$ and $\{\ket{i}_B\}$ as orthonormal bases over $\cH_A$ and $\cH_B$ respectively, the \textit{canonical purification} of a quantum state $\rho_A$ is $\ket{\rho_A} \defeq (\rho_A^{\frac{1}{2}}\otimes\id_B)\left(\sum_i\ket{i}_A\ket{i}_B\right)$. \suppress{For a classical register $A$, we refer to the canonical purification register as a \textit{copy} of classical register $A$.}

A quantum {map} $\cE: \mathcal{L}(\cH_A)\rightarrow \mathcal{L}(\cH_B)$ is a completely positive and trace preserving (CPTP) linear map. A {\em Hermitian} operator $H:\cH_A \rightarrow \cH_A$ is such that $H=H^{\dagger}$. Let $\Lambda_{+}(H)$ denote the set of eigenvectors with positive eigenvalues, i.e. $\Lambda_{+}(H)= \lbrace v \in \cH_A \ :\ Hv=\lambda_v v, \ \lambda_v > 0 \rbrace$. Let $H_+$ be the vector space generated by $\Lambda_{+}(H)$. We say that $H_+$ is the positive part of $H$. A projector $\Pi \in  \mathcal{L}(\cH_A)$ is a Hermitian operator such that $\Pi^2=\Pi$. \suppress {We use projector $\overline{\Pi}$ to mean  $\id-\Pi.$} A {\em unitary} operator $V_A:\cH_A \rightarrow \cH_A$ is such that $V_A^{\dagger}V_A = V_A V_A^{\dagger} = \id_A$. The set of all unitary operators on $\cH_A$ is  denoted by $\mathcal{U}(\cH_A)$. An {\em isometry}  $V:\cH_A \rightarrow \cH_B$ is such that $V^{\dagger}V = \id_A$ and $VV^{\dagger} = \id_B$. A {\em POVM} element is an operator $0 \le M \le \id$. We use the shorthand $\overline{M} \defeq \id - M$, where $\id$ is clear from the context. We use shorthand $M$ to represent $M \otimes \id$, where $\id$ is clear from the context.

\begin{definition}[Classical register in a pure state]\label{def:classicalinpurestate}Let $\X$ be a set. A {\em classical-quantum} (c-q) state $\rho_{XE}$ is of the form \[ \rho_{XE} =  \sum_{x \in \X}  p(x)\ket{x}\bra{x} \otimes \rho^x_E , \] where ${\rho^x_E}$ are states.

Let $\rho_{XEA}$ be a pure state. We call $X$ a classical register in $\rho_{XEA}$, if $\rho_{XE}$ (or $\rho_{XA}$) is a c-q state. We identify random variable $X$ with the register $X$, with $\Pr(X=x) =p(x)$.
\suppress{
In a pure state $\rho_{XEA}$ in which $\rho_{XE}$ (or $\rho_{XA}$) is c-q, we call $X$ a classical register and identify random variable $X$ with it with $\Pr(X=x) =p(x)$.}
\end{definition}

\begin{definition}[Copy of a classical  register]\label{def:copyofaclassicalregister}
Let $\rho_{X\hat{X}E}$ be a pure state with $X$ being a classical register in $\rho_{X\hat{X}E}$ (see Definition~\ref{def:classicalinpurestate}) taking values in $\cX$. Similarly, let $\hat{X}$ be a classical register in $\rho_{X\hat{X}E}$ taking values in $\cX$. Let $\Pi_{\mathsf{Eq}} = \sum_{x \in \cX} \ketbra{x} \otimes \ketbra{x}$ be the \emph{equality} projector acting on the registers $X\hat{X}$. We call $X$ and $\hat{X}$ copies of each other (in the computational basis) if $\tr\left(\Pi_{\mathsf{Eq}} \rho_{X\hat{X}}\right) =1$.
\end{definition}

\begin{definition}[Conditioning] \label{def:conditioning}
Let  
\[ \rho_{XE} =  \sum_{x \in \{0,1\}^n}  p(x)\ket{x}\bra{x} \otimes \rho^x_E , \]
be a c-q state. For an event $\mathcal{S} \subseteq \{0,1\}^n$, define  $$\Pr(\mathcal{S})_\rho \defeq  \sum_{x \in \mathcal{S}} p(x) \quad ; \quad (\rho|X\in \mathcal{S})\defeq \frac{1}{\Pr(\mathcal{S})_\rho} \sum_{x \in \mathcal{S}} p(x)\ket{x}\bra{x} \otimes \rho^x_E.$$
We sometimes shorthand $(\rho|X\in \mathcal{S})$ as $(\rho|\mathcal{S})$ when the register $X$ is clear from the context. 

Let $\rho_{AB}$ be a state with $|A|=n$. We define 
$(\rho|A \in \mathcal{S}) \defeq (\sigma|\mathcal{S})$, where $\sigma_{AB}$ is the c-q state obtained by measuring the register $A$ in $\rho_{AB}$ in the computational basis. In case $\mathcal{S}=\{s\}$ is a singleton set, we shorthand $(\rho|A = s) \defeq \tr_A (\rho|A =s)$.
\end{definition}

\begin{definition}[Extension] \label{def:extension} Let $$\rho_{XE}=  \sum\limits_{x \in \{0,1\}^n}  p(x)\ket{x}\bra{x} \otimes \rho^x_E,$$
be a c-q state. For a function $Z:\cX \rightarrow \cZ$, define the following extension of $\rho_{XE}$, 
\[ \rho_{ZXE} \defeq  \sum_{x\in \cX}  p(x) \ket{Z(x)}\bra{Z(x)} \otimes \ket{x}\bra{x} \otimes  \rho^{x}_E.\]
\end{definition} 

\begin{definition}[Safe maps] \label{def:safe}
We call an isometry $V: \cH_X \otimes \cH_A \rightarrow \cH_X \otimes \cH_B$, {\em safe} on $X$ iff there is a collection of isometries $V_x: \cH_A\rightarrow \cH_B$ such that the following holds.  For all states $\ket{\psi}_{XA} = \sum_x \alpha_x \ket{x}_X \ket{\psi^x}_A$,
$$V  \ket{\psi}_{XA} =  \sum_x \alpha_x \ket{x}_X V_x \ket{\psi^x}_A.$$
We call a CPTP map $\Phi: \mathcal{L}( \cH_X \otimes \cH_A) \rightarrow \mathcal{L}(\cH_X \otimes \cH_B)$, {\em safe} on classical register $X$ iff there is a collection of CPTP maps $\Phi_x: \mathcal{L}(\cH_A)\rightarrow \mathcal{L}(\cH_B)$ such that the following holds.  For all c-q states $\rho_{XA} = \sum_x \Pr(X=x)_{\rho} \ketbra{x} \otimes  \rho^x_A$,
$$\Phi({\rho}_{XA}) =  \sum_x \Pr(X=x)_{\rho} \ketbra{x} \otimes \Phi_x( \rho^x_A).$$
\end{definition}
All isometries (or in general CPTP maps) considered in this paper are safe on classical registers that they act on. CPTP maps applied by adversaries can be assumed w.l.o.g as safe on classical registers, by the adversary first making a (safe) copy of classical registers and then proceeding as before. This does not reduce the power of the adversary. 

For a pure state $\rho_{XEA}$ (with $X$ classical) and a function $Z:\cX \rightarrow \cZ$, define $\rho_{Z\hat{Z}XEA}$ to be a pure state extension of $\rho_{XEA}$ generated via a safe isometry $V: \cH_X \rightarrow \cH_X \otimes \cH_Z \otimes \cH_{\hat{Z}}$ ($Z$ classical with copy $\hat{Z}$).


\begin{definition}
\label{def:infoquant}    
\begin{enumerate}
\item For $p \geq 1$ and matrix $A$,  let $\| A \|_p$ denote the {\em Schatten} $p$-norm.  
\item For $p \geq 1: ~ \| A \|_p  = (\tr(A^\dagger A)^{p/2})^{\frac{1}{p}}.$

\item  For states $\rho,\sigma: \Delta(\rho , \sigma) \defeq \frac{1}{2} \|\rho - \sigma\|_1$.

\item {\bf Fidelity:}  For states $\rho,\sigma: ~\F(\rho,\sigma)\defeq\|\sqrt{\rho}\sqrt{\sigma}\|_1.$ 

\item {\bf Bures metric:}  For states $\rho,\sigma: \Delta_B(\rho,\sigma)\defeq \sqrt{1-\F(\rho,\sigma)}.$ We write $\rho \approx_\eps \sigma$ to denote $\Delta_B(\rho, \sigma) \le \eps$. Being a metric, it satisfies the triangle inequality. 

\item Define $d(X)_\rho \defeq \Delta_B(\rho_X,U_X)$ and  $d(X|Y )_\rho \defeq \Delta_B(\rho_{XY}, U_X \otimes \rho_Y)$. 

\item {\bf Max-divergence (\cite{Datta09}, see also~\cite{JainRS02}):}\label{dmax}  For states $\rho,\sigma$ such that $\supp(\rho) \subset \supp(\sigma)$, $$ \dmax{\rho}{\sigma} \defeq  \min\{ \lambda \in \mathbb{R} :   \rho  \leq 2^{\lambda} \sigma \}.$$ 
\item {\bf Min-entropy and conditional-min-entropy:}  For a state $\rho_{XE}$, the min-entropy of $X$ is defined as,
 $$ \hminone{X}_\rho \defeq - \dmax{\rho_{X}}{\id_X} .$$
 The conditional-min-entropy of $X$, conditioned on $E$, is defined as,
 $$ \hmin{X}{E}_\rho \defeq - \inf_{\sigma_E \in  \mathcal{D}(\cH_{E}) } \dmax{\rho_{XE}}{\id_X \otimes \sigma_E}.$$
 The modified-conditional-min-entropy of $X$, conditioned on $E$, is defined as,
 $$ \hminn{X}{E}_\rho \defeq - \dmax{\rho_{XE}}{\id_X \otimes \rho_E}.$$
 

 \suppress{
\item {\bf Modified-conditional-min-entropy:}  For a state $\rho_{XE}$, the modified-min-entropy of $X$ conditioned on $E$ is defined as, $$ \hminn{X}{E}_\rho = - \dmax{\rho_{XE}}{\id_X \otimes \rho_E}    .$$}

\end{enumerate}
\end{definition}
\suppress{
\begin{fact}[Min-entropy and guessing probability~\cite{KRS09}]\label{fact:guess_minent_op}
For a c-q state $\rho_{XE} \in \mathcal{D}(\cH_{XE})$, the guessing probability is defined as the probability to correctly guess $X$ with the optimal strategy to measure $E$, i.e. \[ p_{guess}(X|E)_\rho =  \sum_x P_X(x) \tr(M_x \rho^x_E) \] where ${M_x}$ is a positive operator-valued measure (POVM) on quantum register $E$. Then the guessing probability is related to the min-entropy by \[ p_{guess}(X|E)_\rho = 2^{- \hmin{X}{E}_\rho } \]
\end{fact}
}
For the facts stated below without citation, we refer the reader to standard text books~\cite{NielsenC00,WatrousQI}.
\begin{fact}[Uhlmann's Theorem~\cite{uhlmann76}]
\label{uhlmann}
Let $\rho_A,\sigma_A\in \mathcal{D}(\cH_A)$. Let $\rho_{AB}\in \mathcal{D}(\cH_{AB})$ be a purification of $\rho_A$ and $\sigma_{AC}\in\mathcal{D}(\cH_{AC})$ be a purification of $\sigma_A$. 
There exists an isometry $V$ (from a subspace of $\cH_C$ to a subspace of $\cH_B$) such that,
\[ \Delta_B\left( \ketbra{\rho}_{AB}, \ketbra{\theta}_{AB}) =  \Delta_B(\rho_A,\sigma_A\right) ,\]
 where $\ket{\theta}_{AB} = (\id_A \otimes V) \ket{\sigma}_{AC}$.
\end{fact}
\begin{fact}[\cite{JainRS02}]
\label{rejectionsampling}
Let $\rho_{A'B}, \sigma_{AB}$ be pure states such that $\dmax{\rho_B}{\sigma_B} \leq k$. Let Alice and Bob share $\sigma_{AB}$. There exists an isometry $V: \cH_A \rightarrow \cH_{A'} \otimes \cH_C$ such that,
\begin{enumerate}
\item  $(V \otimes \id_B) \sigma_{AB}(V \otimes \id_B)^\dagger  = \phi_{A'BC}$, where $C$ is a single qubit register. 
\item Let $C$ be the outcome of measuring $\phi_C$ in the standard basis. Then $\Pr(C=1) \geq 2^{-k}$.
\item Conditioned on outcome $C=1$, the state shared between Alice and Bob is $\rho_{A'B}$.  
\end{enumerate}
\end{fact}

\begin{fact}[\cite{CLW14}]
	\label{fact102}  
	 Let $\cE :    \mathcal{L} (\cH_M ) \rightarrow   \mathcal{L}(\cH_{M'} )$ be a CPTP map and let $\sigma_{XM'} =(\id \otimes \cE)(\rho_{XM}) $. Then,  $$ \hmin{X}{M'}_\sigma  \geq \hmin{X}{M}_\rho  .$$
Above is equality if $\cE$ is a map corresponding to an isometry.
\end{fact}

\begin{fact}[Lemma B.3. in~\cite{DPVR09}]
\label{fact2}  
For a c-q state $\rho_{ABC}$ (with $C$ classical),
$$\hmin{A}{BC}_\rho \geq \hmin{A}{B}_\rho - \vert C \vert.$$

\end{fact}
\suppress{
\begin{fact}[Lemma B.3. in~\cite{DPVR09}]\label{}
     For any state $\rho_{ABZ}$ such that register $Z$ is classical, we have 
     
     \[  \hmin{A}{BZ}_\rho \geq \hmin{A}{B}_\rho - \log ( \dim(\supp(\rho_Z))).\]
\end{fact}\mycomment{Naresh : Use above fact instead of Fact~\ref{fact2} in the paper.}
}

\begin{fact}[]
\label{traceavg}
Let $\rho_{XE},\sigma_{XE}$ be two c-q states. Then,
\begin{itemize}
    \item $ \| \rho_{XE}-\sigma_{XE} \|_1 \geq   \E_{x \leftarrow \rho_X } \| \rho^x_{E}-\sigma^x_{E} \|_1. $
     \item $ \Delta_B( \rho_{XE},\sigma_{XE} ) \geq   \E_{x \leftarrow \rho_X } \Delta_B( \rho^x_{E}, \sigma^x_{E} ). $
\end{itemize}
The above inequalities are equalities iff $\rho_X = \sigma_X$.
\end{fact}

\begin{fact}[\cite{FvdG06}]
\label{fidelty_trace}
Let $\rho,\sigma$ be two states. Then,
\[  1-\F(\rho,\sigma) \leq \Delta(\rho , \sigma) \leq \sqrt{ 1-\F^2(\rho,\sigma)} \quad ; \quad \Delta_B^2(\rho,\sigma) \leq \Delta(\rho , \sigma) \leq  \sqrt{2}\Delta_B(\rho,\sigma).  \]

\end{fact}
\begin{fact}[Data-processing]
\label{fact:data}
Let $\rho, \sigma$  be two states and $\cE$ be a CPTP map. Then 
\begin{itemize}
    \item $ \Delta ( \cE(\rho)  , \cE(\sigma))  \le \Delta (\rho  , \sigma).$    
     \item $ \Delta_B ( \cE(\rho)  , \cE(\sigma))  \le \Delta_B (\rho  , \sigma).$    
    \item  $\dmax{ \cE(\rho) }{ \cE(\sigma) }  \le \dmax{\rho}{ \sigma} .$    
\end{itemize}
Above are equalities if $\cE$ is a map corresponding to an isometry.
\end{fact}

\begin{fact}\label{fact:Conjugation} Let $M,A \in \mathcal{L}(\cH)$. 
If $A \geq 0$ then $M^{\dagger} A M \geq 0$. 
\end{fact}

\begin{fact}
\label{measuredmax}
Let $\rho_{AB} \in \mathcal{D}(\cH_A \otimes \cH_B)$ be a state and $M \in \cL(\cH_B)$ such that $M^\dagger M \leq \id_B$. Let $\hat{\rho}_{AB}= \frac{M \rho_{AB} M^\dagger}{\tr{M \rho_{AB} M^\dagger}}$. Then, 
$$\dmax{\hat{\rho}_A}{\rho_A} \leq \log \left(\frac{1}{\tr{M \rho_{AB} M^\dagger}}\right).$$
\end{fact}
\begin{fact}[Substate Perturbation Lemma (Lemma~$9$ in~\cite{JK21} )] \label{fact:substate_perturbation}
Let $\sigma_{XB}$,  $\psi_X$ and $\rho_B$ be states such that,
\[  \quad   \sigma_{XB} \leq 2^c \left( \psi_X \otimes \sigma_{B}\right) \quad ; \quad \Delta_B\left(\sigma_B, \rho_B\right) \leq \delta_1  .\] 
For any $\delta_0 > 0$, there exists state $\rho^\prime_{XB}$ satisfying
\[\Delta_B\left(\rho^\prime_{XB}, \sigma_{XB} \right)\leq \delta_0 +\delta_1 \quad ;\quad \rho^\prime_{XB} \leq 2^{c+1} \left(1+ \frac{4}{\delta_0^2}  \right) \psi_X \otimes \rho_B \quad ;\quad \rho^\prime_{B}=\rho_{B} \tab\footnote{{\cite{JK21} does not explicitly mention in the statement of the substate perturbation lemma that 
 $\rho^\prime_{B}=\rho_{B}$.
 But it can be easily verified from their proof that this holds.
 }The statement in~\cite{JK21} is more general and is stated for {\em purified distance}, however it holds for any fidelity based distance including the Bures metric.}.\]
\end{fact}

\begin{fact}
\label{fact:close}
Let $\rho_{}, \sigma \in \mathcal{D}(\cH_A)$ be two states  and $M \in \cL(\cH_A)$ such that  $M^\dagger M \leq \id_A$. Then, 
\[ \vert \tr{M \rho_{} M^\dagger} - \tr{M \sigma_{} M^\dagger} \vert  \leq  \frac{\Vert \rho -\sigma \Vert_1}{2}.\]
\end{fact}

\begin{fact}[Gentle Measurement Lemma~\cite{book_Wilde}] \label{fact:gentle_measurement}

Let $\rho \in \mathcal{D}(\cH_A)$ be a state and $M \in \cL(\cH_A)$ such that $M^\dagger M \leq \id_A$ and $\tr(M \rho M^\dagger)
\geq 1-\eps$. Let $\hat{\rho}= \frac{M \rho M^\dagger}{\tr{M \rho M^\dagger}}$. Then, 
$\Delta_B\left(\rho , \hat{\rho}\right) \leq {\sqrt{\eps}}$.
\suppress{

Let $\rho \in \mathcal{D}_{\mathcal{H}}$ and $\Lambda$ be a measurement operator such that $\tr(\Lambda \rho)
\geq 1-\eps$ and $\rho^\prime$ be the post measurement state $\rho^\prime = \dfrac{\sqrt{\Lambda} \rho \sqrt{\lambda}}{\tr{\Lambda \rho}}$. Then,
$\Vert \rho - \rho^\prime\Vert_1 \leq \mathcal{O}(\sqrt{\eps})$.}

\end{fact}

\begin{fact}[Corollary 5.2 in~\cite{CGL15}]\label{fact:samp}
  For any constant $\delta  \in ( 0,1)$, there exist constants $\alpha,\beta < 1/14$ such that for all positive integers $\nu, r, t$, with $r \geq \nu^{\alpha}$ and $t=\cO(\nu^{\beta})$ the following holds.  
  
  There exists a polynomial time computable function 
  $\samp : \{0,1 \}^r \to [\nu]^{t}$, such that for any set $\mathcal{S} \subset [\nu]$ of size $\delta \nu$,
  $$ \Pr( \vert \samp(U_r) \cap \mathcal{S}  \vert \geq 1 ) \geq 1-2^{- \Omega(\nu^{\alpha})}.$$
\end{fact}

\begin{definition} Let $M=2^m$. The inner-product function, $\IP^{n}_{M}: \mathbb{F}_{M}^{n} \times \mathbb{F}_{M}^n \rightarrow \mathbb{F}_{M}$ is defined as follows:\[\IP^{n}_{M}(x,y)=\sum_{i=1}^{n} x_i y_i,\]
where the operations are over the field $\mathbb{F}_{M}.$
\end{definition}

\begin{definition}[Markov-chain]\label{def:markovchain}
A state $\rho_{XEY}$ forms a Markov-chain (denoted $(X-E-Y)_\rho$) iff $\condmutinf{X}{Y}{E}_\rho=0$. 
\end{definition}

\begin{fact}[\cite{H04}]\label{fact:markov}
    A Markov-chain $(X-E-Y)_\rho$ can be decomposed as follows: $$\rho_{XEY} = \sum_{t} \Pr(T=t) \ketbra{t} \otimes \left(\rho^t_{XE_1} \otimes \rho^t_{YE_2}\right),$$ where $T$ is classical register over a  finite alphabet. 
\end{fact}  

\begin{fact}[\cite{anshu2021oneshot}]\label{fact:markov2}
    For a Markov-chain $(X-E-Y)_\rho$, there exists a CPTP map $\Phi:\mathcal{L}( \cH_E ) \rightarrow \mathcal{L}( \cH_E \otimes \cH_Y)$ such that $\rho_{XEY} =({\id_X} \otimes \Phi) \rho_{XE}$.
\end{fact}  
{
\begin{fact}[Corollary 5.5 in \cite{WatrousQI}]
	\label{measurediso}
	Let $\rho_{AB} \in \mathcal{D}(\cH_A \otimes \cH_B)$ be a state and $V_B : \mathcal{L} (\cH_{B}) \rightarrow   \mathcal{L}(\cH_{B'} \otimes \cH_{C})$ be an isometry such that $\vert C \vert =1$. Let $\sigma_{AB'C} = (\id_A \otimes V_B) \rho_{AB} (\id_A \otimes V_B)^\dagger $ and  $\Phi_{AB'} = (\sigma_{AB'C} \vert C=1)$. There exists an operator $M_B$  such that $0 \le  M^\dagger_{B}M_{B} \le \id_{B}$ and 
 \[\Phi_{AB'} =\frac{(\id_A \otimes M_{B} ) \rho_{AB} (\id_A \otimes M_{B} )^\dagger }{\tr(\id_A \otimes M_{B} ) \rho_{AB} (\id_A \otimes M_{B} )^\dagger} \quad ; \quad \Pr[C=1]_{{\sigma}} = \tr \left(M_B \rho_B M_B^\dagger\right). \]
\end{fact} }

\subsection*{Extractors}
Throughout the paper we use extractor to mean seeded extractor unless stated otherwise. 
\begin{definition}[quantum secure extractor]
\label{qseeded}
	An $(n,d,m)$-extractor $\Ext : \{0,1\}^n \times \{0,1\}^d \to \{0,1\}^m$  is said to be $(k,\eps)$-quantum secure if for every state $\rho_{XES}$, such that $\Hmin(X|E)_\rho \geq k$ and $\rho_{XES} = \rho_{XE} \otimes U_d$, we have 
	$$  \| \rho_{\Ext(X,S)E} - U_m \otimes \rho_{E} \|_1 \leq \eps.$$
	In addition, the extractor is called strong if $$  \| \rho_{\Ext(X,S)SE} - U_m \otimes U_d \otimes \rho_{E} \|_1 \leq \eps .$$
	$S$ is referred to as the {\em seed} for the extractor.
	\end{definition}

\begin{fact}[\cite{DPVR09}~\cite{CV16}]
    \label{fact:extractor}\suppress{There exists an explicit $(k,\eps)$-quantum secure strong $(n,d,m)$-extractor $\Ext : \{ 0,1\}^n \times  \{ 0,1\}^d \to  \{ 0,1\}^m$ for parameters  $m= k-4 \log(1/\eps) - \cO(1)$ and $d = \cO( \log^2(n/\eps) \log m )$.} There exists an explicit $(2m,\eps)$-quantum secure strong $(n,d,m)$-extractor $\Ext : \{ 0,1\}^n \times  \{ 0,1\}^d \to  \{ 0,1\}^m$ for parameters  $d = \cO( \log^2(n/\eps) \log m )$.
\end{fact} 

\suppress{
\begin{fact}[\cite{DPVR09}]
    \label{fact:extractor1}
    For any positive integers $n,m$ and $\eps>0$, there exists an explicit $(2m,\eps)$-quantum secure strong $(n,d,m)$-extractor $\Ext : \{ 0,1\}^n \times  \{ 0,1\}^d \to  \{ 0,1\}^m$ for parameters  $d = O( \log^2(n/\eps) \log m )$.
\end{fact} 
}
\suppress{
  \begin{definition}[Map with no fixed points]\label{mapnf}We say that a CPTP map $\cE: \mathcal{L}(\cH_Y \otimes \cH_E )\rightarrow \mathcal{L}(\cH_Y \otimes \cH_{Y'} \otimes \cH_{E'})$ has no fixed points if for all c-q states $\rho_{YE} \in  \cD(\cH_{YE})$ ($Y$ is classical), $\sigma_{YY'E}= \cE( \rho_{YE} )$ is such that $Y,Y'$ are classical registers and $Y' \ne Y$.
\end{definition}}

\begin{definition}[$l\mhyphen\qmas$~\cite{ABJO21}]\label{qmadv}	Let $\tau_{X\hat{X}}$, $\tau_{Y\hat{Y}}$ be the canonical purifications of independent and uniform sources $X, Y$ respectively. Let $\tau_{NM}$ be a pure state. Let 
$$ \theta_{X\hat{X}NMY\hat{Y}}= \tau_{X\hat{X}} \otimes \tau_{NM} \otimes \tau_{Y\hat{Y}}.$$
Let $U : \cH_{X} \otimes \cH_{N} \rightarrow   \cH_{X} \otimes \cH_{N'} \otimes \cH_{A}$ and $V : \cH_Y \otimes \cH_{M} \rightarrow   \cH_{Y} \otimes \cH_{M'} \otimes \cH_{B}$ be isometries
such that registers $A, B$ are single qubit registers. Let $$\rho_{X\hat{X}AN'M'BY\hat{Y}} = (U \otimes V)\theta_{X\hat{X}NMY\hat{Y}}(U \otimes V)^\dagger,$$
and 
 $$l = \log\left( \frac{1}{ \Pr(A=1, B=1)_{\rho}}  \right) \quad ; \quad \sigma_{X\hat{X}N'M'Y\hat{Y}} = (\rho_{X\hat{X}AN'M'BY\hat{Y}} \vert A=1,B=1).$$
 We call $\sigma_{X\hat{X}N'M'Y\hat{Y}}$ an $l\mhyphen\qmas$ .
\end{definition}

\begin{definition}[$(k)\mhyphen\qpas$]\label{qmadvk}
We call a pure state $\sigma_{X\hat{X}NMY\hat{Y}}$, with $(XY)$ classical and $(\hat{X}\hat{Y})$ copy of $(XY)$,  a  $(k) \mhyphen \qpas$ if
\[ \hmin{X}{MY\hat{Y}}_\sigma \geq k \quad ; \quad \sigma_{X\hat{X}NY} = \sigma_{X \hat{X} N} \otimes U_Y.\]
\end{definition}
\begin{definition}[$(k)\mhyphen\nmas$]\label{kqnmadversarydef}
          Let $\sigma_{X\hat{X}NMY\hat{Y}}$ be a $(k) \mhyphen \qpas$. 
          Let $V: \cH_Y \otimes \cH_M \rightarrow \cH_Y \otimes \cH_{Y'} \otimes  \cH_{\hat{Y}'} \otimes \cH_{M'}$ be an isometry such that for $\rho = V \sigma V^\dagger$, we have $Y'$ classical (with copy $\hat{Y}'$) and $\Pr(Y \ne Y')_\rho =1.$ We call state $\rho$ a $(k)\mhyphen\nmas$.
\end{definition}
\begin{remark} In Definition~\ref{kqnmadversarydef} (and in similar such definitions) previous works consider the notion of  \emph{CPTP maps with no fixed points}. However we replace it with the condition $\Pr(Y \ne Y')_\rho=1$,  which suffices for our purposes. 
\end{remark}
We require the non-malleable extractor to extract from every $(k)\mhyphen\nmas$, chosen by the adversary $\nma$ (short for quantum non-malleable adversary). We follow similar convention for $2 \mhyphen$source non-malleable extractors and their extensions to $t\mhyphen$tampering setting.

\begin{definition}[quantum secure non-malleable extractor]\label{nme}
		An $(n,d,m)$-non-malleable extractor $\nmext : \{0,1\}^{n} \times \{0,1\}^{d} \to \{0,1\}^m$ is $(k,\eps)$-secure against $\nma$ 
		if for every  $(k)\mhyphen\nmas$ $\rho$ (chosen by the adversary $\nma$),
	$$  \| \rho_{ \nmext(X,Y)\nmext(X,Y') YY'M'} - U_m \otimes \rho_{ \nmext(X,Y') YY'M'} \|_1 \leq \eps.$$
\end{definition}
\begin{definition}[$(k_1,k_2)\mhyphen\qpas$]\label{qmadvk1k2}
We call a pure state $\sigma_{X\hat{X}NMY\hat{Y}}$, with $(XY)$ classical and $(\hat{X}\hat{Y})$ copy of $(XY)$,  a  $(k_1,k_2)\mhyphen\qpas$ if 
\[ \hmin{X}{MY\hat{Y}}_\sigma \geq k_1 \quad ; \quad \hmin{Y}{NX\hat{X}}_\sigma \geq k_2.\]
\end{definition}
\begin{definition}[$(k_1,k_2)\mhyphen\nmas$]\label{def:2source-qnmadversarydef}
     Let $\sigma_{X\hat{X}NMY\hat{Y}}$ be a $(k_1,k_2)\mhyphen\qpas$.
     Let $U: \cH_X \otimes \cH_N \rightarrow \cH_X \otimes \cH_{X^\prime} \otimes  \cH_{\hat{X}'} \otimes \cH_{N^\prime}$ and $V: \cH_Y \otimes \cH_M \rightarrow \cH_Y \otimes \cH_{Y'} \otimes  \cH_{\hat{Y}'} \otimes \cH_{M'}$ be isometries  such that for $\rho = (U \otimes V)\sigma(U \otimes V)^\dagger,$ we have $(X'Y')$ classical (with copy $\hat{X}'\hat{Y}'$) and, 
       $$\Pr(Y \ne Y^\prime)_\rho =1 \quad or \quad \Pr(X \ne X^\prime)_\rho =1.$$ 
       We call state $\rho$ a $(k_1,k_2)\mhyphen\nmas$.
\end{definition}
\begin{definition}[quantum secure $2$-source non-malleable extractor]\label{def:2nme}
		An $(n,n,m)$-non-malleable extractor $2\nmext : \{0,1\}^{n} \times \{0,1\}^{n} \to \{0,1\}^m$ is $(k_1,k_2,\eps)$-secure against $\nma$ if for every $(k_1,k_2)\mhyphen\nmas$ $\rho$ (chosen by the adversary $\nma$),
	$$  \Vert \rho_{ 2\nmext(X,Y)2\nmext(X^\prime,Y^\prime) Y  Y^\prime M^\prime} - U_m \otimes \rho_{ 2\nmext(X^\prime ,Y^\prime) Y  Y^\prime M^\prime} \Vert_1 \leq \eps. $$ 
\end{definition}
{
\begin{fact}[$\IP$ security against states with $\hminn{\cdot}{\cdot}$ bounds~\cite{ABJO21}] \label{l-qma-needed-fact} Let $n=\frac{n_1}{m}$ and $k_1+k_2-n_1 \geq 2 \log\left(\frac{1}{\eps}\right)+m$. Let $\sigma_{X \hat{X} N' Y \hat{Y} M'}$ be a state with $\vert X \vert = \vert Y\vert = n_1$,  registers $XY$ classical (with copies $\hat{X}\hat{Y}$) and \[\hminn{X}{Y \hat{Y} M}_{\sigma} \geq k_1 \quad ; \quad \hminn{Y}{X \hat{X} N}_{\sigma} \geq k_2 . \]Then
\[\Vert \sigma_{\IP^n_{2^m}(X,Y)XN'} - U_{m} \otimes \sigma_{XN'}  \Vert_1 \leq \eps \quad ; \quad \Vert \sigma_{\IP^n_{2^m}(X,Y)YM'} - U_{m} \otimes \sigma_{YM'}  \Vert_1 \leq \eps.\]
\end{fact}}


\subsection*{Error correcting codes} 
 
 \begin{definition}\label{def:ecc}
     Let $\Sigma$ be a finite set. A mapping $\ecc: \Sigma^k \to \Sigma^n$ is called an error correcting code with relative distance $\gamma$ if for any $x,y \in \Sigma^k$ such that $x \ne y,$ the Hamming distance between $\ecc(x)$ and $\ecc(y)$ is at least $\gamma n.$ The rate of the code denoted by $\delta$, is defined as $\delta \defeq \frac{k}{n}$. The alphabet size of the code is the number of elements in $\Sigma.$
 \end{definition}
 \begin{fact}[\cite{GS95}]\label{fact:ecc}
      Let $p$ be a prime number and $m$ be an even integer. Set $q=p^m$. For every $\delta \in (0,1)$ and for any large enough integer $n$ there exists an efficiently computable linear error correcting code $\ecc: \F^{\delta n}_q \to \F^{ n}_q $ with rate $\delta$ and relative distance $1-\gamma$ such that $$  \delta + \frac{1}{\sqrt{q}-1}\geq \gamma.$$
 \end{fact}

\section{Useful claims and lemmas\label{sec:claims}}

In this section, we prove technical claims and lemmas which will be used throughout the paper.

\begin{claim} \label{claim:traingle_rho_rho_prime}Let $\rho_{ZA}, \rho'_{ZA}$ be states such that $\Delta_B(\rho, \rho') \leq \eps'$. If $d(Z\vert A)_{\rho'} \leq \eps$ then $d(Z\vert A)_{\rho} \leq 2\eps' + \eps$. 
\end{claim}
\begin{proof}Consider,\begin{align*}
        d(Z\vert A)_{\rho} & \leq \Delta_B\left( \rho_{Z A}, \rho'_{Z A}\right)+\Delta_B\left(\rho'_{Z A},  U_{Z} \otimes \rho_{A}\right)& \mbox{(Triangle inequality)}\\
        & \leq \eps' + \Delta_B\left(\rho'_{Z A},  U_{Z} \otimes \rho_{A}\right)  \\
        &\leq  \eps' +\Delta_B\left(\rho'_{Z A}, U_{Z} \otimes \rho'_{A} \right) + \Delta_B \left(   U_{Z} \otimes \rho'_{A} , U_{Z} \otimes \rho_{A} \right) & \mbox{(Triangle inequality)}\\
        & \leq \eps' + \eps + \eps'= 2\eps' + \eps. 
        \end{align*} 
\end{proof}
The above claim holds even when $\Delta_B()$ is replaced with $\Delta()$.
\begin{claim}\label{fact:prefixminentropyfact}
     Let $\rho_{XE} \in \mathcal{D}(\cH_X \otimes \cH_E)$ be a c-q state such that $\vert X \vert =n$ and $\hmin{X}{E}_\rho \geq n-k.$ Let $X_d = \pre(X,d)$ for some integer $k \leq d \leq n$. Then $\hmin{X_d}{E}_\rho \geq d-k.$
\end{claim}
\begin{proof}
Since $\hmin{X}{E}_\rho \geq n-k,$ there exists a state $\sigma_E$ such that $$\dmax{\rho_{XE}}{U_X \otimes \sigma_E} \leq k.$$Using Fact~\ref{fact:data}, we have
\[\dmax{\rho_{X_dE}}{U_{X_d} \otimes \sigma_E} \leq k \implies  \dmax{\rho_{X_dE}}{\id_{X_d} \otimes \sigma_E} \leq k-d.\]Thus, 
\[ \hmin{X_d}{E}_\rho = - \inf_{\tau_E} \dmax{\rho_{X_dE}}{\id_{X_d}\otimes \tau_E} \geq d-k, \] which completes the proof.
\end{proof}
{
\begin{claim}\label{fact:hminpreserves}
Let $\rho_{AB} \in \mathcal{D}(\cH_A \otimes \cH_B)$ be a state and $V_B : \mathcal{L} (\cH_{B}) \rightarrow   \mathcal{L}(\cH_{B'} \otimes \cH_{C})$ be an isometry such that $\vert C \vert =1$. Let $\sigma_{AB'C} = (\id_A \otimes V_B) \rho_{AB} (\id_A \otimes V_B)^\dagger $ and  $\Phi_{AB'} = (\sigma_{AB'C} \vert C=1)$. Then, 
\[ \hminn{A}{B'}_{\Phi} \geq   \hminn{A}{B}_{\rho}. \]
\end{claim}
\begin{proof}
Since $\hminn{A}{B}_\rho = - \dmax{\rho_{AB}}{\id_A \otimes \rho_{B}}$, we have $$\rho_{AB} \leq 2^{-\hminn{A}{B}_\rho}(\id_A \otimes \rho_{B}).$$
By  Fact~\ref{measurediso}, there exists an operators $M_B$ such that $0 \le  M^\dagger_{B}M_{B} \le \id_{B}$ and $$\Phi_{AB'} =\frac{(\id_A \otimes M_{B} ) \rho_{AB} (\id_A \otimes M_{B} )^\dagger }{\tr(\id_A \otimes M_{B} ) \rho_{AB} (\id_A \otimes M_{B} )^\dagger}.$$This further implies,
\[ {\Phi}_{AB'} \leq 2^{-\hminn{A}{B}_\rho} \left(\id_A \otimes \frac{M_{{B}} \rho_{B} M_{{B}}^\dagger}{\tr(M_{{B}} \rho_{B} M_{{B}}^\dagger)}  \right) =  2^{-\hminn{A}{B}_\rho} \left(\id_A \otimes {\Phi}_{B'} \right).\]Thus, 
\[ \hminn{A}{B'}_{{\Phi}} = - \dmax{{ \Phi}_{AB'}}{\id_A \otimes {\Phi}_{B'}}\geq   \hminn{A}{B}_{\rho} . \]
\end{proof}}

\begin{claim}\label{claim:minentropydecrease}
Let $\rho_{ABC} \in \mathcal{D}(\cH_A \otimes \cH_B \otimes \cH_C)$ be a state and $M \in \cL(\cH_C)$ such that $M^\dagger M \leq \id_C$. Let $\hat{\rho}_{ABC}= \frac{M \rho_{ABC} M^\dagger}{\tr{M \rho_{ABC} M^\dagger}}$. Then, 
\[ \hmin{A}{B}_{\hat{\rho}} \geq   \hmin{A}{B}_{\rho} - \log \left(\frac{1}{\tr{M \rho_{ABC} M^\dagger}}\right). \]
\end{claim}
\begin{proof}

Let $\hmin{A}{B}_\rho = - \dmax{\rho_{AB}}{\id_A \otimes \sigma_{B}}$ for some state $\sigma_{B}$. Thus, $$\rho_{AB} \leq 2^{-\hmin{A}{B}_\rho}(\id_A \otimes \sigma_{B}).$$
Since from Fact~\ref{measuredmax}, we have
\[ \hat{\rho}_{AB} \leq \frac{1}{\tr{M \rho_{ABC} M^\dagger}} \rho_{AB}, \]we finally get $\dmax{\hat{\rho}_{AB}}{\id_A \otimes \sigma_{B}} \leq -\hmin{A}{B}_\rho  + \log \left(\frac{1}{\tr{M \rho_{ABC} M^\dagger}}\right).$ Thus, 
\[ \hmin{A}{B}_{\hat{\rho}} = - \inf_{\theta_{B}}\dmax{\hat{ \rho}_{AB}}{\id_A \otimes \theta_{B}}\geq   \hmin{A}{B}_{\rho} - \log \left(\frac{1}{\tr{M \rho_{ABC} M^\dagger}}\right). \]
\end{proof}

{
\begin{claim}[$\IP$ security against $(k_1,k_2)$-$\qpas$] \label{l-qma-needed-fact1} Let $n=\frac{n_1}{m}$ and $k_1+k_2 \geq n_1+m+4+8 \log\left(\frac{1}{\eps}\right)$. Let $\sigma_{X \hat{X} N' Y \hat{Y} M'}$ be a $(k_1,k_2)\mhyphen\qpas$ with $\vert X \vert = \vert Y\vert = n_1$. Let $Z=\IP^n_{2^m}(X,Y)$. Then
\[\Vert \sigma_{ZXN'} - U_{m} \otimes \sigma_{XN'}  \Vert_1 \leq 35\eps \quad ; \quad \Vert \sigma_{ZYM'} - U_{m} \otimes \sigma_{YM'}  \Vert_1 \leq 35\eps.\]
\end{claim}
\begin{proof}
Let state $\rho^{(1)}$ be from Lemma~\ref{lemma:nearby_rho_prime_prime} such that \[\rho^{(1)} \approx_{6 \eps} \sigma^{} \quad ; \quad \hminn{X}{Y \hat{Y} M}_{\rho^{(1)}} \geq k_1 -   2\log \left( \frac{1}{\eps} \right)\quad ; \quad \hminn{Y}{X \hat{X} N}_{\rho^{(1)}} \geq k_2 - 4 -  4\log \left( \frac{1}{\eps} \right). \]  
Using Fact~\ref{l-qma-needed-fact}, we have \[\Vert \rho^{(1)}_{\IP^n_{2^m}(X,Y)XN'} - U_{m} \otimes \rho^{(1)}_{XN'}  \Vert_1 \leq \eps \quad ; \quad \Vert \rho^{(1)}_{\IP^n_{2^m}(X,Y)YM'} - U_{m} \otimes \rho^{(1)}_{YM'}  \Vert_1 \leq \eps.\]
Since $\rho^{(1)} \approx_{6 \eps} \sigma^{} $, using Fact~\ref{fidelty_trace}, we have $\Delta(\rho^{(1)}, \sigma) \leq 6 \sqrt{2} \eps$. 
Using Claim~\ref{claim:traingle_rho_rho_prime} (for $\Delta(.)$ instead of $\Delta_B(.)$), we get 
\[  \Delta(\sigma_{ZXN'}  , U_{m} \otimes \sigma_{XN'} ) \leq 12\sqrt{2} \eps+ \eps/2  \quad ; \quad \Delta(\sigma_{ZYM'} ,U_{m} \otimes \sigma_{YM'} ) \leq  12\sqrt{2} \eps+ \eps/2. \]
We finally get,
\[\Vert \sigma_{ZXN'} - U_{m} \otimes \sigma_{XN'}  \Vert_1 
 \leq 2( 12\sqrt{2} \eps+ \eps/2)\leq 35\eps \quad ; \quad \Vert \sigma_{ZYM'} - U_{m} \otimes \sigma_{YM'}  \Vert_1 \leq 35\eps.\] 
\end{proof}}

\begin{claim}\label{claim:100}
Let $\rho_{XAYB}$ be a pure state. Let $d = |X|$. There exists a pure state $\hat{\rho}_{XAYB}$ such that,
\[ \Delta_B(\hat{\rho}_{XAYB},\rho_{XAYB}) = d(X|YB)_\rho \quad ; \quad  \hmin{Y}{XA}_{\hat{\rho}} = \hmin{Y}{XA}_\rho \quad ; \quad \hat{\rho}_{XYB} = U_{d} \otimes \hat{\rho}_{YB}.\]
\end{claim}
\begin{proof} 
Let $\tau_{XX'}$ be the canonical purification of $\tau_{X} =U_{d}$. Let $\theta_{X_1AYBXX'} = \beta_{X_1AYB} \otimes \tau_{XX'}$ such that $\beta_{X_1AYB} \equiv \rho_{XAYB}$. We use Fact~\ref{uhlmann},  with the following assignment of registers (below the registers on the left are from Fact~\ref{uhlmann} and the registers on the right are the registers in this proof),
 $$(\sigma_A, \rho_A, \sigma_{AC}, \rho_{AB}) \leftarrow (\tau_{X} \otimes \rho_{YB} ,  \rho_{XYB}, \theta_{X_1AYBXX'}, \rho_{XAYB}).$$
 From Fact~\ref{uhlmann}  we get an isometry $V$ such that
\begin{align*}
 \Delta_B \left(    \rho_{XAYB},\hat{\rho}_{XAYB} \right) = \Delta_B( \rho_{XYB}, U_d \otimes \rho_{YB}),
\end{align*}
where,
$$ \hat{\rho}_{XAYB} =  V\left( \theta_{X_1AYBXX'} \right) V^\dagger.$$
 From Fact~\ref{fact102}, 
\[\hmin{Y}{XA}_{\hat{\rho}} = \hmin{Y}{X_1AXX'}_\theta = \hmin{Y}{X_1A}_\beta = \hmin{Y}{XA}_\rho. \]
Noting that isometry $V$ acts trivially on $\theta_{XYB}$, we have $\hat{\rho}_{XYB} =\theta_{XYB}=U_d \otimes \rho_{YB}$. Thus, $\hat{\rho}_{XYB} =U_{d} \otimes \hat{\rho}_{YB}$ which completes the proof.
\end{proof}

\begin{lemma}[Alternating extraction]\label{lem:2}
Let $\theta_{XASB}$ be a pure state with $(XS)$ classical, $\vert X \vert =n, \vert S \vert =d$ and
\[ \hmin{X}{SB}_\theta \geq k \quad ; \quad \Delta_B( \theta_{X A S} , \theta_{X A} \otimes U_d ) \leq \eps^{\prime}. \]
Let $T \defeq \Ext(X,S)$ where $\Ext$ is a $(k,\eps)$-quantum secure strong $(n,d,m)$-extractor. Then, 
\[\Delta_B( \theta_{T B} , U_m \otimes \theta_{B} ) \leq  2\eps' + \sqrt{\eps}.\]
\end{lemma}
\begin{proof}
 We use Fact~\ref{uhlmann},  with the following assignment of registers (below the registers on the left are from Fact~\ref{uhlmann} and the registers on the right are the registers in this proof)
 $$(\sigma_A, \rho_A, \sigma_{AC}, \rho_{AB}) \leftarrow (\theta_{XAS},  \theta_{XA} \otimes U_d, \theta_{XASB}, \beta_{XAS_1B} \otimes \tau_{SS'} ),$$
 where $\tau_{SS'}$ is the canonical purification of $\tau_{S} \equiv U_d$ and $\beta_{XAS_1B} \equiv \theta_{XASB}$. From Fact~\ref{uhlmann}  we get an isometry $V$ (acts trivially on $\theta_{XAS}$) such that,
\begin{align}
     \label{eq1200}
 \Delta_B \left( V \theta_{XASB} V^\dagger,  \beta_{XAS_1B} \otimes \tau_{SS'} \right) = \Delta_B \left(\theta_{XAS}, \theta_{XA} \otimes U_{d} \right) \leq \eps' \enspace.
\end{align}
 Let $\gamma_{TXAS_1BSS'}$ be the state after $T=\Ext(X,S)$ is generated using  $ \beta_{XAS_1B} \otimes \tau_{SS'}$. From Definition~\ref{qseeded}, Fact~\ref{fidelty_trace} and noting that,  $$\hmin{X}{S_1B}_\beta = \hmin{X}{SB}_\theta \geq k \quad ; \quad \gamma_{TSS_1B} \equiv \gamma_{TS'S_1B},$$ 
we get,
\begin{equation} \label{exteq1200}
    \Delta_B\left(\gamma_{TS'S_1B}, U_m  \otimes U_{d} \otimes \beta_{S_1B}  \right) \leq \sqrt{\eps}.
\end{equation}
Consider,
\begin{align*}
    & \Delta_B\left(\theta_{TB}, U_{m} \otimes \theta_B \right) \\
        & = \Delta_B\left(V\theta_{TB}V^\dagger, U_m \otimes V\theta_{B}V^\dagger \right) & \mbox{(Fact~\ref{fact:data})}\\
                & \leq \Delta_B\left(V\theta_{TB}V^\dagger, U_m  \otimes U_{d} \otimes \beta_{S_1B}  \right) + \Delta_B\left(U_m  \otimes U_{d} \otimes \beta_{S_1B}, U_m \otimes V\theta_{B}V^\dagger  \right) & \mbox{(Triangle inequality)}\\
        & \leq \Delta_B\left(V\theta_{TB}V^\dagger, U_m  \otimes U_{d} \otimes \beta_{S_1B}\right) + \eps' & \mbox{(Eq.~\eqref{eq1200} and Fact~\ref{fact:data})}\\
    &\leq  \Delta_B\left(V\theta_{TB}V^\dagger, \gamma_{TS'S_1B}  \right) + \Delta_B\left(\gamma_{TS'S_1B}, U_m  \otimes U_{d} \otimes \beta_{S_1B} \right) +\eps' & \mbox{(Triangle inequality)}\\
    & \leq \Delta_B\left( V \theta_{XASB}V^\dagger,  \beta_{XAS_1B} \otimes \tau_{SS'} \right) + \Delta_B\left(\gamma_{TS'S_1B}, U_m  \otimes U_{d} \otimes \beta_{S_1B} \right)  + \eps'& \mbox{(Fact~\ref{fact:data})}\\
    & \leq 2 \eps' + \sqrt{\eps}.  & \mbox{(Eq.~\eqref{eq1200} and~\eqref{exteq1200})}
\end{align*}
\end{proof}
\begin{lemma}[Min-entropy loss under classical interactive communication]\label{lem:minentropy}
Let $\rho_{XNM}$ be a pure state where Alice holds registers $(XN)$ and Bob holds register $M$, such that register $X$ is classical and
\[ \hmin{X}{M}_\rho \geq k.\]Let Alice and Bob proceed for $t$-rounds, where in each round Alice generates a classical register $R_i$ and sends it to Bob, followed by Bob generating a classical register $S_i$ and sending it to Alice. Alice applies a (safe on $X$) isometry $V^{i}: \cH_X \otimes \cH_{N_{i-1}} \rightarrow \cH_X \otimes \cH_{N'_{i-1}} \otimes \cH_{R_{i}}$ (in round $i$) to generate~\footnote{The isometries in the communication protocols in later sections act as $V^i: \cH_X \rightarrow \cH_X  \otimes \cH_{R_{i}} \otimes  \cH_{\hat{R}_{i}}$.} $R_{i}$. Let 
 $\theta^i_{XN_iM_i}$ be the state at the end of round-$i$, where Alice holds registers $XN_i$ and Bob holds register $M_i$. Then,
 \[ \hmin{X}{M_t}_{\theta^t} \geq k-\sum_{j=1}^{t} \vert R_j\vert .\]
\end{lemma}
\begin{proof}
Proof proceeds by induction on $i$. For $i=0$, the bound follows from initiation (we take $\theta^0 = \rho$). Let us assume the bound for round $i$ \[ \hmin{X}{M_i}_{\theta^i} \geq k-\sum_{j=1}^{i} \vert R_j\vert ,\] and show the bound for round $i+1$. 
Let  $\tau_{XN'_iR_{i+1}M_i}$ be the state after Alice generates $R_{i+1}$. From Fact~\ref{fact2}, we have
$$     \hmin{X}{M_iR_{i+1}}_\tau \geq \hmin{X}{M_i}_\tau - \vert R_{i+1} \vert.$$
Note that since Alice's operations are safe on $X, \tau_{XM_i} = \theta^i_{XM_i}$. Hence,
$$\hmin{X}{M_i}_\tau = \hmin{X}{M_i}_{\theta^i}.$$ From  Fact~\ref{fact102}, we have 
$$  \hmin{X}{M_{i+1}}_{\theta^{i+1}} \geq  \hmin{X}{M_iR_{i+1}}_\tau \geq k-\sum_{j=1}^{i+1} \vert R_j\vert,$$
which shows the desired.
\end{proof}
We now state a lemma which relates $\hminn{}{}$ and $\hmin{}{}$ for a state $\rho$ and a nearby state $\rho'$. The below lemma is also proven in \cite{TRSS10} (we thank the anonymous reviewer for pointing it out), however we include the proof here for completeness.

\begin{lemma}\label{lem:hmin_and_tilde_relation} Let $\rho \in \mathcal{D}(\cH_{AB})$.
There exists $\rho^\prime \in \mathcal{D}(\cH_{AB})$ such that 
\[ \Delta_B(\rho, \rho^\prime) \leq \eps \quad ; \quad \hminn{A}{B}_{\rho} \leq \hmin{A}{B}_{\rho} \leq \hminn{A}{B}_{\rho^\prime} + 2\log\left( \frac{1}{\eps}\right).\]

\end{lemma}
\begin{proof}
    The inequality $\hminn{A}{B}_{\rho} \leq \hmin{A}{B}_{\rho}$ is clear from definitions. 
    
    Let $\hmin{A}{B}_{\rho} = u$ and $\vert A \vert =n$.
    Let  $\sigma_B \in \mathcal{D}(\cH_{B})$ be a state such that $\dmax{\rho_{AB}}{U_{A} \otimes \sigma_B}= n-u$. 
    Set $t= 2 \log\left( \frac{1}{\eps}\right)$.
    Let $\Pi$ denote the projector on $(\sigma_B - 2^t \rho_B)_+$. Hence, $\tr{\left(\Pi \left( \sigma_B - 2^t \rho_B\right)\right)} >0$. 
    This gives us, 
    \[ 2^t \tr{\left( \Pi \rho_B\right)} < \tr{(\Pi \sigma_B}) \leq 1,\] and thus, 
    \begin{gather*} 
        \tr{\left( \Pi \rho_B\right)} < 2^{-t}=\eps^2  \quad ; \quad 
        \tr{\left( \overline{\Pi} \rho_B\right)} >  1- \eps^2. \label{eq:hmin_and_tilda_eqA}
    \end{gather*}
    Also, since $\tr\left(\left(\id_A \otimes \overline{\Pi}\right)  \rho_{AB} \right) = \tr\left(\overline{\Pi}  \rho_{B} \right)$, we have,
    \begin{equation} \label{eq:trace_on_marginal_equality}
        \tr\left(\left(\id_A \otimes \overline{\Pi}\right)  \rho_{AB} \right) > 1- \eps^2.
    \end{equation}Note by construction, $\overline{\Pi} \left( \sigma_B - 2^t \rho_B\right) \overline{\Pi} \leq 0$, and hence,
    \begin{equation} \label{eq:hmin_and_tilda_eqC}
        \overline{\Pi} \sigma_B  \overline{\Pi} \leq   2^t\ \overline{\Pi}  \rho_B \overline{\Pi}.
    \end{equation}
Consider 
\[ \rho^\prime_{AB} = \dfrac{(\id_A \otimes \overline{\Pi})  \rho_{AB} ( \id_A \otimes \overline{\Pi})}{ \tr{\left((\id_A \otimes \overline{\Pi})  \rho_{AB} \right)}} \quad ; \quad \rho^\prime_B =\dfrac{ \overline{\Pi} \rho_B \overline{\Pi} }{\tr{(\overline{\Pi} \rho_B)}}.\]
\suppress{
$\rho^\prime_{AB} = \dfrac{(\id_A \otimes \overline{\Pi})  \rho_{AB} ( \id_A \otimes \overline{\Pi})}{ \tr{\left((\id_A \otimes \overline{\Pi})  \rho_{AB} ( \id_A \otimes \overline{\Pi})\right)}}$.
Note that the corresponding marginal $\rho^\prime_B =\dfrac{ \overline{\Pi} \rho_B \overline{\Pi} }{\tr{(\overline{\Pi} \rho_B \overline{\Pi})}}.$}
Using Fact~\ref{fact:gentle_measurement}, Fact~\ref{fact:data} and Eq.~\eqref{eq:trace_on_marginal_equality}, we have
\begin{gather*} \label{eq:hmin_distance_on_tilda}
    \Delta_B \left( \rho^{\prime}_{B}, \rho_{B} \right) \leq \Delta_B \left( \rho^\prime_{AB}, \rho_{AB} \right) \leq \eps.  
\end{gather*}
Since,  $\dmax{\rho_{AB}}{U_{A} \otimes \sigma_B}= n-u$, we get
\begin{align*}
    \overline{\Pi} \rho_{AB} \overline{\Pi} & \leq 2^{n-u} \cdot U_{A } \otimes \overline{\Pi} \sigma_{B} \overline{\Pi} & (\mbox{Fact~\ref{fact:Conjugation}}) \\
    & \leq 2^{n-u+t} \cdot U_{ A } \otimes \overline{\Pi} \rho_{B} \overline{\Pi} & (\mbox{Eq.~\eqref{eq:hmin_and_tilda_eqC})}
\end{align*}Normalizing by the trace, we get, $\rho^{\prime}_{AB} \leq 2^{n-u+t} \left( U_{ A } \otimes  \rho^\prime_{B}\right)= 2^{t-u} (\id_{A} \otimes \rho^\prime_{B})$, which gives us \[\hminn{A}{B}_{\rho^\prime} \geq \hmin{A}{B}_{\rho} - 2 \log \left( \frac{1}{\eps}\right). \qedhere\]
\end{proof}

{
\begin{lemma}  \label{lemma:nearby_rho_prime_prime} Let $\rho_{X \hat{X} N Y \hat{Y} M}$ be a $(k_1,k_2)\mhyphen\qpas$ such that $\vert X \vert = \vert \hat{X} \vert= \vert Y \vert= \vert \hat{Y} \vert =n$. There exists an $l \mhyphen \qmas$ $\rho^{(1)}_{X \hat{X} N Y \hat{Y} M}$, such that,
\[ \Delta_B(\rho^{(1)}, \rho) \leq {6}\eps \quad and \quad l \leq 2n- k_1 - k_2+ 4 +  6\log \left( \frac{1}{\eps} \right). \]Furthermore, 
\[ \hminn{X}{Y \hat{Y} M}_{\rho^{(1)}} \geq k_1 -   2\log \left( \frac{1}{\eps} \right)\quad ; \quad \hminn{Y}{X \hat{X} N}_{\rho^{(1)}} \geq k_2 - 4 -  4\log \left( \frac{1}{\eps} \right). \]
\end{lemma} }
\begin{proof}
For the ease of notation, let us denote $\tilde{A}= X \hat{X} N$ and $\tilde{B}= Y \hat{Y} M$.
Since, $\hmin{X}{\tilde{B}}_\rho \geq k_1$, using Lemma~\ref{lem:hmin_and_tilde_relation} (on state $\rho_{\tilde{B}X}$) with the assignment of registers $(A,B)  \leftarrow (X,\tilde{B})$, we know that there exists a state $\rho^\prime_{\tilde{B}X}$, such that
    \begin{equation} \label{rho_prime_boundngb}
        \Delta_B\left(\rho_{\tilde{B}X}, \rho^\prime_{\tilde{B}X}\right) \leq \eps \quad; \quad
    \dmax{\rho^\prime_{X\tilde{B}}}{U_{ X } \otimes \rho^\prime_{\tilde{B}}} \leq \vert X\vert-k_1+ 2 \log \left( \frac{1}{\eps} \right)  \defeq c_1. 
    \end{equation}
    Consider a purification of $\rho^{\prime}_{\tilde{B}X}$ denoted as $\rho^{\prime}_{\tilde{B}XE}$. Using Fact~\ref{uhlmann} with the following assignment of registers,
\[\left( \sigma_{A}, \rho_{A}, \sigma_{AC}, \rho_{AB}, \theta_{AB} \right) \leftarrow \left(\rho^{\prime}_{\tilde{B}X}, \rho_{\tilde{B}X},  \rho^{\prime}_{\tilde{B}XE}, \rho_{\tilde{A}\tilde{B}}, \rho^{\prime}_{\tilde{A}\tilde{B}}\right),\] 
there exists a pure state $ \rho^{\prime}_{\tilde{A} \tilde{B}}$ such that,
 \begin{equation} \label{eq:dmax_rho_1ngb}
\Delta_B \left( \rho^{\prime}_{\tilde{A} \tilde{B}} , \rho_{\tilde{A}\tilde{B}} \right) \leq {\eps}  \quad;\quad \dmax{\rho^{\prime}_{X\tilde{B}}}{U_{X} \otimes \rho^\prime_{\tilde{B}}} \leq c_1,
\end{equation}
where the inequalities follow from Eq.~\eqref{rho_prime_boundngb} and noting that isometry taking $\rho^{\prime}_{\tilde{B}XE}$ to $\rho^{\prime}_{\tilde{A}\tilde{B}}$ acts trivially on registers $\tilde{B}X$.
\noindent Similarly,
there exists a pure state $ \rho^{\prime\prime}_{\tilde{A} \tilde{B}}$ such that,
 \begin{equation} \label{eq:dmax_rho_2ngb}
\Delta_B \left( \rho^{\prime\prime}_{\tilde{A} \tilde{B}} , \rho_{\tilde{A}\tilde{B}} \right) \leq {\eps}  \quad;\quad \dmax{\rho^{\prime\prime}_{Y\tilde{A}}}{U_{Y} \otimes \rho^{\prime\prime}_{\tilde{A}}} \leq \vert Y\vert-k_2+2\log \left( \frac{1}{\eps} \right) \defeq c_2.
\end{equation}
Consider the following state:
\[ \theta = \tau_{X \hat{X}} \otimes \rho^\prime_{\tilde{A}\tilde{B}} \otimes \tau_{Y_1 \hat{Y}_1} \]
where $\tau_{X\hat{X}}, \tau_{Y_1\hat{Y_1}}$ are canonical purifications of $\tau_{X} \equiv U_{X}$, $\tau_{Y_1} \equiv U_{Y}$ respectively. Let Alice hold registers $ \tilde{A} \hat{X}$, Bob hold registers $\tilde{B} \hat{Y_1}$ and Referee hold registers $XY_1$. Now using Fact~\ref{rejectionsampling} with the following assignment of registers (below the registers on the left are from Fact~\ref{rejectionsampling} and the registers on the right are the registers in this proof)
    \[\left(\rho_B, \sigma_B, \rho_{A^{\prime}B}, \sigma_{AB} \right)
    \leftarrow 
    \left( \rho^\prime_{X \tilde{B}} ,  \tau_{X} \otimes \rho^\prime_{\tilde{B}}, \rho^\prime_{X\hat{X}N\tilde{B}},\tau_{X \hat{X}} \otimes \rho^\prime_{\tilde{A} \tilde{B}} \right), \]
it follows from Fact~\ref{rejectionsampling} that there exists an isometry $V_{\mathsf{Alice}}: \mathcal{H}_{\tilde{A}\hat{X}} \rightarrow \mathcal{H}_{\hat{X}N} \otimes \mathcal{H}_{C_A}$  such that the following hold:  
\begin{align}
   & \phi_{\tilde{B}X \hat{X} N C_A} = \left( V_{\mathsf{Alice}} \otimes \mathbb{I}_{X \tilde{B}} \right) \left( \rho^\prime_{\tilde{A} \tilde{B}} \otimes \tau_{X \hat{X}}\right) \left(V_{\mathsf{Alice}} \otimes \mathbb{I}_{X \tilde{B}} \right)^\dagger. \label{eq:Alice_Set_1ngb}\\
    &\Pr\left( C_A=1 \right)_{\phi}= p_1 \geq 2^{-c_1} \label{eq:Alice_Set_2ngb}\\
    &\left(\phi \vert C_A=1\right)= \rho^\prime_{\tilde{A} \tilde{B}}\label{eq:Alice_Set_3ngb}.
    \end{align}
Thus starting from state $\theta$, there exists an isometry $V_{\mathsf{Alice}}$ (acting solely on Alice's registers) followed by measuring $C_A$, to get a state which we will denote as $\theta^{(1)}$.
Hence, we get the following:
\begin{align}
    &\phi^{(1)}_{\tilde{B}X \hat{X} N C_A Y_1 \hat{Y}_1} = \left( V_{\mathsf{Alice}} \otimes \mathbb{I}_{X \tilde{B} Y_1 \hat{Y}_1} \right) \theta \left(V_{\mathsf{Alice}} \otimes \mathbb{I}_{X \tilde{B} Y_1 \hat{Y}_1} \right)^\dagger \label{eq:Alice_Set_4ngb}\\
    &\Pr\left( C_A=1 \right)_{\phi^{(1)}}= p_1 \geq 2^{- c_1} \label{eq:Alice_Set_5ngb}\\
    &\theta^{(1)}= \left(\phi^{(1)} \vert C_A=1\right)= \rho^\prime_{\tilde{A} \tilde{B}} \otimes \tau_{Y_1 \hat{Y}_1}\label{eq:Alice_Set_6ngb}.
    \end{align}
Note that Eq.~\eqref{eq:Alice_Set_4ngb}-\eqref{eq:Alice_Set_6ngb} additionally contain $\tau_{Y_1 \hat{Y}_1}$ when compared to Eq.~\eqref{eq:Alice_Set_1ngb}-\eqref{eq:Alice_Set_3ngb}. But as the isometry acts trivially on $\tau_{Y_1 \hat{Y}_1}$, they follow trivially from Eq.~\eqref{eq:Alice_Set_1ngb}-\eqref{eq:Alice_Set_3ngb}. 

\noindent Using Eq.~\eqref{eq:dmax_rho_1ngb}~and Eq.~\eqref{eq:dmax_rho_2ngb} along with triangle inequality, we have
\begin{equation}\label{eq:newscfh}
    \Delta_B \left( \rho^{\prime\prime}_{\tilde{A} \tilde{B}} , \rho^\prime_{\tilde{A}\tilde{B}} \right) \leq 2{\eps}.
\end{equation}Using Fact~\ref{fact:data}, we further have $\Delta_B \left( \rho^{\prime\prime}_{\tilde{A} } , \rho^\prime_{\tilde{A}} \right) \leq 2{\eps}.$
Now,  using Fact~\ref{fact:substate_perturbation} with the following assignment,
    \[ \left( \sigma_{XB}, \psi_X, \rho_B, \rho^{\prime}_{XB}, c,  \delta_0, \delta_1
    \right) \leftarrow \left( \rho^{\prime\prime}_{Y\tilde{A}} , U_{Y}, \rho^{\prime}_{\tilde{A}}, \rho^{(0)}_{Y\tilde{A}},c_2, {\eps}, 2\eps \right)\]
    there exists a state $\rho^{(0)}_{\tilde{A}Y}$ such that,
    \begin{equation*}
        \Delta_B\left(\rho^{(0)}_{\tilde{A} Y} , \rho^{\prime \prime}_{\tilde{A}Y} \right) \leq  {3\eps}  \quad; \quad \rho^{(0)}_{\tilde{A} Y} \leq 2^{c_2+1} \left(1+ \frac{4}{\eps^2}\right) \cdot (U_{Y} \otimes \rho^\prime_{\tilde{A}}) \leq 2^{c'} \cdot (U_{Y} \otimes \rho^\prime_{\tilde{A}})  \quad; \quad \rho^{(0)}_{\tilde{A}} =\rho^\prime_{\tilde{A}},
    \end{equation*} 
where $c' \defeq c_2+4+2 \log\left( \frac{1}{\eps}\right)$. Using Eq.~\eqref{eq:dmax_rho_1ngb}, Eq.~\eqref{eq:dmax_rho_2ngb} and above, we get,
\begin{equation} \label{eq:rho_0_boundsbng}
\Delta_B \left( \rho^{(0)}_{\tilde{A} Y} , \rho^\prime_{\tilde{A}Y} \right) \leq { 5 \eps} \quad;\quad \rho^{(0)}_{\tilde{A} Y} \leq  2^{c'} \cdot (U_{ Y } \otimes \rho^\prime_{\tilde{A}})  \quad; \quad \rho^{(0)}_{\tilde{A}} =\rho^\prime_{\tilde{A}}.
\end{equation}
Consider a purification of $\rho^{(0)}_{\tilde{A}Y}$ denoted as $\rho^{(0)}_{\tilde{A}YE}$. 
Using Fact~\ref{uhlmann} with the following assignment of registers,
\[\left( \sigma_{A}, \rho_{A}, \sigma_{AC}, \rho_{AB}, \theta_{AB} \right) \leftarrow \left(\rho^{(0)}_{\tilde{A}Y}, \rho^\prime_{\tilde{A}Y},  \rho^{(0)}_{\tilde{A}YE}, \rho^\prime_{\tilde{A}\tilde{B}}, \rho^{(1)}_{\tilde{A} \tilde{B}}\right),\] 
there exists a state $ \rho^{(1)}_{\tilde{A} \tilde{B}}$ such that,\begin{equation}\label{eq:ndhfcl}
\Delta_B \left( \rho^{(1)}_{\tilde{A} \tilde{B}} , \rho^\prime_{\tilde{A}\tilde{B}} \right) \leq { {5}\eps}  \quad;\quad \dmax{\rho^{(1)}_{\tilde{A}Y}}{U_{Y} \otimes \rho^\prime_{\tilde{A}}} \leq c^\prime \quad; \quad \rho^{(1)}_{\tilde{A}} =\rho^\prime_{\tilde{A}},
\end{equation}
where the inequalities follow from Eq.~\eqref{eq:rho_0_boundsbng} and noting that isometry taking $\rho^{(0)}$ to $\rho^{(1)}$ acts trivially on registers $\tilde{A}Y$. Consider Fact~\ref{rejectionsampling} with the following assignment of registers,
\[\left(  \rho_B, \sigma_B, \rho_{A'B}, \sigma_{AB}\right) \leftarrow \left( \rho^{(1)}_{\tilde{A}Y}, \rho^\prime_{\tilde{A} } \otimes U_{Y}, \rho^{(1)}_{\tilde{A}\tilde{B}}, \rho^\prime_{\tilde{A} \tilde{B}} \otimes \tau_{Y_1 \hat{Y}_1} \right).\]From Fact~\ref{rejectionsampling}, there exists an isometry $V_{\mathsf{Bob}}: \mathcal{H}_{ \tilde{B} Y_1} \rightarrow \mathcal{H}_{MY_1} \otimes \mathcal{H}_{C_B}$ such that the following hold:
\begin{align}
&\phi^{(2)}_{\tilde{A}M\hat{Y}_1Y_1C_B} = \left(V_{\mathsf{Bob}} \otimes \mathbb{I}_{\tilde{A}\hat{Y}_1} \right) \theta^{(1)} \left(V_{\mathsf{Bob}} \otimes \mathbb{I}_{\tilde{A} \hat{Y}_1 } \right)^\dagger \label{eq:Bob_Set_1bng}\\
&\Pr\left(C_B=1\right)_{\phi^{(2)}} =p_2 \geq 2^{-c^\prime}\label{eq:Bob_Set_2bng}\\
&\rho^{(1)}_{\tilde{A}\tilde{B}} \equiv \left(\phi^{(2)} \vert C_B=1\right)\label{eq:Bob_Set_3bng}
\end{align}
For the ease of notation, let us set $\zeta= \left( V_{\mathsf{Alice}} \otimes V_{\mathsf{Bob}} \right) \theta   \left( V_{\mathsf{Alice}} \otimes V_{\mathsf{Bob}} \right)^\dagger$.
From Eq.~\eqref{eq:Alice_Set_4ngb}-\eqref{eq:Alice_Set_6ngb} and Eq.~\eqref{eq:Bob_Set_1bng}-\eqref{eq:Bob_Set_3bng}, it follows that, 
\begin{align*}
& \rho^{(1)}_{\tilde{A}\tilde{B}} \equiv \left(\zeta  \vert C_A=1, C_B=1 \right) & \mbox(\text{From Eq.~\eqref{eq:Alice_Set_4ngb},\eqref{eq:Alice_Set_6ngb},\eqref{eq:Bob_Set_1bng} and \eqref{eq:Bob_Set_3bng}})\\ 
&\Pr\left( C_A=1, C_B=1\right)_{\zeta} \geq  2^{- c_1} 2^{-c^\prime} & \mbox(\text{From Eq.~\eqref{eq:Alice_Set_2ngb} and \eqref{eq:Bob_Set_2bng}}).
\end{align*}

To summarize, the following properties hold in $ \rho^{(1)}_{\tilde{A}\tilde{B}}$, which completes the proof. 
\begin{itemize}
    \item From construction, it follows that $\rho^{(1)}_{\tilde{A}\tilde{B}}$ is an $l\mhyphen\qmas$ with\[l \leq c_1+c' = 2n- k_1-k_2+4+ 6 \log(1/\epsilon).\]
    \item  $ \rho^{(1)}_{\tilde{A}\tilde{B}} \approx_{6 \epsilon} \rho^{}_{\tilde{A}\tilde{B}} $ follows from Eq.~\eqref{eq:dmax_rho_2ngb},~Eq.~\eqref{eq:ndhfcl} and the triangle inequality.
    \item $\hminn{Y}{\tilde{A}}_{\rho^{(1)}} \geq k_2-4-4 \log(1/\epsilon) $ follows from Eq.~\eqref{eq:ndhfcl}.
    \item  $\hminn{X}{\tilde{B}}_{\rho^{(1)}} \geq \hminn{X}{\tilde{B}Y_1 \hat{Y}_1}_{\theta^{(1)}} = \hminn{X}{\tilde{B}}_{\rho^\prime}\geq k_1-2\log(1/\epsilon) $. Here, the first inequality follows from Claim~\ref{fact:hminpreserves} and the last inequality follows from Eq.~\eqref{rho_prime_boundngb}.    
\end{itemize}    
\end{proof}

\begin{lemma}\label{lem:qmaqpa} Let $\sigma_{X\hat{X}N^\prime M^\prime Y \hat{Y}}$ be an $l\mhyphen\qmas$ such that $\vert X \vert = \vert \hat{X} \vert= \vert Y \vert= \vert \hat{Y} \vert =n$. There exists $k_1,k_2$ such that $k_1 \geq n-l,$ $k_2 \geq n-l$ and $\sigma$ is a $(k_1,k_2)\mhyphen\qpas$. 
\end{lemma}
\begin{proof}Let $\theta_{X\hat{X}NMY\hat{Y}}= \tau_{X\hat{X}} \otimes \tau_{NM} \otimes \tau_{Y\hat{Y}}$ be the initial state as in Definition~\ref{qmadv} (corresponding to an $l\mhyphen\qmas$ $\sigma$). Let  $U : \cH_{X} \otimes \cH_{N} \rightarrow   \cH_{X} \otimes \cH_{N'} \otimes \cH_{A}$ and $V : \cH_Y \otimes \cH_{M} \rightarrow   \cH_{Y} \otimes \cH_{M'} \otimes \cH_{B}$ be isometries as in Definition~\ref{qmadv}. Let  $\rho^{\left(1\right)}= U \theta U^\dagger$. \suppress{Noting isometry $U$ is safe on classical register $X$ and using Fact~\ref{fact102}, we have 
\[\hmin{X}{M Y \hat{Y}}_{\rho^{(1)}}=n \quad ; \quad  \hmin{Y}{N'AX \hat{X}}_{\rho^{(1)}}=n .\]Let $p_1=\Pr(A=1)_{\rho^{(1)}}$ and $\theta^{(1)}= \left( \rho^{(1)} \vert A=1 \right)$.}Noting isometry $U$ is safe on classical register $X$, we have \[\hmin{X}{M Y \hat{Y}}_{\rho^{(1)}}=\hmin{X}{M Y \hat{Y}}_{\theta}=n .\]Let $p_1=\Pr(A=1)_{\rho^{(1)}}$ and $\theta^{(1)}= \left( \rho^{(1)} \vert A=1 \right)$.
Using Claim~\ref{claim:minentropydecrease} with the following assignment (terms on the left are from Claim~\ref{claim:minentropydecrease} and on the right are from here),
\[(A,B, C, \rho, \hat{\rho}) \leftarrow (X,M Y \hat{Y}, A, \rho^{(1)}, \theta^{(1)} ) \]
we get, 
\begin{equation}\label{eq:claim9proof1}
    \hmin{X}{M Y \hat{Y}}_{\theta^{(1)}} \geq \hmin{X}{M Y \hat{Y}}_{\rho^{(1)}} - \log \left( \frac{1}{p_1}\right) = n+ \log(p_1).
\end{equation}
Furthermore, let $\rho^{(2)}= V \theta^{(1)} V^{\dagger}$. 
Again using Fact~\ref{fact102} and noting $V$ is an isometry, we have 
\suppress{\hmin{X}{M^\prime Y \hat{Y} }_{\rho^{(2)}} \geq  \hmin{X}{M^\prime Y \hat{Y} B}_{\rho^{(2)}} = \hmin{X}{M Y \hat{Y}}_{\theta^{(1)}} \quad ; \quad  \hmin{Y}{N^\prime X \hat{X} }_{\rho^{(2)}} = \hmin{Y}{N' X \hat{X}}_{\theta^{(1)}}  .}
\begin{equation}\label{eq:claim9proof2}
    \hmin{X}{M^\prime Y \hat{Y} }_{\rho^{(2)}} \geq  \hmin{X}{M^\prime Y \hat{Y} B}_{\rho^{(2)}} = \hmin{X}{M Y \hat{Y}}_{\theta^{(1)}} .
\end{equation}Let $p_2=\Pr(B=1)_{\rho^{(2)}}$. Note $\sigma= \left( \rho^{(2)} \vert B=1 \right)$ and $l= \log \left(\frac{1}{p_1 \cdot p_2}\right)$. Now we use Claim~\ref{claim:minentropydecrease} with the following assignment (terms on the left are from Claim~\ref{claim:minentropydecrease} and on the right are from here),
\[(A,B, C, \rho, \hat{\rho}) \leftarrow (X,M^\prime Y \hat{Y}, B, \rho^{(2)}, \sigma ) \]
we get $\hmin{X}{Y \hat{Y} M^\prime}_{\sigma} \geq \hmin{X}{Y \hat{Y} M^\prime}_{\rho^{(2)}} + \log(p_2)$. Using Eq.~\eqref{eq:claim9proof1}~and~\eqref{eq:claim9proof2}, we get $$\hmin{X}{Y \hat{Y} M^\prime}_{\sigma} \geq n + \log( p_1 \cdot p_2) =n-l.$$Using similar argument, we get $$\hmin{Y}{X \hat{X} N^\prime}_{\sigma} \geq n-l.$$
\suppress{Similarly with the following assignment, 
\[A,B, C, \rho, \hat{\rho} \leftarrow Y ,N^\prime X \hat{X} A, B, \rho^{(2)}, \sigma  \]
we get,   $\hmin{Y}{X \hat{X} N^\prime A}_{\sigma} \geq \hmin{Y}{X \hat{X} N^\prime A}_{\rho^{(2)}} + \log(L_2)  \geq n+  \log(L_2)  \geq n+ \log(L) = n-l$.}
Thus,  $\sigma$ is a $(k_1,k_2)\mhyphen\qpas$ such that both $k_1,k_2 \geq (n-l)$.
\end{proof}

 \section{A quantum secure non-malleable extractor\label{sec:nmext}}

In this section, we define and prove the quantum security of the non-malleable extractor.
Our non-malleable extractor is based on the constructions by Chattopadhyay, Goyal and Li~\cite{CGL15}.
As stated before, these constructions use the alternating extraction, consisting of a sequence of random variables generated using strong seeded extractors.
In \cite{CGL15}, seeded extractor from \cite{GUV09} was used in alternating extraction. 
However, this extractor is not known to be quantum proof.
In our construction, we use the quantum-proof Trevison extractor, and argue that the process of alternating extraction remains quantum-proof.
After this change, we set our parameters in a very similar manner to that of \cite{CGL15}.
This gives us the following parameters, which hold throughout this section.

 
\subsection*{Parameters}\label{sec:parameters}
Let $\delta>0$ be a small enough constant and $q$ be a prime power. Let  $n,d,d_1,a,v,d_2,s,b,h$ be positive integers and
 $k, \eps', \gamma, \eps > 0$ such that: 
  \[d =  \cO \left(\log^{7} \left(\frac{n}{\eps}\right)\right) \quad ; \quad  v= \frac{d}{\eps} \quad ;\quad  d_1 =\cO \left(\log^2\left(\frac{n}{\eps}\right) \log(\log (v))\right) \quad ; \quad q= \cO\left(\frac{1}{\eps^2}\right) \quad ;  \]
 
\[a=d_1+ \log q \quad ; \quad \gamma = \cO(\eps) \quad ; \quad  2^{\cO(a)}\sqrt{\eps'} = \eps \quad ;  \quad  d_2=\cO\left(\log^2\left(\frac{n}{\eps'}\right) \log d \right) \quad ;      \]

\[s = \cO\left(\log^2\left(\frac{d}{\eps'}\right)\log d \right)  \quad  ; \quad b = \cO\left( \log^2\left(\frac{d}{\eps'}\right) \log d \right) \quad ; \quad h = 10s \quad ;  \quad  k \geq 5d. \]Let \begin{itemize}\label{sec:extparameters}
    \item  $\Ext_0$ be $(2\log v,\eps^2)$-quantum secure $(n,d_1,\log v)$-extractor, 
    \item $\Ext_1$ be $(2b, \eps')$-quantum secure $(d,s,b)$-extractor,
    \item $\Ext_2$ be $(2s, \eps')$-quantum secure $(h,b,s)$-extractor,
    \item $\Ext_3$ be $(2h, \eps')$-quantum secure $(d,b,h)$-extractor,
    \item $\Ext_4$ be $(d/4, \eps^2)$-quantum secure $(d,h,d/8)$-extractor,
    \item $\Ext_5$ be $(2d, \eps')$-quantum secure $(n,d_2,d)$-extractor, \item  $\Ext_6$ be $(\frac{k}{2}, \eps^2)$-quantum secure $(n,d/8,k/4)$-extractor,
 \end{itemize}
 be the quantum secure extractors from Fact~\ref{fact:extractor}.

 Let $\F_q$ be the finite field of size $q$.
 Let $\ecc : \F^d_q \to \F^v_q$ be an error correcting code with relative distance $1-\gamma$ and rate $\eps$ (which exists from~Fact~\ref{fact:ecc} for our choice of parameters). We identify $I$ as an element from $\{1, \ldots, v \}$. By $\ecc(Y)_{I}$, we mean the $I$-th entry of the code-word $\ecc(Y)$, interpreted as a bit string of length $\log q$.

\subsection*{Description of the non-malleable extractor }

At a high-level, our non-malleable extractor construction given by Algorithm~\ref{alg:nmExt} can be broken into three steps:
\begin{itemize}
    \item Advice generation (Step 1 in Algorithm~\ref{alg:nmExt})
    \item Correlation breakers with advice (Step 3 in Algorithm~\ref{alg:nmExt})
    \item Improving the output length (Step 4 in Algorithm~\ref{alg:nmExt}).
\end{itemize}
Correlation breakers  (Algorithm~\ref{alg:AdvCB}) themselves use the flip-flop primitive, given by Algorithm~\ref{alg:FF}. 
Now we expand on each of the above three steps and outline their motivation. 

\textbf{Correlation breakers with advice.} A correlation breaker uses independent randomness to “remove correlation” that may exist between a sequence of random variables.
Let $YY'$ be correlated random variables such that $Y \ne Y'$ with $Y$ having  sufficient min-entropy.
Let $TT'$ be correlated random variables such that $T$ is uniform and independent of $YY'$.
Let $\alpha, \alpha'$ be any two fixed strings of length $a$ such that  $\alpha \ne \alpha'$. Correlation breakers with advice is a function  $\advcb: \lbrace 0,1 \rbrace^d \times \lbrace 0,1 \rbrace^d \times \lbrace 0,1 \rbrace^a \rightarrow \lbrace 0,1 \rbrace^{\frac{d}{8}}$
such that $\advcb(Y,T,\alpha)\advcb(Y',T',\alpha') \approx U_{d/8} \otimes \advcb(Y',T',\alpha')$.
Note that since random variables $YY'$ and $TT'$ are arbitrarily correlated, it is not immediately clear why $\advcb(Y,T,\alpha)\advcb(Y',T',\alpha') $ should be independent.


\textbf{Advice generation.} 
As mentioned above, correlation breakers need two advice strings  $\alpha, \alpha'$  of length $a$ such that  $\alpha \ne \alpha'$.
The job of the Advice generation step is to supply $\advcb$ with this advice.
Let $X$ be a source and $YY'$ be arbitrarily correlated random variables such that $Y \ne Y'$ with $Y$ being uniform.
The goal is to come up with a function $f$ such that ${G}= f(X,Y) \ne f(X,Y') = {G'}$ (with high probability). 
This can be done as follows:
\begin{itemize}
    \item Let $\ecc$ be an error correcting code of constant rate and sufficiently high relative distance (close to 1). Since $Y \ne Y'$, the encodings $\ecc(Y)$ and $\ecc(Y')$ differ at most coordinates. 
    \item Now, take $Y_1$ (a prefix of $Y$) and generate $I = \Ext(X,Y_1)$. Since $X$ is independent of $YY'$, it follows that $I$ is independent of $YY'$.
    Thus $\ecc(Y)_I$ and $\ecc(Y')_I$  are not equal with high probability.
    \item Define $G=Y_1 \circ \ecc(Y)_I$ and $G'=Y'_1 \circ \ecc(Y')_{I'}$.
    If $Y_1 \ne Y'_1$, then $G \ne G'$ trivially.
    Otherwise, $Y_1 = Y'_1$, and thus, $I=I'$ and $\ecc(Y)_I \ne \ecc(Y')_I$ with high probability. 
\end{itemize}
Thus, we have achieved the task of obtaining $G \neq G'$.

\textbf{ Improving the output length.} Advice generator along with correlation breakers already give a non-malleable property, but with logarithmic output length. 
Fortunately, one can show that most of the min-entropy is still intact in the source $X$. 
Thus, one can improve the output length of non-malleable extractor using one additional application of a seeded extractor, which is achieved in Step 4.

\begin{algorithm}
\caption{: $\nmext: \lbrace 0,1 \rbrace ^n\times \lbrace 0,1 \rbrace^d   \rightarrow \lbrace 0,1 \rbrace^{k/4}$}\label{alg:nmExt}
\begin{algorithmic}
\State{}

\noindent \textbf{ Input:}  $X, Y$\\
\suppress{
Let $\advc: \lbrace 0,1 \rbrace^n \times \lbrace 0,1 \rbrace^d \rightarrow \lbrace 0,1 \rbrace^a$ be the advice generator from Lemma~\ref{lemma:block1} and Corollary~\ref{corr:block1}. \\ \\
\noindent Let $\advcb: \lbrace 0,1 \rbrace^d \times \lbrace 0,1 \rbrace^d \times \lbrace 0,1 \rbrace^a \rightarrow \lbrace 0,1 \rbrace^{\frac{d}{8}}$ be the correlation breaker from Theorem~\ref{thm:advcb} \\}


\begin{enumerate}
    \item Advice generator:\quad $Y_1 = \pre(Y,d_1) \quad ; \quad I = \Ext_0(X,Y_1) \quad ;     \quad G=Y_1 \circ \ecc(Y)_{I}$ 
    \item $Y_2=\pre(Y,d_2) \quad ; \quad T=\Ext_5(X,Y_2)$
    \item Correlation breaker with advice:\quad $S=\advcb(Y,T,G)$
    \item $L=\Ext_6(X,S)$
\end{enumerate}

 \noindent \textbf{ Output:} $L$ 
\end{algorithmic}
\end{algorithm}

\suppress{
\begin{algorithm}
\caption{: $ \advc: \lbrace 0,1 \rbrace ^n\times \lbrace 0,1 \rbrace^d   \rightarrow \lbrace 0,1 \rbrace^{a}$}\label{alg:advgen}
\begin{algorithmic}
\State{}

\vspace{0.1cm}

\noindent \textbf{ Input:}  $X, Y$\\
\suppress{
Let $\advc: \lbrace 0,1 \rbrace^n \times \lbrace 0,1 \rbrace^d \rightarrow \lbrace 0,1 \rbrace^a$ be the advice generator from Lemma~\ref{lemma:block1} and Corollary~\ref{corr:block1}. \\ \\
\noindent Let $\advcb: \lbrace 0,1 \rbrace^d \times \lbrace 0,1 \rbrace^d \times \lbrace 0,1 \rbrace^a \rightarrow \lbrace 0,1 \rbrace^{\frac{d}{8}}$ be the correlation breaker from Theorem~\ref{thm:advcb}. \\}


\begin{enumerate}
    \item $Y_1 = \pre(Y,d_1) \quad ; \quad I = \Ext_0(X,Y_1) \quad ; \quad G=Y_1 \circ \ecc(Y)_{I}$
\end{enumerate}

 \noindent \textbf{ Output:} $G$ 
\end{algorithmic}
\end{algorithm}
}

\begin{algorithm}
\caption{: $\advcb: \lbrace 0,1 \rbrace^d \times \lbrace 0,1 \rbrace^d \times \lbrace 0,1 \rbrace^a \rightarrow \lbrace 0,1 \rbrace^{\frac{d}{8}}$}\label{alg:AdvCB}
\begin{algorithmic}
\State{}

\noindent \textbf{ Input: }$Y, T, G$
\suppress{
Let $X \in \lbrace 0,1\rbrace^n,\ Z\in \lbrace 0,1 \rbrace^l$ be random variables with $\alpha \in \lbrace 0,1 \rbrace^a$ as the input. Let $n, a$ be integers, $\eps>0$. Let $s,b,h$ be integers defined as 
\[s = \cO\left(\log^3\left(\frac{n}{\eps}\right)\right)  \quad  ; \quad b = \cO\left( \log^3\left(\frac{l}{\eps}\right) \right) \quad ; \quad h = \Theta(s) \quad ; \quad h \leq l \quad ; \quad n \geq \log^3(l),\]
where we set $l=cah$ for some large enough constant $c$. Note that this choice of $l$ meets the above required conditions. Let $k_x=cab, m=k_x/4$. \\ \\
Let $\Ext_4: \lbrace 0,1 \rbrace^n \times \lbrace 0,1 \rbrace^h  \rightarrow \lbrace 0,1 \rbrace^m$ be $(2m, \eps)$-quantum secure extractor.  \\ \\}

\begin{enumerate}
    \item $Z_0=$Prefix$(T,h)$
    \item For $i=1,2,\ldots,a:$ 
    
     \hspace{1cm}Flip flop: \quad $Z_i=\ff(Y,T,Z_{i-1},G_i)$ 
     \item $S=\Ext_4(Y,Z_a)$

\end{enumerate}

 \noindent \textbf{ Output:} $S$ 
\end{algorithmic}
\end{algorithm} 


\begin{algorithm}
\caption{: $\ff : \lbrace 0,1 \rbrace^d \times \lbrace 0,1 \rbrace^d \times \lbrace 0,1 \rbrace^h \times \lbrace 0,1 \rbrace \rightarrow \lbrace 0,1\rbrace^h$}\label{alg:FF}
\begin{algorithmic}
\State{}

\noindent\textbf{Input:}  $Y, T,  Z,  G$ 
\begin{enumerate}
     
    \item $Z_s=$Prefix$(Z,s)$, $A= \Ext_1 (Y,Z_s),\ C= \Ext_2(Z,A),\ B= \Ext_{1}(Y,C)$
    \item If $G=0$ then $\overline{Z}= \Ext_3(T,A)$ and if $G=1$ then $\overline{Z}= \Ext_3(T,B)$
     \item $\overline{Z}_s=$Prefix$(\overline{Z},s)$, $\overline{A}= \Ext_1 ({Y},\overline{Z}_s),\ \overline{C}= \Ext_2(\overline{Z},\overline{A}),\ \overline{B}= \Ext_{1}({Y},\overline{C})$
    \item If $G=0$, then $O=\Ext_3(T,\overline{B})$ and if $G=1$, then $O=\Ext_3(T,\overline{A})$
\end{enumerate}
 \noindent \textbf{Output:} $O$
\end{algorithmic}
\end{algorithm}

\subsection*{Result}

\suppress{We use the notation $\pre (X,d)$ to denote a random variable which outputs first $d$ bits of random variable output $X=x.$}

 To show the security of $\nmext$, we first explain the correspondence between Algorithm~\ref{alg:nmExt} and Protocol~\ref{prot:block1}. Note that $\nmext$ as defined in Algorithm~\ref{alg:nmExt} is a generation of sequence of random variables until we finally output $L =\nmext(X,Y)$. Our goal is to show that when $\nmext()$ executed on classical registers $(X,Y)$ and $(X,Y')$ in $(k)\mhyphen\nmas$  $\rho_{X\hat{X}NYY'\hat{Y}\hat{Y}'M}$, we have that $$ \| \rho_{LL'YY'M} - U_{k/4} \otimes \rho_{L'YY'M} \|_1 \leq \cO(\eps).$$

As stated before, the main conceptual hurdle in extending the analysis from~\cite{CGL15} to the quantum case, lies in finding the proper framework in which we can express the correlations that arise from quantum side information.
This is necessary since the procedure of alternating extraction is based on repeated application of seeded extractors, which need some sufficient entropy in the source.
In particular, an extraction at Alice's end using some source $X$ would require that $X$ has enough min-entropy given Bob's registers, and hence all the correlations including those that are quantum, need to be accounted for. 
We consider the $ (k_1,k_2)\mhyphen \qpas$ framework to express these quantum correlations. 
At each step in the analysis, we divide the entire state into two parts, one held by Alice and other held by Bob.
This allows us to argue that a relevant register has certain min-entropy given the other party.
We keep track of these registers, their min-entropy and closeness of states in Protocol~\ref{prot:block1}. 
Note that Protocol~\ref{prot:block1} also contains variables such as $Y'$ which are obtained after tampering by the adversary.
Execution of $\nmext$ on such tampered variables results in ``primed" variables. As the purpose of Protocol~\ref{prot:block1} is to keep track of various quantities such as min-entropy and distance at various stages, the exact sequence in which we generate these ``primed" and ``unprimed" variables is highly critical.
Protocol~\ref{prot:block1} gives this exact sequence (along with the analysis as one of the columns).
Thus, these protocols given in the appendix serve as an aid to the security proof of $\nmext$, whose construction is given in Algorithm~\ref{alg:nmExt}.

Note that Protocol~\ref{prot:block1} uses Protocol~\ref{prot:block2} as a subprotocol.
Informally, Protocol~\ref{prot:block2} generates $O$ and $O^\prime$ such that they are independent and are on different parts of the state.
Protocol~\ref{prot:block2} is a \emph{for loop}, which in each iteration, enters one of the six protocols given by
Protocol~\ref{prot:Var_GEN(0,1)analysis}-\ref{prot:Var_GEN(1,1)DiffBefore}; depending on bit values $G$ and $G'$.
The idea here is to output $O$ (extractor output after many rounds of alternating extraction) on Alice's end while Bob already holds $O'$, so that $O$ and all of Bob's registers (including $O'$) are independent.
This is exactly what is achieved by Protocol~\ref{prot:Var_GEN(0,1)analysis} and Protocol~\ref{prot:Var_GEN(1,0)}.
Recall that the advice generation step produces $G \neq G^\prime$ which ensures that at least one of Protocol~\ref{prot:Var_GEN(0,1)analysis} or Protocol~\ref{prot:Var_GEN(1,0)} is run at some point in the loop, giving us the required independence.
Protocol~\ref{prot:Var_GEN(0,0)NotDiffBefore} and Protocol~\ref{prot:Var_GEN(1,1)NotDiffBefore} depict the case until the point where bits of $G$ and $G'$ agree.
At this stage no independence can be gained, which can be seen as these subprotocols output $O$ and $O^\prime$ on the same side (Alice's). 
The rest of the Protocols (Protocol~\ref{prot:Var_GEN(0,0)DiffBefore}  and Protocol~\ref{prot:Var_GEN(1,1)DiffBefore}) ensure that once we gain the independence, it is retained throughout the $n$ iterations of the \emph{for loop}.

At this point, let us clarify some notation regarding Protocol~\ref{prot:block1}.
In Protocol~\ref{prot:block1}, Alice and Bob generate new classical registers using safe isometries on old classical registers. At any stage of Protocol~\ref{prot:block1}, we use $N$ to represent all the registers held by Alice other than the specified registers at that point. Similarly $M$ represents all the registers held by Bob other than the specified registers. At any stage of the protocol, we use $\tilde{A}, \tilde{B}$ to represent all the registers held by Alice and Bob respectively. We use the same convention for communication protocols in later sections as well.

The following theorem shows that the function $\nmext$ as defined in Algorithm~\ref{alg:nmExt} is $(k,\cO(\eps))$-secure against $\nma$ by noting that  $L= \nmext(X,Y)$ and $L'=\nmext(X,Y')$. 

 \suppress{
 Let 
 $a=d_1+ \log q$, where $q= \cO(\frac{1}{\eps^2})$ and $m= \frac{d}{\gamma}$, $\gamma = \cO(\eps)$ (existence follows from Fact~\ref{fact:ecc}). Let $\eps_5=2^{-\cO(a)}\eps$. Let $\Ext_5$ be $(2d, \eps_5)$-quantum secure $(n,d_2,d)$-extractor with seed length $d_2=\cO\left(\log^2\left(\frac{n}{\eps_5}\right) \log d\right)$.
 \suppress{
 \noindent Let $\advc : \{0,1\}^{n} \times \{0,1\}^{d} \to \{0,1\}^a.$
\noindent Let $\advcb: \lbrace 0,1 \rbrace^d \times \lbrace 0,1 \rbrace^d \times \lbrace 0,1 \rbrace^a \rightarrow \lbrace 0,1 \rbrace^{\frac{d}{8}}$. 
\noindent Let $ \ff: \lbrace 0,1 \rbrace^d \times \lbrace 0,1 \rbrace^d \times \lbrace 0,1 \rbrace^h \times \lbrace 0,1 \rbrace \rightarrow \lbrace 0,1\rbrace^h$. \\ }
\noindent Let $s,b,h$ be integers defined as 
\[s = \cO\left(\log^3\left(\frac{d}{\eps'}\right)\right)  \quad  ; \quad b = \cO\left( \log^3\left(\frac{d}{\eps'}\right) \right) \quad ; \quad h = \Theta(s) \quad ; \quad h \leq d ,\]
where $\eps' = 2^{-\Omega(a)}\eps^2$. Let $\Ext_1$ be $(2b, \eps')$-quantum secure $(d,s,b)$-extractor. Let $\Ext_2$ be $(2s, \eps')$-quantum secure $(h,b,s)$-extractor, $\Ext_3$ be $(2h, \eps')$-quantum secure $(d,b,h)$-extractor, $\Ext_4$ be $(d/4, \eps)$-quantum secure $(d,h,d/8)$-extractor and  $\Ext_6$ be $(\frac{k}{2}, \eps)$-quantum secure $(n,d/8,k/4)$-extractor. \\

}

\begin{theorem}[Security of $\nmext$]\label{thm:nmext}
Let $\rho_{X\hat{X}NYY'\hat{Y}\hat{Y}'M}$ be a $(k)\mhyphen\nmas$ with $\vert X \vert = n$ and $\vert Y \vert = d$. Let Protocol~\ref{prot:block1} start with $\rho$. Let $\Lambda$ be the state at the end of the protocol. Then,
$$ \| \rho_{LL'YY'M} - U_{k/4} \otimes \rho_{L'YY'M} \|_1 \leq d(L|\tilde{B})_\Lambda \leq \cO(\eps).$$

\end{theorem}
\begin{proof} The first inequality follows from Fact~\ref{fact:data}.

Since $\nmext$ is comprised of a sequence of applications of seeded extractors, we need to argue that sufficient min-entropy is retained throughout Protocol~\ref{prot:block1} in the sources on which seeded extractors are applied.
To do that, we first argue the total communication in Protocol~\ref{prot:block1} is bounded. For instance, the total communication from Alice to Bob in Protocol~\ref{prot:block1} is at most (from our choice of parameters) 
$$2 \log \left(\frac{d}{\eps}\right) + 6ah + h+ \frac{k}{4} \leq  \left(1/4 + \delta\right) k.$$
This implies, using Lemma~\ref{lem:minentropy}, that throughout Protocol~\ref{prot:block1}, $\hmin{X}{\tilde{B}} \geq (3/4-\delta)k > k/2$. 

Similarly, total communication from Bob to Alice in  Protocol~\ref{prot:block1} is at most $$2d_1 +2d_2 + 2a + 6ab + \frac{d}{4} \leq (1/4+\delta) d.$$
Again using Lemma~\ref{lem:minentropy}, throughout Protocol~\ref{prot:block1}, $\hmin{Y}{\tilde{A}} \geq (3/4-\delta)d$.

Next we need to argue about the state just before invoking correlation breaker with advice (Step 3 in Algorithm~\ref{alg:nmExt}).
Let $\Phi$ be the joint state in Protocol~\ref{prot:block1} after registers $Z_0, Z_0'$ are generated by Alice.
As stated before, we require that $G \ne G'$ with high probability in the state $\Phi$. Furthermore we also need that register $T$ is independent of Bob side registers for the correlation breaker with advice $\advcb$ to function.
Formally, we prove the following two statements in Claim~\ref{lemma:block1}.

\begin{enumerate}
     \item \label{lemma:block1:point3}$\Pr(G=G')_\Phi = \cO(\eps) $.
      \item\label{lemma:block1:point4}$ d(T|\tilde{B})_ \Phi \leq \eps$.
\end{enumerate}
Let $\hat{\Phi}$ be the state obtained from Claim~\ref{claim:100} (by letting $\rho$ in Claim~\ref{claim:100} as $\Phi$ here) such that,
\begin{equation}
 \label{eq:phitilde}   
 \hmin{Y}{\tilde{A}}_{\hat{\Phi}} \geq (3/4-\delta)d \quad ; \quad \hat{\Phi}_{T\tilde{B}}= U_d \otimes \hat{\Phi}_{\tilde{B}} \quad ; \quad  \Delta_B(\hat{\Phi},\Phi ) \leq \eps. 
 \end{equation} 
 Let 
$$\hat{\Phi}^{(\alpha,\alpha')}= \hat{\Phi}| ((G,G')=(\alpha,\alpha')).$$  


 Let $\mathcal{S}_1 \defeq \{(\alpha,\alpha') ~:~ \alpha = \alpha'\}$,  $\mathcal{S}_2 \defeq \{(\alpha,\alpha') ~ :~ \Pr((G,G')=(\alpha,\alpha'))_{\hat{\Phi}} \leq \frac{\eps}{2^{2 \vert G \vert}}\}$ and  $\mathcal{S} \defeq  \mathcal{S}_1 \cup \mathcal{S}_2$.
 Note $\Pr((\alpha,\alpha') \in\mathcal{S}_1)_{ \hat{\Phi}} \leq \Pr((\alpha,\alpha') \in\mathcal{S}_1)_{\Phi} +\eps = \cO(\eps)$  and $\Pr((\alpha,\alpha') \in\mathcal{S}_2)_{\hat{\Phi}} \leq \eps$. From the union bound, $\Pr((\alpha,\alpha') \in\mathcal{S})_{\hat{\Phi}} \leq \cO(\eps).$ 
 
 For every $(\alpha,\alpha') \notin \mathcal{S}_2$, we have (using Fact~\ref{measuredmax} and noting that, in $\hat{\Phi}$, a copy of $(G,G')$ is part of $\tilde{B}$),
 \begin{equation}\label{eq679}
      \hat{\Phi}^{(\alpha,\alpha')}_{Y\tilde{A}} \leq   \frac{\hat{\Phi}_{Y\tilde{A}}  }{\Pr((G,G')=(\alpha,\alpha'))_{\hat{\Phi}}}  \leq  2^{2 \vert G \vert+ \log( \frac{1}{\eps})} \cdot \hat{\Phi}_{Y\tilde{A}}  \enspace.
 \end{equation}
 Eq.~\eqref{eq:phitilde} and~\eqref{eq679} imply that for every $(\alpha,\alpha') \notin \mathcal{S}_2$, we have $$\hmin{Y}{\tilde{A}}_{\hat{\Phi}^{(\alpha,\alpha')}} \geq (3/4-\delta)d-2a - \log \left( \frac{1}{\eps} \right) \geq (3/4-2\delta)d > d/4.$$

 Let $\hat{\Gamma}^{(\alpha,\alpha')}, \hat{\Gamma}, \Gamma$ be the joint states at the end of the Protocol~\ref{prot:block2} (for $i=a$) when starting with the states  $\hat{\Phi}^{(\alpha,\alpha')} ,\hat{\Phi}, \Phi$ respectively. From Claim~\ref{thm:advcb}, we have for every $(\alpha,\alpha') \notin \mathcal{S}$, 
\begin{equation}\label{mainthm:eq1}
     d(Z|\tilde{B})_{\hat{\Gamma}^{(\alpha,\alpha')}}  \leq  \cO(\eps).
 \end{equation}
 Consider (register $Z$ is held by Alice and $\tilde{B} = GG'M$ in the state $\hat{\Gamma}$),  
  \begin{align*}
      d(Z|\tilde{B})_{\hat{\Gamma}} = \Delta_B ( \hat{\Gamma}_{ZGG'M} , U_{h} \otimes \hat{\Gamma}_{GG'M}) &= \E_{(\alpha,\alpha') \leftarrow (G,G')} \Delta_B ( \hat{\Gamma}^{(\alpha,\alpha')}_{Z\tilde{B}} , U_{h} \otimes \hat{\Gamma}^{(\alpha,\alpha')}_{\tilde{B}})  \\
      & \leq \Pr((\alpha,\alpha') \notin \mathcal{S})_{\hat{\Gamma}} \cdot  \cO(\eps) + \Pr((\alpha,\alpha') \in \mathcal{S})_{\hat{\Gamma}}  \\
      & \leq \cO(\eps),
  \end{align*}
  where the first equality follows from Fact~\ref{traceavg}, the first inequality follows from Eq.~\eqref{mainthm:eq1} and the last inequality follows since $\Pr((\alpha,\alpha') \in\mathcal{S})_{\hat{\Gamma}}=\Pr((\alpha,\alpha') \in\mathcal{S})_{\hat{\Phi}} \leq \cO(\eps)$. Since (using Fact~\ref{fact:data} and Eq.~\eqref{eq:phitilde}) $\Delta_B(\hat{\Gamma} ,\Gamma) \leq \Delta_B(\hat{\Phi} ,\Phi) \leq \eps,$ we have, 
  \[d(Z|\tilde{B})_\Gamma = d(Z|\tilde{B})_{\hat{\Gamma}} + \eps = \cO(\eps).\]
  Using arguments as before (involving Lemma~\ref{lem:2}), we have 
  $d(L \vert \tilde{B})_\Lambda \leq \cO(\eps)$. 
\end{proof}
\begin{claim}[Advice generator]\label{lemma:block1}
     $\Pr(G=G')_\Phi = \cO(\eps)$ and $d(T|\tilde{B})_\Phi\leq \eps.$
 \end{claim}

\begin{proof}
    We first prove $\Pr(G=G')_\Phi = \cO(\eps)$. Let $\sigma_{XNIMYY'}$ be the state after Alice has generated register $I$ (before sending to Bob). Let $\beta_{IYY'} = U_{\log v} \otimes \Phi_{YY'}$. We have,
\begin{align}
     \Delta( \Phi_{IYY'} , \beta_{IYY'}) &= \Delta( \sigma_{IYY'} , U_{\log v} \otimes \sigma_{YY'}) & \nonumber\\
     &\leq \Delta( \sigma_{I\tilde{B}} , U_{\log v} \otimes \sigma_{\tilde{B}}) & \mbox{(Fact~\ref{fact:data})}\nonumber \\
     & \leq \sqrt{2} \Delta_B( \sigma_{I\tilde{B}} , U_{\log v} \otimes \sigma_{\tilde{B}}) \nonumber &\mbox{(Fact~\ref{fidelty_trace})}\\
     & \leq \sqrt{2}\eps.  & \mbox{(Lemma~\ref{lem:2})}\label{advgen:eq11}
\end{align}
Consider,
\begin{align*}
   \Pr(G=G')_\Phi &\leq  \Pr(G=G')_\beta + \sqrt{2} \eps \\
   & = \Pr(Y_1 = Y_1')_\beta\Pr(G=G'~|~ Y_1 = Y_1')_\beta + \Pr(Y_1 \neq Y_1')_\beta\Pr(G=G'~|~ Y_1 \neq Y_1')_\beta + \sqrt{2} \eps  \\
      & = \Pr(Y_1 = Y_1')_\beta\Pr(G=G'~|~ Y_1 = Y_1')_\beta + \sqrt{2}\eps  \\
   & \leq \gamma + \sqrt{2}\eps = \cO(\eps). 
\end{align*}
Note that conditioned on $Y_1 = Y_1'$, we have $I = I'$. The first inequality above follows from Eq.~\eqref{advgen:eq11} and noting that the predicate $(G=G')$ is determined from $(I,Y,Y')$. The second equality follows from definition of $G$.  Second inequality follows since $\ecc$ has relative distance $1 - \gamma$. 

\suppress{

For any fixing $(Y,Y')=(y,y')$ such that $y_1 \ne y_1'$, we have  $G_{(Y,Y')=(y,y')}\ne G_{(Y,Y')=(y,y')}$. Now consider for a fixing $(Y,Y')=(y,y')$ such that $y_1 = y_1'$, then we have $I_{(Y,Y')=(y,y')} =I'_{(Y,Y')=(y,y')}$. Note the event $(G = G')_\theta$ can be determined given registers $\theta_{IYY'}$. Since $\ecc$ has a relative distance of $1-\gamma$, for a uniformly chosen $i \in \{1, \ldots, v \}$, the probability that $\ecc(y)_i = \ecc(y')_i$ is bounded by $\gamma.$ Thus from Eq.~\eqref{advgen:eq12}, 
$$\Pr(G=G')_\theta \leq \gamma + \eps .$$

Thus with probability at least $1-\sqrt{\eps^2}$ over $(y,y') \leftarrow (Y,Y')$ it holds that 
    \[ \Ext_0(X,y_1) \approx_{\sqrt{\eps^2}} U_{\log v}\] using standard Markov argument. Together, thus in total we incur the error $\cO(\gamma + \eps)$, i.e.
    $$\Pr(G=G')_\theta = \cO(\gamma + \eps) .$$}

We now prove $d(T|\tilde{B})_\Phi\leq \eps$. Consider Protocol~\ref{prot:block1mod}. Let $(E_A, E_A') \longleftrightarrow (E_B, E_B')$ represent  $2d_1$ distinct EPR pairs each shared between Alice and Bob where $\vert E_A \vert = \vert E_A' \vert= \vert E_B \vert = \vert E_B' \vert= d_1$. 

Let $\tau''$ be the joint state just before Alice receives $Y_2$ and $\tau'$ be the joint state just after Alice generates $T$. Note, $\tau''_{\tilde{A}Y_2} =\tau''_{\tilde{A}} \otimes U_{d_2}$. Hence from Lemma~\ref{lem:2},
\begin{equation}\label{eq2021}
    \Delta_B( \tau'_{T\tilde{B}} ,U_d \otimes  \tau'_{\tilde{B}}) \leq \sqrt{\eps'}.
\end{equation}
Let $\tau$ be the joint state just before Bob  checks $(E_B,E_B')=(Y_1,Y_1')$. Let $C$ be the predicate  $((E_B,E_B')=(Y_1,Y_1'))$. Note \begin{equation}
    \Pr(C=1)_\tau=\Pr((E_B,E_B')=(Y_1,Y_1'))_\tau \geq 2^{-2d_1}. \label{eq:cprob}
\end{equation}
Let $\Phi'$ be the state $\Phi$ with two additional copies of $Y_1 Y_1'$. Note that the state at the end of Protocol~\ref{prot:block1mod}, conditioned on Bob not aborting is $\Phi'$. Consider,
\begin{align*}
   & 2^{-2d_1} \Delta_B( \Phi_{T
   \tilde{B}} ,U_d \otimes  \Phi_{\tilde{B}}) \\ 
   & \leq 2^{-2d_1} \Delta_B( \Phi'_{T
   \tilde{B}} ,U_d \otimes  \Phi'_{\tilde{B}}) & \mbox{(Fact~\ref{fact:data})}\\ 
   & \leq \Pr(C=1)_\tau \Delta_B( \Phi'_{T\tilde{B}} ,U_d \otimes  \Phi'_{\tilde{B}}) & \mbox{(Eq.~\eqref{eq:cprob})}\\
   & \leq \Delta_B( \tau_{T\tilde{B}} ,U_d \otimes  \tau_{\tilde{B}}) & \mbox{(Fact~\ref{traceavg})} \\
   & \leq \Delta_B( \tau'_{T\tilde{B}} ,U_d \otimes  \tau'_{\tilde{B}}) & \mbox{(Fact~\ref{fact:data})} \\
   & \leq \sqrt{\eps'} & \mbox{(Eq.~\eqref{eq2021})}
\end{align*}
which shows the desired.
   \end{proof}

\begin{claim}[Correlation breaker with advice]\label{thm:advcb}
Let Alice and Bob proceed as in Protocol~\ref{prot:block2} with the starting state as $\hat{\Phi}^{(\alpha,\alpha')}$, where $(\alpha, \alpha') \notin \mathcal{S}$. Let $\hat{\Gamma}^{(\alpha,\alpha')}$ be the joint state at the end of the Protocol~\ref{prot:block2} (at $i=a$). Then,
 \[d(Z|\tilde{B})_{\hat{\Gamma}^{(\alpha,\alpha')}} = \cO(\eps). \] 
\end{claim}
\begin{proof} We have 
    \[\hat{\Phi}^{(\alpha,\alpha')}_{Z_0\tilde{B}} = U_h \otimes \hat{\Phi}^{(\alpha,\alpha')}_{\tilde{B}} \quad ; \quad \hmin{Y}{\tilde{A}}_{\hat{\Phi}^{(\alpha,\alpha')}} \geq (3/4-2\delta)d \quad ; \quad  \hmin{T}{\tilde{B}}_{\hat{\Phi}^{(\alpha,\alpha')}} = d.\]
    \suppress{
    Protocol~\ref{prot:Var_GEN(0,1)analysis} corresponds to execution of flip flop procedure when $(\alpha_i,\alpha_i')=(0,1)$, Protocol~\ref{prot:Var_GEN(1,0)} when $(\alpha_i,\alpha_i')=(1,0)$, Protocol~\ref{prot:Var_GEN(0,0)NotDiffBefore} when $(\alpha_i,\alpha_i')=(0,0)$ and $\alpha_j=\alpha_j'$ for $j<i$, Protocol~\ref{prot:Var_GEN(0,0)DiffBefore} when $(\alpha_i,\alpha_i')=(0,0)$ and $\alpha_j \ne \alpha_j'$ for some $j<i$, Protocol~\ref{prot:Var_GEN(1,1)NotDiffBefore} when $(\alpha_i,\alpha_i')=(1,1)$ and $\alpha_j=\alpha_j'$ for $j<i$, Protocol~\ref{prot:Var_GEN(1,1)DiffBefore} when $(\alpha_i,\alpha_i')=(1,1)$ and $\alpha_j \ne \alpha_j'$ for some $j<i$.}
    The total communication from Alice to Bob in Protocol~\ref{prot:block2} is at most $6ah \leq \delta d$. From Lemma~\ref{lem:minentropy}, throughout Protocol~\ref{prot:block2}, we have  $\hmin{T}{\tilde{B}} \geq (1-\delta)d > 2h$. From repeated applications of Claim~\ref{lem:alpha01} we have,
 \[d(Z|\tilde{B})_{\hat{\Gamma}^{(\alpha,\alpha')}} \leq 2^{\cO(a)}\sqrt{\eps'} = \cO(\eps).\] 
   \end{proof}\suppress{
	For the analysis of the flip flop procedure we distinguish between six separate cases. The first two correspond to $\alpha_i \ne \alpha_i'$. We state $6$ different protocols corresponding to each case. Protocol~\ref{prot:Var_GEN(0,1)analysis} corresponds to execution of flip flop procedure when $(\alpha_i,\alpha_i')=(0,1)$, Protocol~\ref{prot:Var_GEN(1,0)} when $(\alpha_i,\alpha_i')=(1,0)$, Protocol~\ref{prot:Var_GEN(0,0)NotDiffBefore} when $(\alpha_i,\alpha_i')=(0,0)$ and $\alpha_j=\alpha_j'$ for $j<i$, Protocol~\ref{prot:Var_GEN(0,0)DiffBefore} when $(\alpha_i,\alpha_i')=(0,0)$ and $\alpha_j \ne \alpha_j'$ for some $j<i$, Protocol~\ref{prot:Var_GEN(1,1)NotDiffBefore} when $(\alpha_i,\alpha_i')=(1,1)$ and $\alpha_j=\alpha_j'$ for $j<i$, Protocol~\ref{prot:Var_GEN(1,1)DiffBefore} when $(\alpha_i,\alpha_i')=(1,1)$ and $\alpha_j \ne \alpha_j'$ for some $j<i$.}
	\begin{claim}[Flip flop]\label{lem:alpha01}
Let $\cP$ be any of the Protocols~\ref{prot:Var_GEN(0,1)analysis},~\ref{prot:Var_GEN(1,0)},~\ref{prot:Var_GEN(0,0)NotDiffBefore},~\ref{prot:Var_GEN(0,0)DiffBefore},~\ref{prot:Var_GEN(1,1)NotDiffBefore} or~\ref{prot:Var_GEN(1,1)DiffBefore} and $i \in [a]$. Let $\alpha$ be the initial joint state in~$\cP$ such that 
	$d(Z|\tilde{B})_\alpha \leq \eta.$  Let $\theta$ be the final joint state at the end of~$\cP$. Then, 
 $d(O|\tilde{B})_\theta \leq \cO(\eta + \sqrt{\eps'})$.
	\end{claim}
	\begin{proof}
We prove the claim when $\cP$ is Protocol~\ref{prot:Var_GEN(0,1)analysis} and $i=1$ and the proof for other cases follows analogously.  
From Fact~\ref{fact:data}, 
$$d(Z_s|\tilde{B})_\alpha \leq d(Z|\tilde{B})_\alpha \leq \eta.$$
\suppress{
Also, from Fact~\ref{fact2} and Fact~\ref{fact102}, we have 
\[ \hmin{T}{YY'Z_s\tilde{M}}_\tau  \geq  \hmin{T}{YY'Z_s\tilde{M}}_\rho- \vert Z_s \vert \geq k_1-s \] 
and \[ \hmin{Y}{TT'ZZ'\tilde{N}}_\tau  \geq  \hmin{Y}{TT'ZZ'\tilde{N}}_\rho \geq k_2. \]}
Let $\gamma$ be the joint state just after Bob generates register $A$. From Lemma~\ref{lem:2}, we have 
$$d(A|\tilde{A})_\gamma \leq 2\eta + \sqrt{\eps'}.$$
Let $\zeta$ be the joint state after Alice sends register $Z'$ to Bob and Bob generates registers $(A'C'B')$. From Fact~\ref{fact:data}, we have $$ d(A|\tilde{A})_\zeta \leq d(A|\tilde{A})_\gamma\leq 2\eta + \sqrt{\eps'}.$$
Let $\beta$ be the joint state just after Alice generates register $\overline{Z}$. From Lemma~\ref{lem:2},  
$$d(\overline{Z}|\tilde{B})_\beta \leq 4\eta + 3\sqrt{\eps'}.$$ Let $\hat{\beta}$ be the state obtained from Claim~\ref{claim:100} (by letting $\rho$ in Claim~\ref{claim:100} as $\beta$ here) such that
\begin{equation}
 \label{eq:beta}   
 \hmin{Y}{\tilde{A}}_{\hat{\beta}}= \hmin{Y}{\tilde{A}}_{\beta} \geq d/4 \quad ; \quad \hat{\beta}_{\overline{Z}\tilde{B}}= U_h \otimes \hat{\beta}_{\tilde{B}} \quad ; \quad  \Delta_B(\hat{\beta} ,\beta) \leq 4\eta + 3\sqrt{\eps'}. 
 \end{equation}
 Let $\theta', \hat{\theta}'$ be the joint states just after Alice generates register $\overline{C}$, proceeding from the states  $\beta, \hat{\beta}$ respectively. Since communication between Alice and Bob after Alice generates register $\overline{Z}$ and before generating $\overline{C}$ is $2s+2b$, from arguments as before involving Lemma~\ref{lem:2} and Lemma~\ref{lem:minentropy},$$d(\overline{C}|\tilde{B})_{\hat{\theta}'} \leq \cO(\eta +\sqrt{\eps'}).$$
 From Eq.~\eqref{eq:beta},  
 $$d(\overline{C}|\tilde{B})_{\theta'} \leq d(\overline{C}|\tilde{B})_{\hat{\theta'}} + \Delta_B(\hat{\beta} ,\beta) = \cO(\eta +\sqrt{\eps'}).$$
 Proceeding till the last round and using similar arguments involving Lemma~\ref{lem:2}, Lemma~\ref{lem:minentropy} and Claim~\ref{claim:100}, we get the desired.
\end{proof}

\suppress{
We call a CPTP map $\Phi: \cH_X \otimes \cH_A \rightarrow \cH_X \otimes \cH_B$, {\em safe} on classical register $X$ iff there is a collection of CPTP maps $\Phi_x: \cH_A\rightarrow \cH_B$ such that the following holds.  For all c-q states $\rho_{XA} = \sum_x \Pr(X=x)_{\rho} \ketbra{x} \otimes  \rho^x_A$,
$$\Phi({\rho}_{XA}) =  \sum_x \Pr(X=x)_{\rho} \ketbra{x} \otimes \Phi_x( \rho^x_A).$$}
We have the following corollary of Theorem~\ref{thm:nmext}.
\begin{corollary}\label{corr:add1}
 Let $\rho_{XEY}$ be a c-q state with registers ($XY$) classical such that  \[ \hmin{X}{E}_\rho \geq k \quad ; \quad \rho_{XEY} =\rho_{XE} \otimes U_d \quad ; \quad \vert X \vert =n.\]Let  $\mathsf{T}: \mathcal{L} (\mathcal{H}_E \otimes \mathcal{H}_{Y}) \rightarrow \mathcal{L} (\mathcal{H}_{E^\prime} \otimes \mathcal{H}_Y \otimes \mathcal{H}_{Y^\prime})$ be a (safe) CPTP map such that for $\sigma_{XE'YY'} =\mathsf{T}(\rho_{XEY})$, we have registers $XYY'$ classical and $\Pr(Y \ne Y')_\sigma=1$. Let the function $\nmext$ be as defined in Algorithm~\ref{alg:nmExt},  $L= \nmext(X,Y)$ and $L'=\nmext(X,Y')$. Then, 
 $$ \| \sigma_{LL'YY'E'} - U_{k/4} \otimes \sigma_{L'YY'E'} \|_1 \leq \cO(\eps).$$
\end{corollary}
\begin{proof}
Let $\rho_{X\hat{X} E \hat{E}Y\hat{Y}}$ be a pure state extension of $\rho_{XEY}$ such that,
\[\rho_{X\hat{X}  \hat{E}EY\hat{Y}} = \rho_{X\hat{X}  \hat{E}E} \otimes \rho_{Y\hat{Y}} \quad ; \quad  \hmin{X}{EY\hat{Y}}_\rho =\hmin{X}{E}_\rho \geq k \quad ; \quad \rho_Y = U_{d},\]
where registers ($XY$) classical (with copies $\hat{X}\hat{Y}$) and $ \rho_{X\hat{X}  \hat{E}E}$ is the canonical purification of $\rho_{XE}$.

For the state $\rho$ with the following assignment (terms on the left are from Definition~\ref{qmadvk1k2} and on the right are from here),
\[(X,\hat{X},N,M,Y,\hat{Y}) \leftarrow (X,\hat{X},\hat{E},E,Y,\hat{Y}),\]$\hmin{X}{EY\hat{Y}}_\rho =\hmin{X}{E}_\rho \geq k$ and $\rho_{\hat{E}X\hat{X}Y}=\rho_{\hat{E}X\hat{X}} \otimes U_d$, we have $\rho$ is a $(k) \mhyphen \qpas$. Let $V: \mathcal{H}_E \otimes \mathcal{H}_{Y} \rightarrow \mathcal{H}_{E^\prime} \otimes \mathcal{H}_{Z} \otimes \mathcal{H}_Y \otimes \mathcal{H}_{Y^\prime} \otimes \mathcal{H}_{ \hat{Y^\prime}}$ be the Stinespring isometry extension~\footnote{Note the Stinespring isometry extension is safe on register $Y$.} of CPTP map $\mathsf{T}$ with additional copy $\hat{Y^\prime}$ of $Y^\prime$, i.e. $\mathsf{T}(\theta)=  \tr_{Z  \hat{Y^\prime}}(V \theta V^\dagger)$ for every c-q state $\theta_{YE}$. Let $\sigma=V \rho V^\dagger$. Note  $\sigma_{XE'YY'} = \tr_{Z\hat{Y^\prime}}(V \rho_{XEY} V^\dagger) = \mathsf{T}(\rho_{XEY})$ and $\sigma$ is a $(k) \mhyphen \nmas$. Using Theorem~\ref{thm:nmext}, we have 
$$ \| \sigma_{LL'YY'E'Z} - U_{k/4} \otimes \sigma_{L'YY'E'Z} \|_1 \leq  \cO(\eps).$$ Using Fact~\ref{fact:data}, we further have
$$ \| \sigma_{LL'YY'E'} - U_{k/4} \otimes \sigma_{L'YY'E'} \|_1 \leq  \cO(\eps),$$
which completes the proof. 
\end{proof}

\section{A quantum secure $2$-source non-malleable extractor\label{sec:2nm}}
In this section, we define and prove the quantum security of $2$-source non-malleable extractor. As specified before, the parameters in our construction are set similarly in line with the construction of~\cite{CGL15} considering the use of quantum secure seeded extractors in the alternating extraction. The following parameters hold throughout this section.
 \subsection*{Parameters}\label{sec:parameters_2nm}
Let $q$ be a prime power and $\delta, \delta_1 >0$ be small enough constants. Let  $n,a,v,s,b,h$ be positive integers and
 $k, \eps', \gamma, \eps > 0$ such that: 
  \[  v= \frac{n}{\eps} \quad ; \quad q= \cO\left(\frac{1}{\eps^2}\right) \quad ; \quad \eps= 2^{- \cO(n^{\delta_1})} \quad ;\]
 
\[a=6k+2 \log q = \mathcal{O}(k) \quad ; \quad \gamma = \cO(\eps) \quad ; \quad  2^{\cO(a)}\sqrt{\eps'} = \eps \quad ;       \]

\[s = \cO\left(\log^2\left(\frac{n}{\eps'}\right)\log n \right)  \quad  ; \quad b = \cO\left( \log^2\left(\frac{n}{\eps'}\right) \log n \right) \quad ; \quad h = 10s \quad ;\quad k = \cO (n^{1/4}) \]
\begin{itemize}\label{sec:extparameters_2nm}
    \item $\IP_1$ be $\IP^{3k/\log v}_{v}$,
    \item $\Ext_1$ be $(2b, \eps')$-quantum secure $(n,s,b)$-extractor,
    \item $\Ext_2$ be $(2s, \eps')$-quantum secure $(h,b,s)$-extractor,
    \item $\Ext_3$ be $(2h, \eps')$-quantum secure $(n,b,h)$-extractor,
    \item $\Ext_4$ be $(n/4, \eps^2)$-quantum secure $(n,h,n/8)$-extractor,
    \item $\IP_2$ be $\IP^{3k^3/h}_{2^h}$,
    \item  $\Ext_6$ be $(\frac{n}{2}, \eps^2)$-quantum secure $(n,n/8,n/4)$-extractor. 
 \end{itemize}
  Let $\ecc : \F^d_q \to \F^v_q$ be an error correcting code with relative distance $1-\gamma$ and rate $\eps$ (which exists from~Fact~\ref{fact:ecc} for our choice of parameters). We identify $R$ as an element from $\{1, \ldots, v \}$. By $\ecc(Y)_{R}$, we mean the $R$-th entry of the code-word $\ecc(Y)$, interpreted as a bit string of length $\log q$.

\subsection*{Description of $2$-source non-malleable extractor}

Similar to $\nmext$, the algorithm for $2$-source non-malleable extractor can also be viewed in three steps:
\begin{itemize}
    \item Advice generation (Step 1 in Algorithm~\ref{alg:2nmExt})
    \item Correlation breakers with advice (Step 3 in Algorithm~\ref{alg:2nmExt}), that are built using the flip-flop primitive (Algorithm~\ref{alg:2FF}). 
    \item Improving the output length (Step 4 in Algorithm~\ref{alg:2nmExt}).
\end{itemize}
The key difference in the constructions of  $2$-source non-malleable extractor and seeded non-malleable extractor is in the Advice generation step which we elaborate below. 
The main reason why advice generation needs to be modified is that none of the sources are uniform and Advice generation step from Step 1 in Algorithm~\ref{alg:nmExt} crucially uses the fact that the second source (seed) is uniform.
One gets around this by using a quantum-secure 2-source extractor (for example, inner product extractor) in place of the Trevisan's extractor.
The argument then follows in similar lines and we state it here briefly for completeness.
 Let $XX'$  and $YY'$ be arbitrarily correlated random variables such that $Y \ne Y'$ and $X \ne X'$. 
Also assume that $X$ and $Y$ both have sufficient min-entropy.
Recall that the goal is to come up with a function such that, with high probability, $G = f(X,Y) \ne f(X',Y') =G^\prime$. 
\begin{itemize}
    \item Let $\ecc$ be a error correcting code of constant rate and relative distance close to $1$. Since $Y \ne Y'$, it is clear that $\ecc(Y)$ and $\ecc(Y')$ differ at most coordinates. Similarly, $\ecc(X)$ and $\ecc(X')$ differ at most coordinates.
    \item Take $Y_1$ (a prefix of $Y$) and $X_1$ (a prefix of $X$) to generate $I = \IP(X_1,Y_1)$. Since $XX'$ is independent of $YY'$, it follows from $2$-source strong extractor properties of $\IP$ that $I$ is independent of $YY'$ (and analogously, independent of $XX'$). Thus $\ecc(Y)_I$ and $\ecc(Y')_I$  are not equal with high probability.
    Similarly, $\ecc(X)_I$ and $\ecc(X')_I$  are not equal with high probability.
    \item Define $G=X_1 \circ \ecc(X)_I \circ Y_1 \circ \ecc(Y)_I$ and $G'=X'_1 \circ \ecc(X')_I \circ Y'_1 \circ \ecc(Y')_{I'}$.
     If $X_1Y_1 \ne X'_1Y'_1$, then $G \ne G'$ trivially.
    Otherwise, $X_1Y_1 = X'_1Y'_1$, then $I=I'$ and $\ecc(Y)_I \ne \ecc(Y')_I$, with high probability.
\end{itemize}

\begin{algorithm}
\caption{: $2\nmext: \lbrace 0,1 \rbrace ^n\times \lbrace 0,1 \rbrace^n   \rightarrow \lbrace 0,1 \rbrace^{n/4}$}\label{alg:2nmExt}
\begin{algorithmic}
\State{}

\noindent \textbf{ Input:}  $X, Y$\\


\begin{enumerate}
    \item Advice generator: \[X_1=\pre(X,3k) \quad  ; \quad Y_1 = \pre(Y,3k) \quad ;
    \quad R= \IP_1(X_1,Y_1) \quad ; \quad  \]
    \[G=X_1 \circ Y_1 \circ \ecc(X)_{R} \circ \ecc(Y)_{R} \]

    \item $X_2=\pre(X,3k^3) \quad ; \quad Y_2=\pre(Y,3k^3)\quad; \quad Z_0=\IP_2(X_2,Y_2)$
    \item Correlation breaker with advice:\quad $S=2\advcb(Y,X,Z_0, G)$
    \item $L=\Ext_6(X,S)$
\end{enumerate}

 \noindent \textbf{ Output:} $L$ 
\end{algorithmic}
\end{algorithm}


\begin{algorithm}
\caption{: $2\advcb: \lbrace 0,1 \rbrace^n \times \lbrace 0,1 \rbrace^n \times \lbrace 0,1\rbrace^h \times \lbrace 0,1 \rbrace^a \rightarrow \lbrace 0,1 \rbrace^{\frac{n}{8}}$}\label{alg:2AdvCB}
\begin{algorithmic}
\State{}

\noindent \textbf{ Input: }$Y, X, Z_0, G$
\suppress{
Let $X \in \lbrace 0,1\rbrace^n,\ Z\in \lbrace 0,1 \rbrace^l$ be random variables with $\alpha \in \lbrace 0,1 \rbrace^a$ as the input. Let $n, a$ be integers, $\eps>0$. Let $s,b,h$ be integers defined as 
\[s = \cO\left(\log^3\left(\frac{n}{\eps}\right)\right)  \quad  ; \quad b = \cO\left( \log^3\left(\frac{l}{\eps}\right) \right) \quad ; \quad h = \Theta(s) \quad ; \quad h \leq l \quad ; \quad n \geq \log^3(l),\]
where we set $l=cah$ for some large enough constant $c$. Note that this choice of $l$ meets the above required conditions. Let $k_x=cab, m=k_x/4$. \\ \\
Let $\Ext_4: \lbrace 0,1 \rbrace^n \times \lbrace 0,1 \rbrace^h  \rightarrow \lbrace 0,1 \rbrace^m$ be $(2m, \eps)$-quantum secure extractor.  \\ \\}

\begin{enumerate}
    \item For $i=1,2,\ldots,a:$ 
    
     \hspace{1cm}Flip flop: \quad $Z_i=2\ff(Y,X,Z_{i-1},G_i)$ 
     \item $S=\Ext_4(Y,Z_a)$

\end{enumerate}

 \noindent \textbf{ Output:} $S$ 
\end{algorithmic}
\end{algorithm} 


\begin{algorithm}
\caption{: $2\ff : \lbrace 0,1 \rbrace^n \times \lbrace 0,1 \rbrace^n \times \lbrace 0,1 \rbrace^h \times \lbrace 0,1 \rbrace \rightarrow \lbrace 0,1\rbrace^h$}\label{alg:2FF}
\begin{algorithmic}
\State{}

\noindent\textbf{Input:}  $Y, X,  Z,  G$ 
\begin{enumerate}
     
    \item $Z_s=$Prefix$(Z,s)$, $A= \Ext_1 (Y,Z_s),\ C= \Ext_2(Z,A),\ B= \Ext_{1}(Y,C)$
    \item If $G=0$ then $\overline{Z}= \Ext_3(X,A)$ and if $G=1$ then $\overline{Z}= \Ext_3(X,B)$
     \item $\overline{Z}_s=$Prefix$(\overline{Z},s)$, $\overline{A}= \Ext_1 ({Y},\overline{Z}_s),\ \overline{C}= \Ext_2(\overline{Z},\overline{A}),\ \overline{B}= \Ext_{1}({Y},\overline{C})$
    \item If $G=0$, then $O=\Ext_3(X,\overline{B})$ and if $G=1$, then $O=\Ext_3(X,\overline{A})$
\end{enumerate} \noindent \textbf{Output:} $O$
\end{algorithmic}
\end{algorithm}

\subsection*{Result}
The following theorem shows that the function $2\nmext$ as defined in Algorithm~\ref{alg:2nmExt} is $(n-k,n-k,\cO(\eps))$-secure against $\nma$ by noting that  $L= \nmext(X,Y)$ and $L'=\nmext(X^\prime,Y^\prime)$.
\begin{theorem}[Security of $2\nmext$] \label{thm:2nmext}

Let $\rho_{X\hat{X}X^\prime \hat{X}' NYY'\hat{Y}\hat{Y}'M}$ be a $(n-k,n-k)\mhyphen\nmas$ with $\vert X \vert = \vert Y \vert = n$. Let Protocol~\ref{prot:2nmExt_full} start with $\rho$. Let $\Lambda$ be the state at the end of the protocol.\suppress{
Let $\rho_{X X^\prime N Y Y^\prime M}$ be an  $(n-k,n-k)\mhyphen\nmas$ such that $\vert X \vert =\vert Y \vert =n$. Let Alice and Bob proceed according to Protocol~\ref{prot:2nmExt_full} starting with $\rho$ (Alice holds $X X^\prime N$ and Bob holds $Y Y^\prime M$). Let $\Psi_{LNL'YY'M}$ (registers $(LN)$ with Alice and registers $(L'YY'M)$ with Bob) be the state reached at the end of the protocol.} Then,
\suppress{$$ \Vert \rho_{ 2\nmext(X,Y)2\nmext(X^\prime,Y^\prime) Y  Y^\prime M^\prime} - U_{n/4} \otimes \rho_{ 2\nmext(X^\prime ,Y^\prime) Y  Y^\prime M^\prime} \Vert_1 \leq d(L \vert \tilde{B})_\Lambda \leq \cO( \eps).$$ }

\[ \Vert \rho_{ L L^{\prime} Y  Y^\prime M} - U_{n/4} \otimes \rho_{ L^\prime Y  Y^\prime M} \Vert_1 \leq d(L \vert \tilde{B})_\Lambda \leq \cO( \eps).\]
	
\end{theorem}
\begin{proof} The first inequality follows from Fact~\ref{fact:data}.
\suppress{Note that the starting state $\rho$ is symmetric in Alice and Bob's sides.
So one can get a similar statement by replacing Alice and Bob's roles.
We show only one side and the second one follows analogously.}
Most of our arguments here are similar to the case of seeded extractor; so we note the modifications that we need to take care of in case of \textsf{2nmExt}.
\suppress{
 Note that the total communication from Alice to Bob in Protocol~\ref{prot:2nmExt_full} is at most (from our choice of parameters) 
$$6k+2 \log q+ 6ah + h+ \frac{n}{4} \leq  \left(1/4 + \delta\right) n.$$
This implies (using Lemma~\ref{lem:minentropy}) that throughout Protocol~\ref{prot:2nmExt_full}, $\hmin{X}{\tilde{B}} \geq (3/4-\delta)n-k$. 

The total communication from Bob to Alice in  Protocol~\ref{prot:2nmExt_full} is at most $$46k + 2a + 6ab + \frac{n}{4} \leq (1/4+\delta) d.$$
Again using Lemma~\ref{lem:minentropy}, throughout Protocol~\ref{prot:2nmExt_full}, $\hmin{Y}{\tilde{A}} \geq (3/4-\delta)d$.}

First note that using Lemma~\ref{lem:minentropy}, throughout Protocol~\ref{prot:2nmExt_full}, $X$ and $Y$ have enough conditional min-entropy left for necessary extractions since the total communication from Alice to Bob is at most (from our choice of parameters) 
$$6k+2 \log q+ 6ah + h+ \frac{n}{4} \leq  \left(1/4 + \delta\right) n$$
 and the total communication from Bob to Alice is at most $$6k +6k^3 + 2a + 6ab + \frac{n}{4} \leq (1/4+\delta) n.$$Thus, at any state $\varrho$ in Protocol~\ref{prot:2nmExt_full},
$\hmin{X}{\tilde{B}}_\varrho \geq n-k-\left(1/4 + \delta\right) n \geq \left(3/4 -2 \delta\right) n \geq n/2$. Similarly, $\hmin{Y}{\tilde{A}}_\varrho \geq \left(3/4 -2\delta\right)n \geq n/2$.

We start with a state $\rho_{X X' N Y Y' M}$ such that $\hmin{X}{\tilde{B}}_{\rho} \geq n-k$ and $\hmin{Y}{\tilde{A}}_{\rho} \geq n-k$. From Claim~\ref{fact:prefixminentropyfact}, we have,
\[\hmin{X_1}{\tilde{B}}_{\rho} \geq 3k- k =2k \quad ; \quad \hmin{Y_1}{\tilde{A}}_{\rho} \geq 3k-k=2k .\]
Now from Claim~\ref{l-qma-needed-fact1} with the below assignment of registers (and noting registers $(XX',YY' )$ are included in $(\tilde{A},\tilde{B})$ respectively),
\[(Z,X,Y, \sigma) \leftarrow (R,X_1, Y_1, \rho) \quad ; \quad (k_1,k_2,m,n_1,\eps ) \leftarrow (2k,2k, \log(n/\eps), 3k, \eps^2) \]we have,
\[ \Delta( \rho_{RYY'}, U_R \otimes \rho_{YY'} ) \leq \cO(\eps^2) \quad ; \quad \Delta( \rho_{RXX'}, U_R \otimes \rho_{XX'} ) \leq  \cO(\eps^2). \]
Using Fact~\ref{fidelty_trace}, we get 
\[ \Delta_B( \rho_{RYY'}, U_R \otimes \rho_{YY'} ) \leq  \cO(\eps) \quad ; \quad \Delta_B( \rho_{RXX'}, U_R \otimes \rho_{XX'} ) \leq  \cO(\eps). \] 

\suppress{
\mycomment{ Suppressed the following and used fact 18 directly. Using Lemma~\ref{lemma:nearby_rho_prime_prime}, there exists an $l$-$\qmas$\  $\rho^{(1)}$ such that
\[\Delta_B(\rho^{(1)},\rho) \leq \mathcal{O}(\eps) \quad ; \quad l \leq 6k-4k+4+\cO \left( \log \left( \frac{1}{\eps}\right) \right) \leq 2.1k. \]Using Fact~\ref{l-qma-needed-fact}, for our choice of parameters, it follows that
\[d(R \vert YY')_{\rho^{(1)}} \leq \cO(\eps) \quad ; \quad d(R \vert XX')_{\rho^{(1)}} \leq \mathcal{O}(\eps). \]
Now from Claim~\ref{claim:traingle_rho_rho_prime} with the below assignment of registers (and $(\rho, \rho') \leftarrow (\rho, \rho^{(1)})$),
\[(Z,A ) \leftarrow (R, YY') \quad ; \quad (Z,A ) \leftarrow (R, XX') \]we have,
\[d(R \vert YY')_{\rho}\leq  \cO(\eps) \quad ; \quad d(R \vert XX')_{\rho} \leq \cO(\eps). \]}}Let $\kappa$ be the state just before Bob sends $Y_2$. Note that till then, communication from Alice to Bob and Bob to Alice is at most $7k$ each. Hence, by Lemma~\ref{lem:minentropy}, $\hmin{X}{\tilde{A}}_{\kappa} \geq n-8k$ and $\hmin{Y}{\tilde{B}}_\kappa \geq n-8k$; which implies (from Claim~\ref{fact:prefixminentropyfact}), $\hmin{X_2}{\tilde{A}}_{\kappa} \geq 3k^3-8k \geq 2k^3$ and $\hmin{Y_2}{\tilde{B}}_\kappa \geq 2k^3$ respectively using Fact~\ref{fact:prefixminentropyfact}. Let $\eta$ be the state just after Alice generates $Z_0$.
Using similar argument as before involving  Claim~\ref{l-qma-needed-fact1}, we have $d(Z_0 \vert \tilde{B})_{\eta} \leq  \mathcal{O}(\eps)$.\suppress{ Modified - Using similar argument as before involving Lemma~\ref{lemma:nearby_rho_prime_prime},  Fact~\ref{l-qma-needed-fact} and Claim~\ref{claim:traingle_rho_rho_prime}, we have $d(Z_0 \vert \tilde{B})_{\eta} \leq  \mathcal{O}(\eps)$.  }

\suppress{
Again, by Lemma~\ref{lemma:nearby_rho_prime_prime}, there exists a state $\kappa^{(1)}$ which is an $l$-\qma\  state such that $\Delta_B(\kappa, \kappa^{(1)}) \leq \cO(\eps)$ and $l \leq  6 k^3 -4 k^3 +4 + \cO\left(\log \frac{1}{\eps}\right) \leq 2.1k^3.$ For the given choice of parameters, from Fact~\ref{l-qma-needed-fact}, we get $d(Z\vert \tilde{B})_{\kappa^{(1)}} \leq \mathcal{O}(\eps)$, which in turn, using Claim~\ref{claim:traingle_rho_rho_prime}, gives $d(Z \vert \tilde{B})_{\kappa} \leq  \mathcal{O}(\eps)$. }
Note that the state obtained in Protocol~\ref{prot:2nmExt_full}, just before Protocol~\ref{prot:block2} starts as a subroutine, is similar to the state obtained in Protocol~\ref{prot:block1} (just before Protocol~\ref{prot:block2} starts as a subroutine) with the below assignment of registers,
\[(Z_0, T, Y ) \leftarrow (Z_0, X, Y) .\]Here the variables on the left are from Protocol~\ref{prot:block1} and variables on the right are from Protocol~\ref{prot:2nmExt_full}. The proof then proceeds using similar arguments as~Theorem~\ref{thm:nmext} involving Lemma~\ref{lem:2}, Lemma~\ref{lem:minentropy}, Claim~\ref{claim:100} after noting Claim~\ref{lemma:block12t}.

We can verify the following claim regarding the state $\Phi$ (the state obtained in Protocol~\ref{prot:2nmExt_full}, just before Protocol~\ref{prot:block2} starts as a subroutine) using similar arguments as proof of Claim~\ref{lemma:block1}.
\begin{claim}\label{lemma:block12t}
1. $\Pr(G= G^\prime)_{\Phi} = \mathcal{O(\eps)}$ \quad and \quad 2. $d(Z_0\vert \tilde{B})_\Phi \leq \mathcal{O}(\eps)$. 
\end{claim}
Since we have either $\Pr(Y \neq Y^\prime)=1$ or $\Pr(X \neq X^\prime)=1$, we show it for the first case and second case will follow analogously.
Now note that the event $G=G^\prime$ is a sub-event of $G_1=G_1^\prime$ where $G_1=Y_1 \circ \ecc(Y)_R$ and $G_1^\prime= Y_1^\prime \circ \ecc(Y^\prime)_{R^\prime}$.
Thus we get,
$\Pr(G = G^\prime)_\Phi \leq \Pr(G_1 = G_1^\prime)_\Phi$.
Rest of the argument follows similar lines to Claim~\ref{lemma:block1}. 

This completes the proof.
\end{proof}

We have the following corollary of Theorem~\ref{thm:2nmext}.
\begin{corollary}\label{corr:add2}
Let the function $2 \nmext$ be as defined in Algorithm~\ref{alg:2nmExt}. $2 \nmext$ is an $(n-k,n-k,\cO(\eps))$-quantum secure $2$-source non-malleable extractor against $\qia$.
\end{corollary}
\begin{proof}
Let  $\rho_{XE_1E_2Y}$ be a state (for $k_1=k_2=n-k$), $\mathsf{T}_1$ and  $\mathsf{T}_2$ be CPTP maps as defined in Definition~\ref{intro2nmextiadvmodel}.
Let $\rho_{X\hat{X} E_1 \hat{E_1}E_2 \hat{E_2}Y\hat{Y}}$ be a pure state extension of $\rho_{XE_1E_2Y}$ such that,
\[\rho_{X\hat{X} E_1 \hat{E_1}E_2 \hat{E_2}Y\hat{Y}}= \rho_{X\hat{X}  \hat{E_1}E_1} \otimes \rho_{Y\hat{Y} \hat{E_2}E_2} \quad ; \quad  \hmin{X}{E_1}_\rho \geq n-k \quad ; \quad \hmin{Y}{E_2}_\rho \geq n-k ,\]
where registers ($XY$) are classical (with copies $\hat{X}\hat{Y}$), $\rho_{X\hat{X}  \hat{E_1}E_1} $ is canonical purification of $\rho_{XE_1}$ and $\rho_{Y\hat{Y} \hat{E_2}E_2}$ is canonical purification of $\rho_{YE_2}.$

Let $U: \mathcal{H}_{E_2} \otimes \mathcal{H}_{X} \rightarrow \mathcal{H}_{E_2^\prime} \otimes \mathcal{H}_{Z_2} \otimes \mathcal{H}_X \otimes \mathcal{H}_{X^\prime} \otimes \mathcal{H}_{ \hat{X^\prime}}$ be the Stinespring isometry extension~\footnote{Note the Stinespring isometry extension is safe on register $X$.} of CPTP map $\mathsf{T}_1$ with additional copy $\hat{X^\prime}$ of $X^\prime$, i.e. $\mathsf{T}_1(\theta)=  \tr_{Z_2  \hat{X^\prime}}(U \theta U^\dagger)$ for every c-q state $\theta_{XE_2}$. Similarly let $V: \mathcal{H}_{E_1} \otimes \mathcal{H}_{Y} \rightarrow \mathcal{H}_{E_1^\prime} \otimes \mathcal{H}_{Z_1} \otimes \mathcal{H}_Y \otimes \mathcal{H}_{Y^\prime} \otimes \mathcal{H}_{ \hat{Y^\prime}}$ be the Stinespring isometry extension of CPTP map $\mathsf{T}_2$ with additional copy $\hat{Y^\prime}$ of $Y^\prime$.  Since $\rho_{XE_1Y\hat{Y}\hat{E}_2} =\rho_{XE_1} \otimes \rho_{Y\hat{Y}\hat{E}_2},$ we have 
$$\hmin{X}{Y\hat{Y}\hat{E}_2E_1}_\rho = \hmin{X}{E_1}_\rho \geq n-k.$$ Similarly since $\rho_{YE_2X\hat{X}\hat{E}_1} =\rho_{YE_2} \otimes \rho_{X\hat{X}\hat{E}_1},$ we have 
$$\hmin{Y}{X\hat{X}\hat{E}_1E_2}_\rho = \hmin{Y}{E_2}_\rho \geq n-k.$$Thus $\rho$ is a $(n-k,n-k) \mhyphen \qpas$, with the following assignment (terms on the left are from Definition~\ref{qmadvk1k2} and on the right are from here),
\[(X,\hat{X},N,M,Y,\hat{Y}) \leftarrow (X,\hat{X},\hat{E}_1E_2,\hat{E}_2E_1,Y,\hat{Y}).\]Let $\sigma=(U \otimes V)\rho (U \otimes V)^\dagger$. Note $\sigma_{XX'E_1'E_2'YY'} =(\mathsf{T}_1 \otimes \mathsf{T}_2) (\rho_{XE_1E_2Y})$ and $\sigma$ is a $(n-k,n-k) \mhyphen \nmas$. Using Theorem~\ref{thm:2nmext}, we have 
$$ \| \sigma_{2  \nmext (X,Y)2  \nmext (X',Y')YY'E_1'\hat{E}_2Z_1} - U_{n/4} \otimes \sigma_{2  \nmext (X',Y')YY'E_1'\hat{E}_2Z_1} \|_1 \leq  \cO(\eps).$$
\suppress{
and $$ \| \sigma_{2  \nmext (X,Y)2  \nmext (X',Y')XX'\hat{E}_1E_2'Z_2} - U_{n/4} \otimes \sigma_{2  \nmext (X',Y')XX'\hat{E}_1E_2'Z_2} \|_1 \leq  \cO(\eps).$$}
Using Fact~\ref{fact:data}, we further have $$ \| \sigma_{2  \nmext (X,Y)2  \nmext (X',Y')YY' E_1^\prime} - U_{n/4} \otimes \sigma_{2  \nmext (X',Y')YY' E_1^\prime} \|_1 \leq  \cO(\eps).$$
 \suppress{and $$ \| \sigma_{2  \nmext (X,Y)2  \nmext (X',Y')XX' E_2^\prime} - U_{n/4} \otimes \sigma_{2  \nmext (X',Y')XX' E_2^\prime} \|_1 \leq  \cO(\eps)$$} which completes the proof. \end{proof}

We have the following additional corollary of Theorem~\ref{thm:2nmext}.
\begin{corollary}\label{corr:add3}
Let the function $2 \nmext$ be as defined in Algorithm~\ref{alg:2nmExt}. $2 \nmext$ is an $(n-k,n-k,\cO(\eps))$-quantum secure $2$-source non-malleable extractor against $\qmra$.
\end{corollary}
\begin{proof}
Let  $\rho_{XEY}$ be a state (for $k_1=k_2=n-k$), $\mathsf{T}_1$ and  $\mathsf{T}_2$ be CPTP maps as defined in Definition~\ref{intro:def:2nmextmarkov}.  
Let $\rho_{X\hat{X}T\hat{T} E_1 \hat{E_1}E_2 \hat{E_2}Y\hat{Y}}$ be a pure state extension of $\rho_{XEY} \equiv \rho_{XE_1TE_2Y}$ such that,
\[\rho_{X\hat{X}T\hat{T} E_1 \hat{E_1}E_2 \hat{E_2}Y\hat{Y}} = \sum_{t}\sqrt{\Pr(T=t)} \ket{tt}_{T\hat{T}} \ket{\rho}_{X\hat{X}E_1 \hat{E_1}E_2\hat{E_2}Y\hat{Y}}^t \quad ; \]
\[\hmin{X}{E}_\rho \geq n-k \quad ; \quad \hmin{Y}{E}_\rho \geq n-k , \]registers ($XYT$) are classical (with copies $\hat{X}\hat{Y}\hat{T}$), $\ket{\rho}_{X\hat{X}E_1 \hat{E_1}E_2\hat{E_2}Y\hat{Y}}^t = \ket{\rho}_{X\hat{X}E_1 \hat{E_1}}^t \otimes \ket{\rho}_{E_2\hat{E_2}Y\hat{Y}}^t$ is the pure state extension of $\rho^t_{XE_1} \otimes \rho^t_{YE_2}$ with $ \ket{\rho}_{X\hat{X}E_1 \hat{E_1}}^t, \ket{\rho}_{E_2\hat{E_2}Y\hat{Y}}^t$ canonical purifications of  $\rho^t_{XE_1}, \rho^t_{YE_2}$ respectively.  Since $E \equiv E_1TE_2$, using Fact~\ref{fact102}, we have 
\begin{equation}\label{eq:corr3111}
     \hmin{X}{E_1T}_\rho \geq \hmin{X}{E}_\rho \geq n-k.
\end{equation}Similarly, $\hmin{Y}{E_2T}_\rho \geq n-k.$ Note,
\begin{equation}\label{eq:corr3112}
    \rho_{XY\hat{Y}\hat{E}_2E_1 \hat{T}} \equiv \rho_{XY\hat{Y}{E}_2E_1 {T}}.
\end{equation}
The first equivalence follows since for every $T=\hat{T}=t$, $\ket{\rho}_{E_2\hat{E_2}Y\hat{Y}}^t$ is the 
canonical purification of $\rho_{YE_2^t}$ implying $\rho_{YE_2^t}=\rho_{Y\hat{E}_2^t}$. Consider,
\begin{align*}
   \hmin{X}{Y\hat{Y}\hat{E}_2E_1 \hat{T}}_\rho &= \hmin{X}{Y\hat{Y}E_1TE_2}_\rho  \\
   & = \hmin{X}{TE_1}_\rho  \\
   & \geq n-k. 
\end{align*}First equality follows from~Eq.~\eqref{eq:corr3112}, second equality follows from~Fact~\ref{fact102}, noting  for every $T=t$, ${\rho}_{XY\hat{Y}E_1E_2}^t ={\rho}_{XE_1}^t \otimes {\rho}_{Y\hat{Y}E_2}^t$ and first inequality follows from~Eq.~\eqref{eq:corr3111}. Similarly, $\hmin{Y}{X\hat{X}\hat{E}_1E_2 {T}}_\rho \geq n-k.$ Thus, $\rho$ is an $(n-k,n-k) \mhyphen \qpas$, with the following assignment (terms on the left are from Definition~\ref{qmadvk1k2} and on the right are from here),
\[(X,\hat{X},N,M,Y,\hat{Y}) \leftarrow (X,\hat{X},\hat{E}_1E_2T,\hat{E}_2E_1\hat{T},Y,\hat{Y}).\]

Let $U: \mathcal{H}_{E_2} \otimes \mathcal{H}_{X}\otimes \mathcal{H}_{T} \rightarrow \mathcal{H}_{E_2^\prime} \otimes \mathcal{H}_{Z_2} \otimes \mathcal{H}_X \otimes \mathcal{H}_{X^\prime} \otimes \mathcal{H}_{ \hat{X^\prime}}\otimes \mathcal{H}_{T}$ be the Stinespring isometry extension~\footnote{Note the Stinespring isometry extension is safe on classical registers.} of CPTP map $\mathsf{T}_1$ with additional copy $\hat{X^\prime}$ of $X^\prime$, i.e. $\mathsf{T}_1(\theta)=  \tr_{Z_2  \hat{X^\prime}}(U \theta U^\dagger)$ for every state $\theta$. Similarly let $V: \mathcal{H}_{E_1} \otimes \mathcal{H}_{Y}\otimes \mathcal{H}_{\hat{T}} \rightarrow \mathcal{H}_{E_1^\prime} \otimes \mathcal{H}_{Z_1} \otimes \mathcal{H}_Y \otimes \mathcal{H}_{Y^\prime} \otimes \mathcal{H}_{ \hat{Y^\prime}}\otimes \mathcal{H}_{\hat{T}}$ be the Stinespring isometry extension of CPTP map $\mathsf{T}_2$ with additional copy $\hat{Y^\prime}$ of $Y^\prime$ (and treating register $T$ as $\hat{T}$ since  $\hat{T} \equiv T$).  Let $\sigma=(U \otimes V)\rho (U \otimes V)^\dagger$. Note   $\sigma_{XX'E_1'TE_2'YY'} =(\mathsf{T}_1 \otimes \mathsf{T}_2) (\rho_{XEY})$. Thus, $\sigma$ is an $(n-k,n-k) \mhyphen \nmas$. Thus, using Theorem~\ref{thm:2nmext}, we have 
$$ \| \sigma_{2  \nmext (X,Y)2  \nmext (X',Y')YY'E_1'\hat{E}_2Z_1\hat{T}} - U_{n/4} \otimes \sigma_{2  \nmext (X',Y')YY'E_1'\hat{E}_2Z_1\hat{T}} \|_1 \leq  \cO(\eps).$$
\suppress{
and $$ \| \sigma_{2  \nmext (X,Y)2  \nmext (X',Y')XX'\hat{E}_1E_2'Z_2T} - U_{n/4} \otimes \sigma_{2  \nmext (X',Y')XX'\hat{E}_1E_2'Z_2T} \|_1 \leq  \cO(\eps).$$}
Using Fact~\ref{fact:data}, we further have $$ \| \sigma_{2  \nmext (X,Y)2  \nmext (X',Y')YY' E_1^\prime \hat{T}} - U_{n/4} \otimes \sigma_{2  \nmext (X',Y')YY' E_1^\prime \hat{T}} \|_1 \leq  \cO(\eps).$$\suppress{
and $$ \| \sigma_{2  \nmext (X,Y)2  \nmext (X',Y')XX'  E_2^\prime T} - U_{n/4} \otimes \sigma_{2  \nmext (X',Y')XX'  E_2^\prime T} \|_1 \leq  \cO(\eps).$$}The desired follows noting $T \equiv \hat{T}$ in $\sigma$ which completes the proof.
\end{proof}

\subsection*{Acknowledgment}
We thank Divesh Aggarwal and Maciej Obremski for introducing us to the problem, sharing their insights on the classical constructions and several other helpful discussions. 

The work of NGB was done while he was a graduate student at the Centre for Quantum Technologies. 

This work is supported by the National Research Foundation, through Grants NRF-NRFF2013-13, NRF2021-QEP2-02-P05 and the VanQuTe Grant NRF2017-NRF-ANR004; the Prime Minister’s Office and the Ministry of Education,
Singapore, under the Research Centres of Excellence program and Grant MOE2012-T3-1-009.

\appendix
\section{A quantum secure $t$-non-malleable extractor}
\label{sec:tnmext}

\begin{definition}[$(t;k)\mhyphen\nmas$]\label{kqnmadversarydeft}
     Let $\sigma_{X\hat{X}NMY\hat{Y}}$ be a $(k) \mhyphen \qpas$. Let $V: \cH_Y \otimes \cH_M \rightarrow \cH_Y \otimes \cH_{Y^{[t]}}  \otimes  \cH_{\hat{Y}^{[t]}}  \otimes \cH_{M'}$ be an isometry such that for
     $\rho = V\sigma V^\dagger,$ we have $Y^{[t]}$ classical (with copy $\hat{Y}^{[t]}$) and
      $\forall i \in [t]: ~\Pr(Y \neq Y^i)_\rho =1.$ We call $\rho$ a $(t;k)\mhyphen\nmas$.
\end{definition}
\begin{definition}[quantum secure $t$-non-malleable extractor]\label{tnme}
		An $(n,d,m)$-non-malleable extractor $t\mhyphen\nmext : \{0,1\}^{n} \times \{0,1\}^{d} \to \{0,1\}^m$ is $(t;k,\eps)$-secure against $\nma$ if for every $(t;k)\mhyphen\nmas$ $\rho$ (chosen by the adversary $\nma$),
	$$  \| \rho_{ t\mhyphen\nmext(X,Y)t\mhyphen\nmext(X,Y^1) \ldots t\mhyphen\nmext(X,Y^t) YY^{[t]}M'} - U_m \otimes \rho_{t\mhyphen\nmext(X,Y^1) \ldots t\mhyphen\nmext(X,Y^t) YY^{[t]}M'} \|_1 \leq \eps.$$
\end{definition}

We define the parameters we use in the construction of quantum secure $t$-non-malleable extractor as follows. These parameters hold throughout this section.
 \subsection*{Parameters}\label{sec:parameterst}
Let $ \delta,  \delta_3>0$ be small enough constants and  $\delta_1, \delta_2 < \frac{1}{14}$ be constants chosen according to Fact~\ref{fact:samp}. Let  $n,n_1,d,d_1,d_2,a,v,s,b,h,t$ be positive integers and
 $k, \eps, \eps' , \eps''> 0$ such that: 
  \[d =  \cO \left(\log^{7} \left(\frac{n}{\eps}\right)\right) \quad ; \quad  v=5d \quad ;\quad n_1 \geq v^{\delta_1} \quad ; \quad  d_1 =\cO \left(\log^2\left(\frac{nt^2}{\eps^2}\right) \log d \right) \quad ; \quad a=d_1+ \cO( v^{\delta_2} ) \quad ; \]
 
\[t = \min \{ \cO(d^{\delta_3}), 2^{\cO(d^{\delta_1})-\log \left(\frac{1}{\eps}\right)} \} \quad ; \quad  2^{\cO(a)}\sqrt{\eps'} = \eps \quad ;  \quad  d_2=\cO\left(\log^2\left(\frac{n}{\eps''}\right) \log d \right) \quad ;  \quad \eps'' = 2^{-2(t+1)d_1} \eps^2  \quad ; \]

\[q=\cO(1) \quad ; \quad s = \cO\left(\log^2\left(\frac{d}{\eps'}\right)\log d \right)  \quad  ; \quad b = \cO\left( \log^2\left(\frac{d}{\eps'}\right) \log d \right) \quad ; \quad h = 10ts \quad ;  \quad  k \geq 5d. \]Let \begin{itemize}\label{sec:extparameterst}
    \item  $\Ext_0$ be $(2n_1,\eps^2/t^2)$-quantum secure $(n,d_1,n_1)$-extractor, 
    
    \item $\Ext_1$ be $(2b, \eps')$-quantum secure $(d,s,b)$-extractor,
    \item $\Ext_2$ be $(2s, \eps')$-quantum secure $(h,b,s)$-extractor,
    \item $\Ext_3$ be $(2h, \eps')$-quantum secure $(d,b,h)$-extractor,
    \item $\Ext_4$ be $(d/4t, \eps^2)$-quantum secure $(d,h,d/8t)$-extractor,
     \item $\Ext_5$ be $(2d, \eps'')$-quantum secure $(n,d_2,d)$-extractor,\item  $\Ext_6$ be $(k/4t, \eps^2)$-quantum secure $(n,d/8t,k/8t)$-extractor,
 \end{itemize}
 be the quantum secure extractors from Fact~\ref{fact:extractor}.

\subsection*{Definition of $t$-non-malleable extractor} 
Let $\F_q$ be the finite field of size $q$. Let $\ecc : \F^d_q \to \F^v_q$ be an error correcting code with relative distance $\frac{1}{10}$ and rate $\frac{1}{5}$ (which exists from~Fact~\ref{fact:ecc} for our choice of parameters) for this section. Let $\samp : \{0,1 \}^r \to [v]^{t_1}$ be the sampler function from Fact~\ref{fact:samp} where $t_1=\cO(v^{\delta_2})$ and $r \geq v^{\delta_1}$. We identify the output of $\samp$ as $t_1$ samples from the set $[v]$.  By $\ecc(Y)_{\samp(I)}$, we mean the $\samp(I)$ entries of codeword $\ecc(Y)$, interpreted as a bit string. 


\begin{algorithm}
\caption{: $t\mhyphen\nmext: \lbrace 0,1 \rbrace ^n\times \lbrace 0,1 \rbrace^d   \rightarrow \lbrace 0,1 \rbrace^{k/8t}$}\label{alg:nmExtt}
\begin{algorithmic}
\State{}

\noindent \textbf{ Input:}  $X, Y$\\ 
\begin{enumerate}
    \item \label{alg:nmext:point1t} $t$-advice generator:\[ Y_1 = \pre(Y,d_1) \quad ; \quad I = \Ext_0(X,Y_1) \quad ; \quad G=Y_1 \circ \ecc(Y)_{\samp(I)}\]
    \item \label{alg:nmext:point2t} $Y_2=$Prefix$(Y,d_2)$ \quad ; \quad $T=\Ext_5(X,Y_2)$ 
    \item \label{alg:nmext:point3t} Correlation breaker with advice:\quad $S=\advcb(Y,T,G)$\Comment{Algorithm~\ref{alg:AdvCB}}
     \item \label{alg:nmext:point4t} $L=\Ext_6(X,S)$ 
\end{enumerate}

 \noindent \textbf{ Output:} $L$ 
\end{algorithmic}
\end{algorithm}

\suppress{

\begin{algorithm}
\caption{: $t$-$\advc: \lbrace 0,1 \rbrace ^n\times \lbrace 0,1 \rbrace^d   \rightarrow \lbrace 0,1 \rbrace^{a}$}\label{alg:advgent}
\begin{algorithmic}
\State{}

\noindent Let $\samp : \{0,1 \}^r \to [v]^{t_1}$ be the sampler function from Fact~\ref{fact:samp} where $t_1=\cO(v^{\delta_2})$ and $r \geq v^{\delta_1}$. We identify the output of $\samp$ as $t_1$ samples from the set $[v]$.  By $\ecc(Y)_{\samp(I)}$, we mean the $\samp(I)$ entries of codeword $\ecc(Y)$, interpreted as a bit string.

\vspace{0.1cm}

\noindent \textbf{ Input:}  $X, Y$\\

\noindent \ Perform the following steps:

\begin{enumerate}
    \item $Y_1 = \pre(Y,d_1)$
    \item $I = \Ext_0(X,Y_1)$
    \item $G=Y_1 \circ \ecc(Y)_{\samp(I)}$
 
\end{enumerate}

 \noindent \textbf{ Output:} $G$ 
\end{algorithmic}
\end{algorithm}
}

\subsection*{Result} 
 
 The following theorem shows that the function $t\mhyphen\nmext$ as defined in Algorithm~\ref{alg:nmExtt} is $(t;k,\cO(\eps))$-secure against $\nma$ by noting that  $L=t\mhyphen\nmext(X,Y)$ and $L^i=t\mhyphen\nmext(X,Y^i)$ for every $i \in [t]$. 
 
\begin{theorem}[Security of $t\mhyphen\nmext$]\label{thm:nmextt}
Let $\rho_{X\hat{X}NYY^{[t]} \hat{Y} \hat{Y}^{[t]} M}$ be a $(t;k)\mhyphen\nmas$. Let Alice and Bob proceed with Protocol~\ref{prot:block1t} starting with $\rho$ . Let $\Lambda$  be the state at the end of the protocol. Then,
$$\Vert \| \rho_{LL^{[t]}YY^{[t]}M} - U_{k/8t} \otimes \rho_{L^{[t]}YY^{[t]}M} \Vert_1 \leq  d(L|\tilde{B})_\Lambda \leq \cO(\eps).$$

\end{theorem}

\begin{proof}
The first inequality follows from Fact~\ref{fact:data}.
Note that the total communication from Alice to Bob in Protocol~\ref{prot:block1t} is at most (from our choice of parameters) 
$$(t+1)n_1 + 6ah (t+1) + h+ (t+1)\frac{k}{8t} \leq  \left(1/4 + \delta\right) k.$$
This implies (using Lemma~\ref{lem:minentropy}) that throughout Protocol~\ref{prot:block1t}, $\hmin{X}{\tilde{B}} \geq (3/4-\delta)k > \frac{k}{2}$. 

The total communication from Bob to Alice in  Protocol~\ref{prot:block1t} is at most $$(t+1)d_1 +(t+1)d_2 + (t+1)a + (t+1)6ab + (t+1) \frac{d}{8t} \leq (1/4+\delta) d.$$
Again using Lemma~\ref{lem:minentropy}, throughout Protocol~\ref{prot:block1t}, $\hmin{Y}{\tilde{A}} \geq (3/4-\delta)d$.

The proof then proceeds using similar arguments as~Theorem~\ref{thm:nmext} involving Lemma~\ref{lem:2}, Lemma~\ref{lem:minentropy}, Claim~\ref{claim:100} after noting Claim~\ref{lemma:block1t}.
\end{proof}

\begin{claim}[$t$-advice generator]\label{lemma:block1t}
Let $\Phi$ be the joint state after registers $Z_0, Z_{0}^{[t]}$ are generated by Alice. Then,
\[ \Pr( \forall i \in [t]:~(G\ne G^{i}))_\Phi \geq 1-\cO(\eps) \quad and \quad d(T|\tilde{B})_\Phi\leq  \eps. \]
\suppress{
\begin{enumerate}
     \item \label{lemma:block1:point3t}With probability at least $1-\cO(\eps)$, $(G\ne G^{i})_\Phi $ for every $i \in [t]$. 
      \item\label{lemma:block1:point4t}$ d(T|\tilde{B})_\Phi\leq  \eps$.
\end{enumerate}}
 \end{claim}
\begin{proof}
    We first prove $\Pr( \forall i \in [t]:~ (G\ne G^{i}))_\Phi \geq 1-\cO(\eps)$. Let $\sigma_{XNIMYY^{[t]}}$ be the state after Alice has generated register $I$. Let $\beta_{IYY^{[t]}} = U_{n_1} \otimes \Phi_{YY^{[t]}}$. We have,
\begin{align}
     \Delta( \Phi_{IYY^{[t]}} , \beta_{IYY^{[t]}}) &= \Delta( \sigma_{IYY^{[t]}} , U_{n_1} \otimes \sigma_{YY^{[t]}}) & \nonumber\\
     &\leq \Delta( \sigma_{I\tilde{B}} , U_{n_1} \otimes \sigma_{\tilde{B}}) & \mbox{(Fact~\ref{fact:data})}\nonumber \\
     & \leq \sqrt{2} \Delta_B( \sigma_{I\tilde{B}} , U_{n_1} \otimes \sigma_{\tilde{B}}) \nonumber &\mbox{(Fact~\ref{fidelty_trace})}\\
     & \leq \sqrt{2}\eps/t.  & \mbox{(Lemma~\ref{lem:2})}\label{advgen:eq11t}
\end{align}
Fix an integer $i \in [t]$ and consider,
\begin{align*}
   \Pr(G=G^{i})_\Phi &\leq  \Pr(G=G^i)_\beta + \sqrt{2} \eps/t \\
   & = \Pr(Y_1 = Y_1^i)_\beta\Pr(G=G^i~|~ Y_1 = Y_1^i)_\beta + \Pr(Y_1 \neq Y_1^i)_\beta\Pr(G=G^i~|~ Y_1 \neq Y_1^i)_\beta + \sqrt{2} \eps/t  \\
      & = \Pr(Y_1 = Y_1^i)_\beta\Pr(G=G^i~|~ Y_1 = Y_1^i)_\beta + \sqrt{2}\eps/t  \\
   & \leq 2^{- \Omega(n_1)} + \sqrt{2}\eps/t \\
   & \leq \cO \left(\frac{\eps}{t}\right). 
\end{align*}
Note that conditioned on $Y_1 = Y_1^i$, we have $I = I^i$. The first inequality above follows from Eq.~\eqref{advgen:eq11t} and noting that the predicate $(G=G^i)$ is determined from $(I,Y,Y^i)$. The second equality follows from definition of $G$. 
Let $\mathcal{S}_i \defeq \{ j \in [v] : \ecc(Y)_j \ne \ecc(Y^i)_j \}$. Second inequality follows since $\ecc$ has relative distance $ \frac{1}{10}$ and considering $\samp$  with $(r,\delta, \nu, \mathcal{S})$ in Fact~\ref{fact:samp} as $(v^{\delta_1},\frac{1}{10}, v, \mathcal{S}_i)$ here. Third inequality follows by our choice of parameters. Now the desired follows from the union bound. 

Using arguments similar to the proof of Point~\ref{lemma:block1:point4} of Claim~\ref{lemma:block1}, we get,
$$ d(T|\tilde{B})_\Phi\leq 2^{(t+1)d_1}\sqrt{\eps''} \leq \eps .$$
\end{proof}

\section{A quantum secure $2$-source $t$-non-malleable extractor}
\label{sec:2tnm}
\begin{definition}[$(t;k_1,k_2)\mhyphen\nmas$]\label{def:2tsource-qnmadversarydef}
     Let $\sigma_{{X}\hat{X}NMY\hat{Y}}$ be a $(k_1,k_2)\mhyphen\qpas$. Let $U: \cH_X \otimes \cH_N \rightarrow \cH_X \otimes \cH_{X^{[t]}}  \otimes  \cH_{\hat{X}^{[t]}}  \otimes \cH_{N'}$ and $V: \cH_Y \otimes \cH_M \rightarrow \cH_Y \otimes \cH_{Y^{[t]}}  \otimes  \cH_{\hat{Y}^{[t]}}  \otimes \cH_{M'}$ be isometries such that for
     $\rho = (U \otimes V)\sigma (U \otimes V)^\dagger,$ we have $X^{[t]}Y^{[t]}$ classical (with copy $\hat{X}^{[t]}\hat{Y}^{[t]}$) and
      \begin{gather*}
           \forall i\in [t],\ \Pr(Y \ne Y^i)_\rho =1 \quad \text{or} \quad \Pr(X \ne X^i)_\rho =1. 
       \end{gather*}We call $\rho$ a $(t;k_1,k_2)\mhyphen\nmas$.
\end{definition}
\begin{definition}[quantum secure $2$-source  $t$-non-malleable extractor]\label{def:2tsourcenme}
		An $(n,n,m)$-non-malleable extractor $t\mhyphen2\nmext : \{0,1\}^{n} \times \{0,1\}^{n} \to \{0,1\}^m$ is $(t;k_1,k_2,\eps)$-secure against $\nma$ if for every $(t;k_1,k_2)\mhyphen\nmas$ $\rho$ (chosen by the adversary $\nma$),
\suppress{	$$  \| \rho_{ t\mhyphen2\nmext(X,Y)t\mhyphen2\nmext(X^1,Y^1) \ldots t\mhyphen2\nmext(X^t,Y^t) XX^{[t]}N'} - U_m \otimes \rho_{t\mhyphen2\nmext(X^1,Y^1) \ldots t\mhyphen2\nmext(X^t,Y^t) XX^{[t]}N'} \|_1 \leq \eps,$$and}
	$$  \| \rho_{ t\mhyphen2\nmext(X,Y)t\mhyphen2\nmext(X^1,Y^1) \ldots t\mhyphen2\nmext(X^t,Y^t) YY^{[t]}M'} - U_m \otimes \rho_{t\mhyphen2\nmext(X^1,Y^1) \ldots t\mhyphen2\nmext(X^t,Y^t) YY^{[t]}M'} \|_1 \leq \eps.$$
\end{definition}
 \subsection*{Parameters}\label{sec:parameters_2t_sourcenm}
 Let $\delta,\delta_3>0$ be small enough constants.
 Let $\delta_1,\delta_2< \frac{1}{14}$ be constants chosen according to Fact~\ref{fact:samp}, $n,v,n_1,t,a,s,b,h>0$ be positive integers and  $k,\varepsilon,\varepsilon^\prime>0$ be such that, 
 
  \[  v= 5n \quad ; \quad  n_1 =v^{\delta_1} \quad ; \quad \eps= 2^{- \cO\left(n^{\delta_3}\right)} \quad ; \quad t\leq n^{\delta_3}\quad; \]
 
\[k = \cO (n^{1/4}) \quad ; \quad a=6k+2 \cO( v^{\delta_2}) = \mathcal{O}(k) \quad ;  \quad  2^{\cO(a)}\sqrt{\eps'} = \eps \quad ;       \]

\[s = \cO\left(\log^2\left(\frac{n}{\eps'}\right)\log n \right)  \quad  ; \quad b = \cO\left( \log^2\left(\frac{n}{\eps'}\right) \log n \right) \quad ; \quad h = 10ts \quad. \]
\begin{itemize}\label{sec:extparameters_2tnm}
    \item $\IP_1$ be $\IP^{3k/n_1}_{2^{n_1}}$,
    \quad ; \quad $\IP_2$ be $\IP^{3k^3/h}_{2^h}$,
    \item $\Ext_1$ be $(2b, \eps')$-quantum secure $(n,s,b)$-extractor,
    \item $\Ext_2$ be $(2s, \eps')$-quantum secure $(h,b,s)$-extractor,
    \item $\Ext_3$ be $(2h, \eps')$-quantum secure $(n,b,h)$-extractor,
    \item $\Ext_4$ be $(\frac{n}{4t}, \eps^2)$-quantum secure $\left(n,h,\frac{n}{8t}\right)$-extractor,
    \item  $\Ext_6$ be $\left(\frac{n}{2t}, \eps^2\right)$-quantum secure $(n,\frac{n}{8t},\frac{n}{4t})$-extractor. 
 \end{itemize}
 
 \subsection*{Definition of $2$-source $t$-non-malleable extractor}
 Let $\F_q$ be the finite field of size $q=\cO(1)$. Let $\ecc : \F^n_q \to \F^v_q$ be an error correcting code with relative distance $\frac{1}{10}$ and rate $\frac{1}{5}$ (which exists from~Fact~\ref{fact:ecc} for our choice of parameters) for this section. Let $\samp : \{0,1 \}^r \to [v]^{t_1}$ be the sampler function from Fact~\ref{fact:samp} where $t_1=\cO(v^{\delta_2})$ and $r \geq v^{\delta_1}$. We identify the output of $\samp$ as $t_1$ samples from the set $[v]$.  By $\ecc(Y)_{\samp(I)}$, we mean the $\samp(I)$ entries of codeword $\ecc(Y)$, interpreted as a bit string.
 
 \begin{algorithm}
\caption{: $t\mhyphen2\nmext: \lbrace 0,1 \rbrace ^n\times \lbrace 0,1 \rbrace^n   \rightarrow \lbrace 0,1 \rbrace^{n/4t}$}\label{alg:2tnmExt}
\begin{algorithmic}
\State{}

\noindent \textbf{ Input:}  $X, Y$\\


\begin{enumerate}
    \item Advice generator: \[X_1=\pre(X,3k) \quad  ; \quad Y_1 = \pre(Y,3k) \quad ;
    \quad R= \IP_1(X_1,Y_1) \quad ; \quad  \]
    \[G=X_1 \circ Y_1 \circ \ecc(X)_{\samp(R)} \circ \ecc(Y)_{\samp(R)} \]

    \item $X_2=\pre(X,3k^3) \quad ; \quad Y_2=\pre(Y,3k^3)\quad; \quad Z_0=\IP_2(X_2,Y_2)$
    \item Correlation breaker with advice:\quad $S=2\advcb(Y,X,Z_0, G)$
    \item $L=\Ext_6(X,S)$
\end{enumerate}

 \noindent \textbf{ Output:} $L$ 
\end{algorithmic}
\end{algorithm}
\subsection*{Result} 
 The following theorem shows that the function $t\mhyphen2\nmext$ as defined in Algorithm~\ref{alg:2tnmExt} is $(t;k_1,k_2,\cO(\eps))$-secure against $\nma$ by noting that  $L=t\mhyphen2\nmext(X,Y)$ and $L^i=t\mhyphen2\nmext(X^i,Y^i)$ for every $i \in [t]$.
 Note that $2\advcb$ in Algorithm~\ref{alg:2tnmExt} is same as the one in Algorithm~\ref{alg:2AdvCB} except for parameters and extractors which are to be used as mentioned in this section.  

\begin{theorem}\label{thm:nmext2t}
Let $\rho_{{X X^{[t]} \hat{X} \hat{X}^{[t]} N YY^{[t]} \hat{Y} \hat{Y}^{[t]} M}}$ be a $(t;n-k,n-k)\mhyphen\nmas$. Let Alice and Bob proceed with Protocol~\ref{prot:2tnmExt_full} starting with $\rho$. 
Let $\Lambda$ be the state at the end of the protocol. Then,
\[
   \Vert \rho_{ L L^{[t]} Y  Y^{[t]} M} - U_{n/4t} \otimes \rho_{L^{[t]} Y  Y^{[t]} M} \Vert_1 \leq   d(L|\tilde{B})_\Lambda \leq \cO(\eps).
\] 
\end{theorem}
\begin{proof} First inequality follows from Fact~\ref{fact:data}.

 Note that the total communication from Alice to Bob in Protocol~\ref{prot:2tnmExt_full}  is at most
 \[(t+1) (n_1+3k + 6ah) +h + \frac{n}{8t}(t+1)  \leq \left(\frac{1}{4}+ \delta \right) n. \]
 Similarly, total communication from Bob to Alice is at most 
 \[(3k+a+6ab)(t+1)+ 3k^3 (t+1)+ \frac{n}{8t}(t+1) \leq \left(\frac{1}{4}+\delta\right)n.\]
 Hence, using Lemma~\ref{lem:minentropy}, at any stage $\varrho$ in Protocol~\ref{prot:2tnmExt_full}, we have,
$\hmin{X}{\tilde{B}}_\varrho \geq n- k- (1/4 + \delta)n \geq n/2$ and similarly, $\hmin{Y}{\tilde{A}}_\varrho \geq n/2$. 
Thus, both Alice and Bob have enough entropy throughout the protocol for necessary extractions.

We start with a state $\rho_{X X^{[t]} N Y Y^{[t]} M}$ such that $\hmin{X}{\tilde{B}}_{\rho} \geq n-k$ and $\hmin{Y}{\tilde{A}}_{\rho} \geq n-k$. From Fact~\ref{fact:prefixminentropyfact}, we have,
\[\hmin{X_1}{\tilde{B}}_{\rho} \geq 3k- k =2k \quad ; \quad \hmin{Y_1}{\tilde{A}}_{\rho} \geq 3k-k=2k .\]

Now from Claim~\ref{l-qma-needed-fact1} with the below assignment of registers (and noting registers $(XX^{[t]},YY^{[t]})$ are included in $(\tilde{A},\tilde{B})$ respectively),
\[(Z,X,Y, \sigma) \leftarrow (R,X_1, Y_1, \rho) \quad ; \quad (k_1,k_2,m,n_1,\eps ) \leftarrow (2k,2k, n_1, 3k,  (\eps/t)^2) \]we have,
\[ \Delta( \rho_{RYY^{[t]}}, U_R \otimes \rho_{YY^{[t]}} ) \leq  \cO((\eps/t)^2) \quad ; \quad \Delta( \rho_{RXX^{[t]}}, U_R \otimes \rho_{XX^{[t]}} ) \leq  \cO( (\eps/t)^2). \]
Using Fact~\ref{fidelty_trace}, we get 
\[ \Delta_B( \rho_{RYY^{[t]}}, U_R \otimes \rho_{YY^{[t]}} ) \leq   \cO( \eps/t) \quad ; \quad \Delta_B( \rho_{RXX^{[t]}}, U_R \otimes \rho_{XX^{[t]}} ) \leq   \cO( \eps/t). \]

\suppress{
\mycomment{ Replacing the below argument: using Lemma~\ref{lemma:nearby_rho_prime_prime}, there exists an $l\mhyphen \qmas$  $\rho^{(1)}$ such that
\[\Delta_B(\rho^{(1)},\rho) \leq \mathcal{O}(\eps/t) \quad ; \quad l \leq 6k-4k+4+\cO \left( \log \left( \frac{t}{\eps}\right) \right) \leq 2.1k. \]Using Fact~\ref{l-qma-needed-fact}, for our choice of parameters, it follows that
\[d(R \vert Y Y^{[t]})_{\rho^{(1)}} \leq \cO(\eps/t) \quad ; \quad d(R \vert X X^{[t]})_{\rho^{(1)}} \leq \mathcal{O}(\eps/t). \]Now from Claim~\ref{claim:traingle_rho_rho_prime} with the below assignment of registers (and $(\rho, \rho') \leftarrow (\rho, \rho^{(1)})$),
\[(Z,A ) \leftarrow (R, Y Y^{[t]}) \quad ; \quad (Z,A ) \leftarrow (R, X X^{[t]})\]
we have,
\[d(R \vert Y Y^{[t]})_{\rho}\leq  \cO(\eps/t) \quad ; \quad d(R \vert XX^{[t]})_{\rho} \leq \cO(\eps/t).\]}}Let $\kappa$ be the state just before Bob sends $Y_2$. 
Note that till then, communication from Alice to Bob and Bob to Alice is at most $\mathcal{O}\left(tk\right) \leq \mathcal{O}\left( k^2\right)$ each.
Hence, by Lemma~\ref{lem:minentropy}, $\hmin{X}{\tilde{B}}_{\kappa} \geq n-k-\mathcal{O}\left( k^2 \right) $ and thus $\hmin{X_2}{\tilde{B}}_\kappa \geq 3 k^3-k-\mathcal{O}\left( k^2 \right) \geq 2k^3.$ Similarly,  $\hmin{Y_2}{\tilde{A}}_\kappa \geq 3 k^3-k-\mathcal{O}\left( k^2 \right) \geq 2k^3.$ Let $\eta$ be the state just after Alice generates $Z_0$. Using similar argument as before involving Claim~\ref{l-qma-needed-fact1}, we have $d(Z_0 \vert \tilde{B})_{\eta} \leq  \mathcal{O}(\eps)$. \suppress{Modified: Using similar arguments as before involving Lemma~\ref{lemma:nearby_rho_prime_prime}, Fact~\ref{l-qma-needed-fact} and Claim~\ref{claim:traingle_rho_rho_prime}, we have $d(Z_0 \vert \tilde{B})_{\eta} \leq  \mathcal{O}(\eps)$.}

Rest of the proof follows similar lines to that of Theorem~\ref{thm:2nmext} with the following change in the Claim~\ref{lemma:block12t} as follows:

\begin{claim}
Let $\Phi$ be the joint state after registers $Z_0, Z_0^{[t]}$ are generated by Alice.
Then,
\[\Pr(\forall i \in [t], \  G \neq G^i )_\Phi \geq 1- \mathcal{O}(\eps) \quad \text{and} \quad d(Z_0 \vert \tilde{B})_\Phi \leq \cO(\eps).\]
\end{claim}

The proof of the above claim follows from that of Claim~\ref{lemma:block1t}, after noting that $G=G^i$ is a sub-event of both $G_a=G_a^i$ and $G_b = G_b^i$ where $G_a= X_1 \circ \ecc(X)_{\samp(R)}$, $G_a^i= X_1^i \circ \ecc(X^i)_{\samp(R^i)}$, $G_b= Y_1 \circ \ecc(Y)_{\samp(R)}$ and $G_b^i= Y_1^i \circ \ecc(Y^i)_{\samp(R^i)}$.

This completes our proof.
\end{proof}


\newpage

\newgeometry{top=1in,bottom=0.1cm,left=1in}
\section{Communication Protocols \label{sec:communication_protocols}}
\section*{Protocols for $\nmext$}
\pagenumbering{gobble}

\begin{Protocol}[htb]
	\begin{center}
	\scalebox{0.96}{	
		\begin{tabular}{l l l l}
			Alice:  $(X,\hat{X},N)$ &  & Bob: $(Y,Y',\hat{Y},\hat{Y}',M)$ &  Analysis \\
			\hline\\
			& &  $ Y_1 = \pre(Y,d_1)$ & $d( Y_1 \vert \tilde{A} ) = {0}$\\ \\
			$I = \Ext_0(X,Y_1)$& $ Y_1 \longleftarrow Y_1$ & & $d( I \vert \tilde{B} ) \leq {\eps}$\\ \\
	
			$I' = \Ext_0(X,Y'_1) $	& $   Y'_1 \longleftarrow Y'_1$ & $ Y'_1 = \pre(Y',d_1)$&$d( I \vert \tilde{B} ) \leq {\eps}$\\ \\
				
			 & $ I \longrightarrow I$ & $G=Y_1 \circ \ecc(Y)_{I}$ &\\ \\

			
			 & $I' \longrightarrow I'$ & $ G'=Y'_1 \circ \ecc(Y')_{I'}$  \\ \\

				$T = \Ext_5(X,Y_2) $  & $Y_2 \longleftarrow Y_2$ & $ Y_2 = \pre(Y,d_2)$ & $d( T \vert \tilde{B} ) \leq \eps$\\ \\
			$T' = \Ext_5(X,Y'_2) $ & $Y'_2 \longleftarrow Y'_2$ & $ Y'_2 = \pre(Y',d_2)$ & 
			\\ \\
			
			$Z_0 = \pre(T,h) $	& $ G \longleftarrow G$ \\
			&& \\
				$Z_0' = \pre(T',h) $	& $ G' \longleftarrow G'$ &&$d( T \vert \tilde{B} ) \leq \eps$\\\\
		
			&& \\
			\hline
			&& \\
			
			Alice: $(T,T',Z_0,Z_0',G,G',N)$&& Bob: $(Y,Y',G,G',M)$\\
		
			\hline
			&& \\ 
			& Protocol~\ref{prot:block2}~$( T,T',Z_0,Z_0',$&\\
        	&$G,G',N,Y,Y',G,G',M)$ & \\
			&& \\
				Alice: $(X,Z,N)$ &  & Bob: $(Y,Y',Z',M)$ & \\
			\hline
			&& \\
			 $L' =\Ext_6(X,S')$&$S' \longleftarrow S'$ &  $S' =\Ext_4(Y',Z')$ &$d( Z \vert \tilde{B} ) \leq \cO(\eps)$ \\
		    && \\
		    & $ Z \longrightarrow Z$ &  $S =\Ext_4(Y,Z)$ &$d( S \vert \tilde{A} ) \leq \cO(\eps)$ \\
		    && \\
		    & $ L' \longrightarrow L'$ & &$d( S \vert \tilde{A} ) \leq \cO(\eps)$  \\
		    && \\
		    $L =\Ext_6(X,S)$	&$S \longleftarrow S$ &  &$d( L \vert \tilde{B} ) \leq \cO(\eps)$ \\
			&& \\
			\hline
			&& \\
			
			Alice: $(L,N)$&& Bob: $(L',Y,Y',M)$\\
		\end{tabular}
		{\small {\caption{\label{prot:block1}
					 $(X,\hat{X}, N, Y, Y',\hat{Y},\hat{Y}',M)$.}}
		}}
	\end{center}
\end{Protocol}

\begin{Protocol}[htb]

\vspace{0.1in}
For $i=1,2,\ldots,a:$

 \begin{itemize}
        \item Protocol~\ref{prot:Var_GEN(0,1)analysis}~$( T,T',Z,Z',G,G',N, Y,Y',G,G',M)$ for $(\alpha_i,\alpha_i')=(0,1)$.
        \item Protocol~\ref{prot:Var_GEN(1,0)}~$( T,T',Z,Z',G,G',N, Y,Y',G,G',M)$ for $(\alpha_i,\alpha_i')=(1,0)$.
        \item Protocol~\ref{prot:Var_GEN(0,0)NotDiffBefore}~$( T,T',Z,Z',G,G',N, Y,Y',G,G',M)$ for $(\alpha_i,\alpha_i')=(0,0)$ and $\alpha_j=\alpha_j'$ for $j<i$.
        \item Protocol~\ref{prot:Var_GEN(0,0)DiffBefore}~$( T,T',Z,Z',G,G',N, Y,Y',G,G',M)$ for $(\alpha_i,\alpha_i')=(0,0)$ and $\alpha_j \ne \alpha_j'$ for some $j<i$.
        \item Protocol~\ref{prot:Var_GEN(1,1)NotDiffBefore}~$( T,T',Z,Z',G,G',N, Y,Y',G,G',M)$ for $(\alpha_i,\alpha_i')=(1,1)$ and $\alpha_j=\alpha_j'$ for $j<i$.
        \item Protocol~\ref{prot:Var_GEN(1,1)DiffBefore}~$( T,T',Z,Z',G,G',N, Y,Y',G,G',M)$ for $(\alpha_i,\alpha_i')=(1,1)$ and $\alpha_j \ne \alpha_j'$ for some $j<i$.
        
    \end{itemize}

	\quad \quad	$(Z,Z') = (O,O')$.

	\vspace{0.25cm}
	
		{\small {\caption{\label{prot:block2}
					 $( T,T',Z,Z',G,G',N, Y,Y',G,G',M)$.}}
		}
\end{Protocol}

\newpage

\begin{Protocol}[htb]
	\begin{center}
	\scalebox{0.88}{
		\begin{tabular}{l l l l}
			Alice:  $(X,N)$ &  & ~~~~~~~~~~~~Bob: $(Y,Y',M)$ & Analysis\\
			
			\hline\\
			& $(E_A, E_A') \longleftrightarrow (E_B, E_B')$ \\ 
			& & $ Y_1 = \pre(Y,d_1)$ &\\
			&& &\\
			& &  $ Y'_1 = \pre(Y',d_1)$
			\\
			$I = \Ext_0(X,E_A)$ &  && $d(I \vert \tilde{B}) \leq \eps$  \\ \\
			 &  $ I \longrightarrow I$ &  &  \\ \\
			$I' = \Ext_0(X,E_A') $&   $I' \longrightarrow I'$&  & \\ \\

			 $T = \Ext_5(X,Y_2) $  & $ Y_2 \longleftarrow Y_2$ & $ Y_2 = \pre(Y,d_2)$ & $d(T \vert \tilde{B}) \leq \sqrt{\eps'}$\\ \\
			
			$T' = \Ext_5(X,Y'_2)$& $Y'_2 \longleftarrow Y'_2$ & $ Y'_2 = \pre(Y',d_2)$ &  \\ \\
		
				$Z_0 = \pre(T,h)$&  $ G \longleftarrow G$ & $  G=Y_1 \circ \ecc(Y)_{I}$ &  \\ \\
		
				$Z_0' = \pre(T',h) $	&   $ G' \longleftarrow G'$ &  			 $ G'=Y'_1 \circ \ecc(Y')_{I'}$& $d(T \vert \tilde{B}) \leq \sqrt{\eps'}$  \\ \\
			&& \textbf{If:} $(E_B,E_B') \ne (Y_1,Y_1')$  \\
			&&\textbf{abort}&\\
			&& \textbf{else:}  \textbf{continue} & $d(T \vert \tilde{B}) \leq  2^{2d_1}\sqrt{\eps'} \leq \eps$ \\
		    && \\
			\hline
			&& \\
				Alice: $(T,T',Z_0,Z_0',G,G',N)$&&Bob: $(Y,Y',G,G',M)$
		\end{tabular}
		{\small {\caption{\label{prot:block1mod}
			Modified advice generator.}}
		}}
	\end{center}
\end{Protocol}

\begin{Protocol}
	\begin{center}
		\begin{tabular}{l l r r}
			Alice:  $(T,T^\prime,Z, Z^\prime, G,G',N)$ &  & ~~~~~~~~~~~~Bob: $(Y, Y^\prime,G,G',M)$ & $\quad$ Analysis \\
			\hline\\
			$ Z_s =$ Prefix$(Z,s)$  & &  &$d( Z_s \vert \tilde{B} ) \leq {\eta}$\\ \\
			 & $Z_s \longrightarrow Z_s$ & $A= \Ext_1(Y,Z_s)$ &$ d(A \vert \tilde{A} )  \leq { \cO(\eta + \sqrt{\eps'})}$ \\ \\

			& $Z^\prime \longrightarrow Z^\prime$ & 			 $A^\prime = \Ext_1(Y^\prime, Z_s^\prime)$ & \\ & &  $C^\prime= \Ext_2(Z^\prime, A^\prime)$ & \\ & &  $B^\prime= \Ext_1(Y^\prime, C^\prime)$ &$d (A \vert \tilde{A} ) \leq  {\cO(\eta + \sqrt{\eps'})}$\\ \\
			
			$\overline{Z}= \Ext_3(T,A)$ &$A \longleftarrow A$ && \\ 
				$\overline{Z}_s= \pre(\overline{Z},s)$& & &$d (\overline{Z}_s  \vert  \tilde{B}  ) \leq {\cO(\eta + \sqrt{\eps'})}$ \\ \\
	        
	        
			$\overline{Z}^\prime =\Ext_3(T^\prime, B^\prime)$ & $ B^\prime \longleftarrow B^\prime$ & &$d (\overline{Z}_s  \vert  \tilde{B}  ) \leq {\cO(\eta + \sqrt{\eps'})}$ \\ 
				$\overline{Z}^\prime_s= \pre(\overline{Z}',s)$& & & \\ \\
			& $\overline{Z}_s \longrightarrow \overline{Z}_s $ & $\generateAbar $&$d (\overline{A}  \vert  \tilde{A}  ) \leq {\cO(\eta + \sqrt{\eps'})}$ \\ \\
			
			&$\overline{Z}^\prime_s \longrightarrow \overline{Z}^\prime_s $&$\generateAAbar$& $d (\overline{A}  \vert  \tilde{A}  ) \leq {\cO(\eta + \sqrt{\eps'})}$ \\ \\
			
			$\generateTbar$ & $ \overline{A} \longleftarrow \overline{A} $ &  &$d (\overline{C} \vert \tilde{B} ) \leq {\cO(\eta + \sqrt{\eps'})}$ \\ \\
			 
			 	$O^\prime = \Ext_3(T^\prime, \overline{A}^\prime)$&$\overline{A}^\prime \longleftarrow \overline{A}^\prime$&  &$d(\overline{C} \vert  \tilde{B} ) \leq {\cO(\eta + \sqrt{\eps'})}$\\ 
			 &&\\

			 & $\overline{C} \longrightarrow \overline{C}$  & $\overline{B}= \Ext_1(Y,\overline{C})$ & $d (\overline{B} \vert \tilde{A} ) \leq {\cO(\eta + \sqrt{\eps'})}$\\ \\
			
	 & $O^\prime \longrightarrow O^\prime$ & &$d(\overline{B} \vert \tilde{A} ) \leq {\cO(\eta + \sqrt{\eps'})}$\\ \\



			$O= \Ext_3(T,\overline{B})$ & $\overline{B} \longleftarrow \overline{B}$ & &$d(O \vert  \tilde{B} ) \leq {\cO(\eta + \sqrt{\eps'})}$\\

		\end{tabular}
		{\small {\caption{\label{prot:Var_GEN(0,1)analysis}
					 $( T,T',Z,Z',G,G',N, Y,Y',G,G',M)$.}}
		}
	\end{center}
\end{Protocol}

\begin{Protocol}
	\begin{center}
		\begin{tabular}{l l r r}
			Alice:  $(T,T^\prime,Z, Z^\prime,G,G',N)$ &  & ~~~~~~~~~~~~Bob: $(Y, Y^\prime,G,G',M)$ & $\quad$ Analysis \\
			
			\hline\\
			$ Z_s =$ Prefix$(Z,s)$  &  & &$d(Z_s \vert \tilde{B} ) \leq {\eta}$\\ \\
			& $Z_s \longrightarrow Z_s$ & $A= \Ext_1(Y,Z_s)$ &$d( A \vert \tilde{A} ) \leq {\cO(\eta + \sqrt{\eps'})}$\\ \\
			$Z'_s =\pre(Z',s)$ & $Z_s^\prime \longrightarrow Z_s^\prime$ &  $\generateAA$&$d( A \vert \tilde{A} ) \leq {\cO(\eta + \sqrt{\eps'})}$	\\ \\ 	
			
				$\generateT$ & $A \longleftarrow A$ & & $d( C \vert \tilde{B} ) \leq {\cO(\eta + \sqrt{\eps'})}$\\ \\
			 $\overline{Z}^\prime= \Ext_3(T^\prime,A^\prime)$ & \sendAArl & &$d( C \vert \tilde{B} ) \leq {\cO(\eta + \sqrt{\eps'})}$\\ \\ 
			& \sendTlr & $\generateB$ &$d( B \vert \tilde{A} ) \leq {\cO(\eta + \sqrt{\eps'})}$\\ \\
		
		& \sendYYbarlr &$\generateAAbar$  &\\ 
		 & & $\generateTTbar$ &\\
		 & & $\generateBBbar$ &$d( B \vert \tilde{A} ) \leq {\cO(\eta + \sqrt{\eps'})}$\\ \\
			
			 $\overline{Z}= \Ext_3(T,B)$ & \sendBrl &  &$d( \overline{Z}_s \vert \tilde{B} ) \leq {\cO(\eta + \sqrt{\eps'})}$\\ \\

			$O^\prime= \Ext_3(T^\prime, \overline{B}^\prime)$	& \sendBBbarrl &  &$d( \overline{Z}_s \vert \tilde{B} ) \leq {\cO(\eta + \sqrt{\eps'})}$\\ \\
			
			&  \sendYSbarlr& $\generateAbar$ &$d( \overline{A} \vert \tilde{A} ) \leq {\cO(\eta + \sqrt{\eps'})}$\\ \\
			
		& $O^\prime \longrightarrow O^\prime$& &$d( \overline{A} \vert \tilde{A} ) \leq {\cO(\eta + \sqrt{\eps'})}$ \\ \\

		$O= \Ext_3(T,\overline{A})$& \sendAbarrl & &$d( O \vert \tilde{B} ) \leq {\cO(\eta + \sqrt{\eps'})}$\\ \\
		\end{tabular}
		{\small {\caption{\label{prot:Var_GEN(1,0)}
				$( T,T',Z,Z',G,G',N, Y,Y',G,G',M)$.}}
		}
	\end{center}
\end{Protocol}

\begin{Protocol}
	\begin{center}
		\begin{tabular}{l l r r}
			Alice:  $(T,T^\prime,Z, Z^\prime,G,G',N)$ &  & ~~~~~~~~~~~~Bob: $(Y, Y^\prime,G,G',M)$ & $\quad$ Analysis \\
			
			\hline\\
			$ Z_s =$ Prefix$(Z,s)$  & & &$d(Z_s \vert \tilde{B} ) \leq {\eta}$\\ \\
			
				 & $Z_s \longrightarrow Z_s$ &$A= \Ext_1(Y,Z_s)$ &$d(A \vert \tilde{A} ) \leq {\eta}$\\ \\
				 
				 $ Z'_s =$ Prefix$(Z',s)$ 	& $Z_s^\prime \longrightarrow Z_s^\prime$ & $\generateAA$ 	&$d( A \vert \tilde{A} ) \leq {\cO(\eta + \sqrt{\eps'})}$\\ \\ 
				 	
			 $\overline{Z}= \Ext_3(T,A)$ & $A \longleftarrow A$ &  &\\ 
			 
			 	 $ \overline{Z}_s = \pre(\overline{Z},s)$  &  &  &$d( \overline{Z_s} \vert \tilde{B} ) \leq {\cO(\eta + \sqrt{\eps'})}$\\ \\
				 
			$\overline{Z}^\prime= \Ext_3(T^\prime,A^\prime)$&  \sendAArl & &$d( \overline{Z_s} \vert \tilde{B} ) \leq {\cO(\eta + \sqrt{\eps'})}$\\ \\
		 
			 & \sendYSbarlr &  $\generateAbar$ &$d( \overline{A} \vert \tilde{A} ) \leq {\cO(\eta + \sqrt{\eps'})}$\\ \\
			
			$\overline{Z}^\prime_s= \pre(\overline{Z}^\prime,s)$  & $\overline{Z}^\prime_s \longrightarrow \overline{Z}^\prime_s$  & $\generateAAbar$ &$d( \overline{A} \vert \tilde{A} ) \leq {\cO(\eta + \sqrt{\eps'})}$ \\ \\

		$\generateTbar$& \sendAbarrl & 	&$d( \overline{C} \vert \tilde{B} ) \leq {\cO(\eta + \sqrt{\eps'})}$\\ \\
		
		$\generateTTbar$	& \sendAAbarrl&  &$d( \overline{C} \vert \tilde{B} ) \leq {\cO(\eta + \sqrt{\eps'})}$\\ \\
			
			
		 & \sendTbarlr & $\generateBbar$&$d( \overline{B} \vert \tilde{A} ) \leq {\cO(\eta + \sqrt{\eps'})}$\\ \\

		& \sendTTbarlr & $\generateBBbar$&$d( \overline{B} \vert \tilde{A} ) \leq {\cO(\eta + \sqrt{\eps'})}$\\ \\
		
		$O= \Ext_3(T,\overline{B})$ & \sendBbarrl&  &$d( O \vert \tilde{B} ) \leq {\cO(\eta + \sqrt{\eps'})}$\\ \\

	  $O^\prime= \Ext_3(T^\prime, \overline{B}^\prime)$& \sendBBbarrl	& &$d( O \vert \tilde{B} ) \leq {\cO(\eta + \sqrt{\eps'})}$\\

		\end{tabular}
		{\small {\caption{\label{prot:Var_GEN(0,0)NotDiffBefore}
					 $( T,T',Z,Z',G,G',N, Y,Y',G,G',M)$.}}
		}
	\end{center}
\end{Protocol}
\begin{Protocol}
	\begin{center}
		\begin{tabular}{l l r r}
			Alice:  $(T,T^\prime,Z,G,G',N)$ &  & ~~~~~~~~~~~~Bob: $(Y, Y^\prime,Z',G,G',M)$ & $\quad$ Analysis \\
			
			\hline\\
			
			 $ Z_s =$ Prefix$(Z,s)$ 	&   & &$d( Z_s \vert \tilde{B} ) \leq {\eta}$ \\ \\
			 
		     	$\overline{Z}^\prime= \Ext_3(T^\prime, A^\prime)$ & \sendAArl   & $\generateAA$ &$d( Z_s \vert \tilde{B} ) \leq {\eta}$ \\ \\
		     	
			 & $Z_s \longrightarrow Z_s$ &  $A= \Ext_1(Y,Z_s)$&$d( A \vert \tilde{A} ) \leq {\cO(\eta + \sqrt{\eps'})}$\\ \\
			 
			  
			  	  & $ \overline{Z}^\prime \longrightarrow \overline{Z}^\prime $ &$\generateAAbar$ &\\ 
			  	  && $\generateTTbar$& \\ 
			  	  &&$\generateBBbar$&$d( A \vert \tilde{A} ) \leq {\cO(\eta + \sqrt{\eps'})}$ \\ \\
			 
			$\overline{Z}= \Ext_3(T,A)$ & $A \longleftarrow A$ & &\\ 
			 $\overline{Z}_s = \pre(\overline{Z},s)$&&&$d( \overline{Z}_s \vert \tilde{B} ) \leq {\cO(\eta + \sqrt{\eps'})}$ \\ \\
			
			 	  $O^\prime= \Ext_3(T^\prime, \overline{B}^\prime)$	& \sendBBbarrl &  &$d( \overline{Z}_s \vert \tilde{B} ) \leq {\cO(\eta + \sqrt{\eps'})}$ \\ \\
			

			  &  $ \overline{Z}_s \longrightarrow \overline{Z}_s$ & $\generateAbar$ &$d( \overline{A} \vert \tilde{A} ) \leq {\cO(\eta + \sqrt{\eps'})}$\\ \\


		 $\generateTbar$& \sendAbarrl & &$d(\overline{C}  \vert \tilde{B} ) \leq {\cO(\eta + \sqrt{\eps'})}$\\ \\

		 & \sendTbarlr & $\generateBbar$  &$d( \overline{B} \vert \tilde{A} ) \leq {\cO(\eta + \sqrt{\eps'})}$\\ \\

	     & 	$O^\prime \longrightarrow O^\prime$ & &$d( \overline{B} \vert \tilde{A} ) \leq {\cO(\eta + \sqrt{\eps'})}$ \\ \\
	     
		$O= \Ext_3(T,\overline{B})$& \sendBbarrl &&$d( O \vert \tilde{B} ) \leq {\cO(\eta + \sqrt{\eps'})}$ \\ \\

		\end{tabular}
		{\small {\caption{\label{prot:Var_GEN(0,0)DiffBefore}
					 $( T,T',Z,Z',G,G',N, Y,Y',G,G',M)$.}}
		}
	\end{center}
\end{Protocol}
\begin{Protocol}
	\begin{center}
		\begin{tabular}{l l r r}
			Alice:  $(T,T^\prime,Z, Z^\prime,G,G',N)$ &  & ~~~~~~~~~~~~Bob: $(Y, Y^\prime,G,G',M)$ & $\quad$ Analysis \\
			
			\hline\\
			$ Z_s =$ Prefix$(Z,s)$  &  & &$d( Z_s \vert \tilde{B} ) \leq {\eta}$\\ \\
			 & $Z_s \longrightarrow Z_s$ &$A= \Ext_1(Y,Z_s)$  &$d( A \vert \tilde{A} ) \leq {\cO(\eta + \sqrt{\eps'})}$\\ \\
			
			$ Z'_s =$ Prefix$(Z',s)$ 	& $Z^\prime_s \longrightarrow Z^\prime_s$ & $\generateAA$ &$d( A \vert \tilde{A} ) \leq {\cO(\eta + \sqrt{\eps'})}$	\\ \\ 
				
			$\generateT$& $A \longleftarrow A$ &  &$d( C \vert \tilde{B} ) \leq {\cO(\eta + \sqrt{\eps'})}$ \\ \\
			 $\generateTT$& \sendAArl& &$d( C \vert \tilde{B} ) \leq {\cO(\eta + \sqrt{\eps'})}$\\ \\

		    & \sendTlr & $\generateB$ &$d( B \vert \tilde{A} ) \leq {\cO(\eta + \sqrt{\eps'})}$\\ \\ 
		  
		   &\sendTTlr &$\generateBB$ &$d( B \vert \tilde{A} ) \leq {\cO(\eta + \sqrt{\eps'})}$\\ \\
		   
		   $\overline{Z}= \Ext_3(T,B)$&\sendBrl & & \\ 
		   
		   \generateYSbar&&&$d( \overline{Z}_s \vert \tilde{B} ) \leq {\cO(\eta + \sqrt{\eps'})}$ \\ \\
		   
		   $\overline{Z}^\prime= \Ext_3(T^\prime,B^\prime)$ &\sendBBrl & &$d( \overline{Z}_s \vert \tilde{B} ) \leq {\cO(\eta + \sqrt{\eps'})}$\\ \\

		   & \sendYSbarlr &  $\generateAbar$ &$d( \overline{A} \vert \tilde{A} ) \leq {\cO(\eta + \sqrt{\eps'})}$\\ \\
		   
		 	\generateYSSbar&  \sendYSSbarlr& $\generateAAbar$&$d( \overline{A} \vert \tilde{A} ) \leq {\cO(\eta + \sqrt{\eps'})}$\\ \\

	$O=\Ext_3(T,\overline{A}) $& \sendAbarrl& &$d( O \vert \tilde{B} ) \leq {\cO(\eta + \sqrt{\eps'})}$ \\ \\

			$O^\prime= \Ext_3(T^\prime, \overline{A}^\prime)$ & \sendAAbarrl& &$d( O \vert \tilde{B} ) \leq {\cO(\eta + \sqrt{\eps'})}$\\ \\

		\end{tabular}
		{\small {\caption{\label{prot:Var_GEN(1,1)NotDiffBefore}
					 $( T,T',Z,Z',G,G',N, Y,Y',G,G',M)$. }}
		}
	\end{center}
\end{Protocol}
\begin{Protocol}
	\begin{center}
		\begin{tabular}{l r r r}
			Alice:  $(T,T^\prime,Z,G,G',N)$ &  & ~~~~~~~~~~~~Bob: $(Y, Y^\prime,Z',G,G',M)$ & $\quad$ Analysis \\
			
			\hline\\
			$Z_s = \pre(Z,s)$ & &  &$d( Z_s \vert \tilde{B} ) \leq {\eta}$\\ \\ 
			&&	$Z^\prime_s = \pre(Z^\prime,s)$& \\
				&  & 			 $A^\prime = \Ext_1(Y^\prime, Z_s^\prime)$ & \\ & &  $C^\prime= \Ext_2(Z^\prime, A^\prime)$ & \\ & &  $B^\prime= \Ext_1(Y^\prime, C^\prime)$ &\\ \\
				
			
		  $\overline{Z}^\prime= \Ext_3(T^\prime, B^\prime)$& \sendBBrl& &$d( Z_s \vert \tilde{B} ) \leq {\eta}$\\ \\
		 
		  & \sendYSlr& $\generateA$ &$d( A \vert \tilde{A} ) \leq {\cO(\eta + \sqrt{\eps'})}$\\ \\
		  
		  $\overline{Z}_s^\prime=\pre( \overline{Z}^\prime,s) $	& \sendYSSbarlr&$\generateAAbar$ &$d( A \vert \tilde{A} ) \leq {\cO(\eta + \sqrt{\eps'})}$ \\ \\
		  	
		 $\generateT$& \sendArl&  &$d(C \vert \tilde{B} ) \leq {\cO(\eta + \sqrt{\eps'})}$ \\ \\
	
	  $O^\prime= \Ext_3(T^\prime, \overline{A}^\prime)$& \sendAAbarrl& &$d( C \vert \tilde{B} ) \leq {\cO(\eta + \sqrt{\eps'})}$ \\ \\ 
	 
		 & \sendTlr& $\generateB$ &$d( B \vert \tilde{A} ) \leq {\cO(\eta + \sqrt{\eps'})}$ \\ \\
		 
			 	 & $O^\prime \longrightarrow O^\prime$ & &$d( B \vert \tilde{A} ) \leq {\cO(\eta + \sqrt{\eps'})}$\\ \\
		
		 $\overline{Z}= \Ext_3(T,B)$& \sendBrl& & \\ 
		$\overline{Z}_s=$Prefix$(\overline{Z},s)$&&&$d( \overline{Z}_s \vert \tilde{B} ) \leq {\cO(\eta + \sqrt{\eps'})}$ \\ \\
		  &\sendYSbarlr &$\generateAbar$ &$d( \overline{A} \vert \tilde{A} ) \leq {\cO(\eta + \sqrt{\eps'})}$\\ \\
		  
		 $O= \Ext_3(T, \overline{A})$& \sendAbarrl &  &$d(O \vert \tilde{B} ) \leq {\cO(\eta + \sqrt{\eps'})}$\\ \\

		\end{tabular}
		{\small {\caption{\label{prot:Var_GEN(1,1)DiffBefore}
					 $( T,T',Z,Z',G,G',N, Y,Y',G,G',M)$.}}
		}
	\end{center}
\end{Protocol}
\begin{changemargin}{0cm}{0cm}
\section*{Protocols for $2\nmext$}
{Protocol~\ref{prot:block2}' is exactly same as Protocol~\ref{prot:block2} except replacing extractors and parameters as in Section~\ref{sec:2nm}.} 
\begin{Protocol}[H] 
	\begin{center}
	\scalebox{0.88}{
		\begin{tabular}{l c c r}
			Alice:  $(X, \hat{X}, X^\prime, \hat{X^\prime}, N)$ &  & ~~~~~~~~~~~~Bob: $(Y, \hat{Y}, Y^\prime, \hat{Y^\prime}, M)$ & $\quad$ Analysis \\
			\hline\\
			$X_1=\pre(X,3k)$&  &  &$\hmin{X_1}{\tilde{B}} \geq 2k$\\ 
			&&$Y_1=\pre(Y,3k)$ & $\hmin{Y_1}{\tilde{A}} \geq 2k$ \\ \\
		&$X_1 \longrightarrow X_1$ &$R=\mathsf{IP}_1(X_1,Y_1)$& $d(R \vert XX') \leq \cO(\eps)$\\
		& &&\\
		$R=\mathsf{IP}_1(X_1,Y_1)$ & $Y_1 \longleftarrow Y_1$& & $d(R \vert YY') \leq \cO(\eps)$ \\
		$V= \ecc(X)_R$ & & $W= \ecc(Y)_R$ & \\
		 & $V \longrightarrow V$ & $G= X_1 \circ Y_1 \circ V \circ W$& \\ \\
		 $X_1^\prime=\pre(X^\prime,3k)$&  \\ 
		 &&$Y_1^\prime=\pre(Y^\prime,3k)$ & \\
		&$X_1^\prime \longrightarrow X_1^\prime$ &$R^\prime=\mathsf{IP}_1(X_1^\prime,Y_1^\prime)$& \\
		& &$W^\prime= \ecc(Y^\prime)_{R^\prime}$ &\\ \\
		$R^\prime=\mathsf{IP}_1(X_1^\prime,Y_1^\prime)$& $Y_1^\prime \longleftarrow Y_1^\prime$ & & \\ \\
	
		 	$V^\prime= \ecc(X^\prime)_{R^\prime}$& $V^\prime \longrightarrow V^\prime$ &$G^\prime= X_1^\prime \circ Y_1^\prime \circ V^\prime \circ W^\prime$ & \\ \\
		  & &  $Y_2=\pre(Y,3k^3)$& $\hmin{Y_2}{\tilde{A}} \geq 2k^3$\\ 
		 $X_2=\pre(X,3k^3)$ & & & $\hmin{X_2}{\tilde{B}} \geq 2k^3$
		 \\
		 $Z_0=\IP_2(X_2, Y_2) $ &$Y_2 \longleftarrow Y_2$ & & $d(Z_0 \vert \tilde{B}) \leq \cO(\eps)$ \\
		 & &  $Y_2^\prime=\pre(Y^\prime,3k^3)$
		& 
		 \\$X_2^\prime=\pre(X^\prime,3k^3)$
		 \\
		 $Z_0^\prime=\IP_2(X^\prime_2, Y^\prime_2) $ &$Y_2^\prime \longleftarrow Y_2^\prime$ & & \\ \\
		 &$\left(G,G^\prime\right) \longleftarrow \left(G,G^\prime\right) $&& $d(Z_0 \vert \tilde{B}) \leq \cO(\eps)$ \\ \\
		
		 	Alice: $(X,X',Z_0,Z_0',G,G',N)$ &  & Bob: $(Y,Y',G,G',M)$ & \\
		  \hline
	 &&&\\
		 	& Protocol~\ref{prot:block2}'~$( X,X',Z_0,Z_0',$&\\
        	&$G,G',N,Y,Y',G,G',M)$ & \\

        		&& \\
				Alice: $(X,X',Z,N)$ &  & Bob: $(Y,Y',Z',M)$ & \\ 
			\hline
			&& \\
			$L' =\Ext_6(X,S')$&$S' \longleftarrow S'$ &  $S' =\Ext_4(Y',Z')$ &$d( Z \vert \tilde{B} ) \leq \cO(\eps)$ \\
		    && \\
		    & $ Z \longrightarrow Z$ &  $S =\Ext_4(Y,Z)$ &$d( S \vert \tilde{A} ) \leq \cO(\eps)$ \\
		    && \\
		     & $ L' \longrightarrow L'$ & &$d( S \vert \tilde{A} ) \leq \cO(\eps)$  \\
		    && \\
		    $L =\Ext_6(X,S)$	&$S \longleftarrow S$ &  &$d( L \vert \tilde{B} ) \leq \cO(\eps)$ \\
			\hline \\
		  Alice($L, N$)                & & Bob($L^\prime, Y, Y^{\prime}, M$) &                                   
		\end{tabular}
		{\small {\caption{ \label{prot:2nmExt_full}
					 $( X, \hat{X},X^\prime,\hat{X}', N, Y,\hat{Y}, Y^\prime,\hat{Y}',M )$.}}
		}}
	\end{center}
	
\end{Protocol}
\end{changemargin}

\suppress{
\begin{Protocol} 
	\begin{center}
		\begin{tabular}{l c c r}
			Alice:  $(X,X^\prime, N)$ &  & ~~~~~~~~~~~~Bob: $(Y, Y^\prime,M)$ & Analysis \\
			\hline\\
			& \quad Protocol~\ref{prot:2nmExt} ($X, X^\prime, N, Y, Y^\prime, M$) & & \\
			& & & $d(Z_0 \vert \tilde{B}) \leq \mathcal{O}(\eps)$\\
		\hline
			&& \\
			& Protocol~\ref{prot:block2}~$( X,X',Z_0,Z_0',$&\\
        	&$G,G',N,Y,Y',G,G',M)$ & \\
        	
        		&& \\
				Alice: $(X,X',Z,N)$ &  & Bob: $(Y,Y',Z',M)$ & \\ 
			\hline
			&& \\
			$L' =\Ext_6(X,S')$&$S' \longleftarrow S'$ &  $S' =\Ext_4(Y',Z')$ &$d( Z \vert \tilde{B} ) \leq \cO(\eps)$ \\
		    && \\
		    & $ Z \longrightarrow Z$ &  $S =\Ext_4(Y,Z)$ &$d( S \vert \tilde{A} ) \leq \cO(\eps)$ \\
		    && \\
		     & $ L' \longrightarrow L'$ & &$d( S \vert \tilde{A} ) \leq \cO(\eps)$  \\
		    && \\
		    $L =\Ext_6(X,S)$	&$S \longleftarrow S$ &  &$d( L \vert \tilde{B} ) \leq \cO(\eps)$ \\

		\end{tabular}
		{\small {\caption{ \label{prot:2nmExt_full}
					 $\mathsf{2nmExt}:( X,X^\prime, N, Y, Y^\prime,M )$.}}
		}
	\end{center}
	
\end{Protocol}
}
\section*{Protocols for $t\mhyphen\nmext$}
\begin{Protocol}[H]
	\begin{center}
		\scalebox{0.88}{
		\begin{tabular}{l l l l}
			Alice:  $(X,\hat{X}, N)$ &  & Bob: $(M,Y,\hat{Y}, Y^{[t]}, \hat{Y}^{[t]})$ & $\quad$ Analysis \\
			\hline\\
			& &  $ Y_1 = \pre(Y,d_1)$ & $d( Y_1 \vert \tilde{A} ) = {0}$\\ \\
			$I = \Ext_0(X,Y_1)$& $ Y_1 \longleftarrow Y_1\quad$ & & $d( I \vert \tilde{B} ) \leq {\eps/t}$\\ \\
	
			$I^{1} = \Ext_0(X,Y^{i}_1) $& $  Y^{1}_1 \longleftarrow Y^{1}_1$ & $ Y^{1}_1 = \pre(Y^{1},d_1)$&$d( I \vert \tilde{B} ) \leq {\eps/t}$ \\ 
			
			\quad \quad \vdots &  \quad \quad \vdots & \quad \quad \vdots & \quad \quad \vdots \\ 
			
				$I^{t} = \Ext_0(X,Y^{t}_1) $	& $  Y^{t}_1 \longleftarrow Y^{t}_1$ & $ Y^{t}_1 = \pre(Y^{t},d_1)$&$d( I \vert \tilde{B} ) \leq {\eps/t}$\\ \\

			 & $ I \longrightarrow I$ & $G=Y_1 \circ \ecc(Y)_{ \samp(I)}$ &\\ \\

			
			 & $I^{1} \longrightarrow I^{1}$ & $ G^{1}=Y^{1}_1 \circ \ecc(Y^{1})_{\samp(I^{1})}$  \\ 
			 
			  & \quad \quad \vdots & \quad \quad \vdots &  \\ 
			 
			  & $I^{t} \longrightarrow I^{t}$ & $ G^{t}=Y^{t}_1 \circ \ecc(Y^{t})_{\samp(I^{t})}$  \\ \\
			  
			$T = \Ext_5(X,Y_2) $  & $Y_2 \longleftarrow Y_2$ & $ Y_2 = \pre(Y,d_2)$ & $d( T \vert \tilde{B} ) \leq \eps$\\ \\

			$T^{1} = \Ext_5(X,Y^{1}_2) $ & $Y^{1}_2 \longleftarrow Y^{1}_2$ & $ Y^{1}_2 = \pre(Y^{1},d_2)$ & \\ 
			
			\quad \quad \vdots& \quad \quad \vdots &\quad \quad \vdots&  \\ 
			
			$T^{t} = \Ext_5(X,Y^{t}_2) $ & $Y^{t}_2 \longleftarrow Y^{t}_2$ & $ Y^{t}_2 = \pre(Y^{t},d_2)$ & \\  \\
			
				$Z_0 = \pre(T,h) $& $ G \longleftarrow G\quad$ &&\\ \\
			
				$Z_0^{1} = \pre(T^{1},h) $& $ G^{1} \longleftarrow G^{1}$ & &\\
			\quad \quad \vdots& \quad \quad \vdots &&  \\  
			
			$Z_0^{t} = \pre(T^{t},h) $& $ G^{t} \longleftarrow G^{t}$ & &$d( T \vert \tilde{B} ) \leq \eps$\\ \\

			\hline
			&& \\
			
			Alice: $(T,T^{[t]},Z_0,$&&Bob: $(Y,Y^{[t]},$ & \\
		
				 $Z_0^{[t]},G,G^{[t]},N)$&&$G,G^{[t]},M)$\\
			\hline
			&& \\
			& Protocol~\ref{prot:block2t}~$( T,T^{[t]},Z_0,Z_0^{[t]},$&\\
        	&$G,G^{[t]},N,Y,Y^{[t]},G,G^{[t]},M)$ & \\
        	
        		&& \\
				Alice: $(X,Z,N)$ &  & Bob: $(Y,Y^{[t]},Z^{[t]},M)$ & $\quad$\\
			\hline
			&& \\
			$L^1 =\Ext_6(X,S^1)$&$S^1 \longleftarrow S^1$ &  $S^1 =\Ext_4(Y^1,Z^1)$ &$d( Z \vert \tilde{B} ) \leq \cO(\eps)$ \\ 
			
			\quad \quad \vdots& \quad \quad  \vdots& \quad \quad \vdots& \quad \quad \vdots\\
				$L^t =\Ext_6(X,S^t)$&$S^t \longleftarrow S^t$ &  $S^t =\Ext_4(Y^t,Z^t)$ &$d( Z \vert \tilde{B} ) \leq \cO(\eps)$ \\ \\
		   
		    & $ Z \longrightarrow Z$ &  $S =\Ext_4(Y,Z)$ &$d( S \vert \tilde{A} ) \leq \cO(\eps)$ \\ \\
		    
		     & $ L^1 \longrightarrow L^1$ & &$d( S \vert \tilde{A} ) \leq \cO(\eps)$  \\
		    & \quad \quad  \vdots& &\quad \quad \vdots \\ 
		     & $ L^t \longrightarrow L^t$ & &$d( S \vert \tilde{A} ) \leq \cO(\eps)$  \\ \\
		    $L =\Ext_6(X,S)$	&$S \longleftarrow S$ &  &$d( L \vert \tilde{B} ) \leq \cO(\eps)$ \\
			\hline \\
			Alice($L, N$)                & & Bob($L^{[t]}, Y, Y^{[t]}, M$) & 
		\end{tabular}
		{\small {\caption{\label{prot:block1t}
					 $(X, \hat{X}, N, M, Y, \hat{Y}, Y^{[t]}, \hat{Y}^{[t]})$.}}
		}}
	\end{center}
\end{Protocol}

\newpage

\begin{Protocol}[htb]

\vspace{0.1in}
For $i=1,2,\ldots,a:$ \\

Let $\ind_i = \{ j : G_i = G_i^j \}$ and $\indbar_i = [t]\setminus \ind_i$. Let $\ind^0_i = \{ j : (G_i = G_i^j) \ \wedge \ (G_k = G_k^j  \quad \text{for every} \quad k<i) \}$ and $\ind^1_i = \{ j : (G_i = G_i^j) \ \wedge \ (G_k \ne G_k^j  \quad \text{for some} \quad k<i) \}$. 
	
 \begin{itemize}
        \item Protocol~\ref{prot:Var_GEN-analysis0}~$( T,T^{[t]},Z,Z^{[t]},G,G^{[t]},N, Y,Y^{[t]},G,G^{[t]},M)$ for $(G_i=0)$.
        \item Protocol~\ref{prot:Var_GEN-analysis1}~$( T,T^{[t]},Z,Z^{[t]},G,G^{[t]},N, Y,Y^{[t]},G,G^{[t]},M)$ for $(G_i=1)$.
    
    \end{itemize}

	\quad \quad	$(Z,Z^1, \ldots, Z^t) = (O,O^1, \ldots O^t)$.

	\vspace{0.25cm}
	
		{\small {\caption{\label{prot:block2t}
					 $( T,T^{[t]},Z,Z^{[t]},G,G^{[t]},N, Y,Y^{[t]},G,G^{[t]},M)$.}}
		}
\end{Protocol}



\begin{Protocol}
	\begin{center}
		\begin{tabular}{l l l l}

			Alice:  $(T,T^{[t]},Z,Z^{[t]},G,G^{[t]},M) $ &  & Bob: $(Y, Y^{[t]},Z^{\ind^1_i},G,G^{[t]},M)$ & $\quad$ Analysis \\
			\hline
			  	
			$ Z_s =$ Prefix$(Z,s)$  & &  &$d( Z_s \vert \tilde{B} ) \leq {\eta}$\\ \\
			
			  $\overline{Z}^q= \Ext_3(T^q, A^q)$& $A^q \longleftarrow A^q$ & $A^q = \Ext_1(Y^q,Z_s^q)$ &$d( Z_s \vert \tilde{B} ) \leq {\eta}$ \\ \\
			 
			 & $Z_s \longrightarrow Z_s$ & $A= \Ext_1(Y,Z_s)$ &$ d(A \vert \tilde{A} )  \leq { \cO(\eta + \sqrt{\eps'})}$ \\ \\

			& $Z^p_s \longrightarrow Z^p_s$ & 			 $A^p = \Ext_1(Y^p, Z_s^p)$ & $d (A \vert \tilde{A} ) \leq  {\cO(\eta + \sqrt{\eps'})}$\\ \\
			
		& $ \overline{Z}^q \longrightarrow \overline{Z}^q $ & $ \overline{A}^q=\Ext_1(Y^q,\overline{Z}^q_s)$&\\ 
			&  & $ \overline{C}^q=\Ext_1(\overline{Z}^q,\overline{A}^q)$&\\ 
			 
			 	&  & $ \overline{B}^q=\Ext_1(Y^q,\overline{C}^q)$&$d( A \vert \tilde{A} ) \leq {\cO(\eta + \sqrt{\eps'})}$\\ \\

			 	& $Z^r \longrightarrow Z^r$ & 			 $A^r = \Ext_1(Y^r, Z_s^r)$ & \\ & &  $C^r= \Ext_2(Z^r, A^r)$ & \\ & &  $B^r= \Ext_1(Y^r, C^r)$ &$d (A \vert \tilde{A} ) \leq  {\cO(\eta + \sqrt{\eps'})}$\\ \\
			
			$\overline{Z}= \Ext_3(T,A)$ &$A \longleftarrow A$ &&$d (\overline{Z}_s  \vert  \tilde{B}  ) \leq {\cO(\eta + \sqrt{\eps'})}$ \\ \\ 
		
	        		$\overline{Z}^p= \Ext_3(T^p,A^p)$&  $A^p \longleftarrow A^p$ &  &$d( \overline{Z}_s \vert \tilde{B} ) \leq {\cO(\eta + \sqrt{\eps'})}$\\  \\
	        		
		   $O^q= \Ext_3(T^q, \overline{B}^q)$	& $\overline{B}^q \longleftarrow \overline{B}^q$ &  &$d( \overline{Z}_s \vert \tilde{B} ) \leq {\cO(\eta + \sqrt{\eps'})}$ \\ \\
		 
			$\overline{Z}^r =\Ext_3(T^r, B^r)$ & $ B^r \longleftarrow B^r$ & &$d (\overline{Z}_s  \vert  \tilde{B}  ) \leq {\cO(\eta + \sqrt{\eps'})}$ \\ \\
			
			& $\overline{Z}_s \longrightarrow \overline{Z}_s $ & $\generateAbar $&$d (\overline{A}  \vert  \tilde{A}  ) \leq {\cO(\eta + \sqrt{\eps'})}$ \\ \\
			
				& $\overline{Z}^p_s \longrightarrow \overline{Z}^p_s $ & $\overline{A}^p=\Ext(Y^p,\overline{Z}^p_s)$&$d (\overline{A}  \vert  \tilde{A}  ) \leq {\cO(\eta + \sqrt{\eps'})}$ \\ \\


			&$\overline{Z}^r_s \longrightarrow \overline{Z}^r_s $&$\overline{A}^r = \Ext_1(Y^r,\overline{Z}^r_s)$& $d (\overline{A}  \vert  \tilde{A}  ) \leq {\cO(\eta + \sqrt{\eps'})}$ \\ \\
			
			$\generateTbar$ & $ \overline{A} \longleftarrow \overline{A} $ &  &$d (\overline{C} \vert \tilde{B} ) \leq {\cO(\eta + \sqrt{\eps'})}$  \\ \\
			
				$\overline{C}^p=\Ext_2(\overline{Z}^p, \overline{A}^p)$ & $ \overline{A}^p \longleftarrow \overline{A}^p $ &  &$d (\overline{C} \vert \tilde{B} ) \leq {\cO(\eta + \sqrt{\eps'})}$ \\  \\

			 $O^r = \Ext_3(T^r, \overline{A}^r)$&$\overline{A}^r \longleftarrow \overline{A}^r$&  &$d(\overline{C} \vert  \tilde{B} ) \leq {\cO(\eta + \sqrt{\eps'})}$\\ \\

			 & $\overline{C} \longrightarrow \overline{C}$  & $\overline{B}= \Ext_1(Y,\overline{C})$ & $d (\overline{B} \vert \tilde{A} ) \leq {\cO(\eta + \sqrt{\eps'})}$\\ \\
			
			 & $\overline{C}^p \longrightarrow \overline{C}^p$  & $\overline{B}^p= \Ext_1(Y^p,\overline{C}^p)$ & $d (\overline{B} \vert \tilde{A} ) \leq {\cO(\eta + \sqrt{\eps'})}$\\ \\
			 
		 & $O^q \longrightarrow O^q$ & &$d( \overline{B} \vert \tilde{A} ) \leq {\cO(\eta + \sqrt{\eps'})}$ \\ \\

		 & $O^r \longrightarrow O^r$ & &$d(\overline{B} \vert \tilde{A} ) \leq {\cO(\eta + \sqrt{\eps'})}$\\ \\

			$O= \Ext_3(T,\overline{B})$ & $\overline{B} \longleftarrow \overline{B}$ & &$d(O \vert  \tilde{B} ) \leq {\cO(\eta + \sqrt{\eps'})}$\\ \\
			
				  $O^p= \Ext_3(T^p, \overline{B}^p)$& $\overline{B}^p \longleftarrow \overline{B}^p$	& &$d( O \vert \tilde{B} ) \leq {\cO(\eta + \sqrt{\eps'})}$\\

		\end{tabular}
		{\small {\caption{\label{prot:Var_GEN-analysis0}
					 $( T,T^{[t]},Z,Z^{[t]},G,G^{[t]},N, Y,Y^{[t]},G,G^{[t]},M)$. \\ Execute $\forall p \in \ind^0_i $, $\forall q \in \ind^1_i$,  $\forall r \in \indbar_i $.}}
		}
	\end{center}
\end{Protocol}

\begin{Protocol}
	\begin{center}
		\begin{tabular}{l l l l}
			Alice:  $(T,T^{[t]},Z,Z^{[t]},G,G^{[t]},N) $ &  & Bob: $(Y, Y^{[t]},Z^{\ind^1_i},G,G^{[t]},M)$ & $\quad$ Analysis \\
			\hline
			$ Z_s =$ Prefix$(Z,s)$  &  & &$d( Z_s \vert \tilde{B} ) \leq {\eta}$\\ 
			& &$A^q = \Ext_1(Y^q,Z_s^q)$ &\\ 
			& &$C^q = \Ext_1(Z^q,A^q)$ &\\

		  $\overline{Z}^q= \Ext_3(T^q, B^q)$& $B^q \longleftarrow B^q$&$B^q = \Ext_1(Y^q,C^q)$ &$d( Z_s \vert \tilde{B} ) \leq {\eta}$\\ \\

			& $Z_s \longrightarrow Z_s$ & $A= \Ext_1(Y,Z_s)$ &$d( A \vert \tilde{A} ) \leq {\cO(\eta + \sqrt{\eps'})}$\\ \\
			
			& $Z^p_s \longrightarrow Z^p_s$ & $A^p= \Ext_1(Y^p,Z^p_s)$ &$d( A \vert \tilde{A} ) \leq {\cO(\eta + \sqrt{\eps'})}$\\ \\
			
			  $\overline{Z}_s^q=\pre( \overline{Z}^q,s) $	& $\overline{Z}_s^q \longrightarrow \overline{Z}_s^q$&$ \overline{A}^q = \Ext_1(Y^q,\overline{Z}_s^q)$ &$d( A \vert \tilde{A} ) \leq {\cO(\eta + \sqrt{\eps'})}$ \\ \\

			& $Z_s^r \longrightarrow Z_s^r$ &$A^r = \Ext_1(Y^r,Z_s^r)$ &$d( A \vert \tilde{A} ) \leq {\cO(\eta + \sqrt{\eps'})}$	\\ \\ 	
			
				$\generateT$ & $A \longleftarrow A$ & & $d( C \vert \tilde{B} ) \leq {\cO(\eta + \sqrt{\eps'})}$\\ \\
				
					$C^p = \Ext_2(Z^p,A^p)$& $A^p \longleftarrow A^p$ & & $d( C \vert \tilde{B} ) \leq {\cO(\eta + \sqrt{\eps'})}$\\ \\
					
					 $O^q= \Ext_3(T^q, \overline{A}^q)$ & $\overline{A}^q \longleftarrow \overline{A}^q$& &$d( C \vert \tilde{B} ) \leq {\cO(\eta + \sqrt{\eps'})}$ \\ \\

			  $\overline{Z}^r= \Ext_3(T^r,A^r)$& $A^r \longleftarrow A^r$ &  &$d( C \vert \tilde{B} ) \leq {\cO(\eta + \sqrt{\eps'})}$\\ \\ 
			& \sendTlr & $\generateB$ &$d( B \vert \tilde{A} ) \leq {\cO(\eta + \sqrt{\eps'})}$\\ \\
			
				& $C^p \longrightarrow C^p$ & $B^p=\Ext_1(Y^p,C^p)$ &$d( B \vert \tilde{A} ) \leq {\cO(\eta + \sqrt{\eps'})}$\\ \\
				
				 & $O^q \longrightarrow O^q$ & &$d( B \vert \tilde{A} ) \leq {\cO(\eta + \sqrt{\eps'})}$\\ \\

		& $\overline{Z}^r \longrightarrow \overline{Z}^r$ &$ \overline{A}^r = \Ext_1(Y^r,\overline{Z}^r_s)$  &\\ 
		 & & $\overline{C}^r = \Ext_1(\overline{Z}^r,\overline{A}^r)  $ &\\
		 & & $\overline{B}^r = \Ext_1(Y^r,\overline{C}^r)$ &$d( B \vert \tilde{A} ) \leq {\cO(\eta + \sqrt{\eps'})}$\\ \\
			
			 $\overline{Z}= \Ext_3(T,B)$ & \sendBrl &  &$d( \overline{Z}_s \vert \tilde{B} ) \leq {\cO(\eta + \sqrt{\eps'})}$\\ \\

			 $\overline{Z}^p= \Ext_3(T^p,B^p)$ & $B^p \longleftarrow B^p$ &  &$d( \overline{Z}_s \vert \tilde{B} ) \leq {\cO(\eta + \sqrt{\eps'})}$\\ \\
			
			$O^r= \Ext_3(T^r, \overline{B}^r)$& $\overline{B}^r \longleftarrow \overline{B}^r$ &  &$d( \overline{Z}_s \vert \tilde{B} ) \leq {\cO(\eta + \sqrt{\eps'})}$\\ \\
			
			&  \sendYSbarlr& $\generateAbar$ &$d( \overline{A} \vert \tilde{A} ) \leq {\cO(\eta + \sqrt{\eps'})}$\\ \\
			
				& $\overline{Z}^p_s \longrightarrow \overline{Z}^p_s$& $\overline{A}^p=\Ext_1(Y^p,\overline{Z}^p_s)$ &$d( \overline{A} \vert \tilde{A} ) \leq {\cO(\eta + \sqrt{\eps'})}$\\ \\

			& $O^r \longrightarrow O^r$& &$d( \overline{A} \vert \tilde{A} ) \leq {\cO(\eta + \sqrt{\eps'})}$ \\ \\

		$O= \Ext_3(T,\overline{A})$& \sendAbarrl & &$d( O \vert \tilde{B} ) \leq {\cO(\eta + \sqrt{\eps'})}$\\ \\
		
			$O^p= \Ext_3(T^p,\overline{A}^p)$& $\overline{A}^p \longleftarrow \overline{A}^p$ & &$d( O \vert \tilde{B} ) \leq {\cO(\eta + \sqrt{\eps'})}$\\ \\
			
		\end{tabular}
		{\small {\caption{\label{prot:Var_GEN-analysis1}
					 $( T,T^{[t]},Z,Z^{[t]},G,G^{[t]},N, Y,Y^{[t]},G,G^{[t]},M)$. \\ Execute $\forall p \in \ind^0_i $, $\forall q \in \ind^1_i$,  $\forall r \in \indbar_i $.}}
		}
	\end{center}
\end{Protocol}
\begin{changemargin}{0cm}{0cm}

\section*{Protocols for $t\mhyphen 2\nmext$}
{Protocol~\ref{prot:block2t}' is same as Protocol~\ref{prot:block2t} except replacing extractors and parameters as in Section~\ref{sec:2tnm}.}
\begin{Protocol}[H] 
\centering
		\scalebox{0.88}{
		\begin{tabular}{l l l l}
			Alice: $(  X, \hat{X},X^{[t]},\hat{X}^{[t]}, N)$ &  &Bob: $( Y, \hat{Y},Y^{[t]},\hat{Y}^{[t]},M)$ & $\quad$ Analysis \\
			\hline\\
			$X_1=\pre(X,3k)$&  &  &$\hmin{X_1}{\tilde{B}} \geq 2k$\\ 
			&&$Y_1=\pre(Y,3k)$ & $\hmin{Y_1}{\tilde{A}} \geq 2k$ \\ 
		&$X_1 \longrightarrow X_1$ &$R=\mathsf{IP}_1(X_1,Y_1)$& $d(R \vert XX^{[t]}) \leq \cO(\eps/t)$\\
		& & &\\
		$R=\mathsf{IP}_1(X_1,Y_1)$ & $Y_1 \longleftarrow Y_1$& & $d(R \vert YY^{[t]}) \leq \cO(\eps/t)$ \\ 
		$V= \ecc(X)_{\samp(R)}$ & & $W= \ecc(Y)_{\samp(R)}$& \\ 
		 & $V \longrightarrow V$ & $G= X_1 \circ Y_1 \circ V \circ W$& \\ \\
		 $X_1^1=\pre(X^1,3k)$& &  $Y_1^1=\pre(Y^1,3k)$&  \\ 
		 \hspace{1cm} \vdots & &  \vdots& \\ 
		 $X_1^t=\pre(X^t,3k)$& &  $Y_1^t=\pre(Y^t,3k)$&  \\ 
		&$(X_1^1, \ldots  X_1^t)  \longrightarrow (X_1^1, \ldots,  X_1^t)$ \\
		\\ & &  $R^1=\IP_1(X_1^1, Y_1^1)$ &\\
		&& \vdots &\\
		 & &  $R^t=\IP_1(X_1^t, Y_1^t)$ &\\
		 & $(Y_1^1,\ldots,Y_1^t) \longleftarrow (Y_1^1,\ldots,Y_1^t)$& & 	\\
		  $R^1=\IP_1(X_1^1, Y_1^1)$ & & &\\
		 \hspace{1cm} \vdots & & &\\
		   $R^t=\IP_1(X_1^t, Y_1^t)$ & & &
		   \\ $V^1= \ecc(X^1)_{\samp(R^1)}$& &$W^1= \ecc(Y^1)_{\samp(R^1)}$\\
		   \hspace{1cm} \vdots & &  \vdots& \\
		 $V^t= \ecc(X^t)_{\samp(R^t)}$& &$W^1= \ecc(Y^t)_{\samp(R^t)}$\\
		 & $(V^1,\ldots,V^t) \longrightarrow (V^1,\ldots,V^t)$& & \\
		 & & $G^1= X_1^1 \circ Y_1^1 \circ V^1 \circ W^1$ & \\
		 \ & &  \vdots& \\
		 & & $G^t= X_1^t \circ Y_1^t \circ V^t \circ W^t$ & \\
		  \\
		  & &  $Y_2=\pre(Y,3k^3)$& \\ 
		 $X_2=\pre(X,3k^3)$ & & & 
		 \\
		 $Z_0=\IP_2(X_2, Y_2) $ &$Y_2 \longleftarrow Y_2$ & & $d(Z_0 \vert \tilde{B}) \leq \cO(\eps)$ \\
		 	  $X_2^1=\pre(X^1,3k^3)$& &  $Y_2^1=\pre(Y^1,3k^3)$& \\
		\hspace{1cm}$\vdots$ & & $\vdots$ & \\
		   $X_2^t=\pre(X^t,3k^3)$& &  $Y_2^t=\pre(Y^t,3k^3)$& \\
		   &$(Y_2^1,\ldots,Y_2^t) \longleftarrow (Y_2^1,\ldots,Y_2^t)$&& \\
		   $Z_0^1=\IP_2(X_2^1,Y_2^1) $ &&& \\
		   \hspace{1cm}$\vdots$ &&&\\
		   $Z_0^t=\IP_2(X_2^t,Y_2^t) $ &&&  \\ 
		   & $G \longleftarrow G$ & & \\
		   \\
		  &  $(G^1,\ldots,G^t) \longleftarrow (G^1,\ldots,G^t)$&& \\ \\ \hline \\
		  	Alice: $(X,X^{[t]},Z_0,$&&Bob: $(Y,Y^{[t]},$ & \\
		
				 $Z_0^{[t]},G,G^{[t]},N)$&&$G,G^{[t]},M)$
		\end{tabular}
		{\small {\caption{ \label{prot:2tnmExt}
					 $( X, \hat{X},X^{[t]},\hat{X}^{[t]}, N, Y,\hat{Y}, Y^{[t]},\hat{Y}^{[t]},M )$.}}
		}}
\end{Protocol}
\end{changemargin}

\begin{Protocol} 
	\begin{center}
	\scalebox{0.88}{
		\begin{tabular}{lccr}
			Alice:  $(X, \hat{X}, X^{[t]}, \hat{X}^{[t]}, N)$ &  & ~~~~~~~~~~~~Bob: $(Y, \hat{Y}, Y^{[t]}, \hat{Y}^{[t]}, M)$ & Analysis \\
			\hline\\
			& \quad Protocol~\ref{prot:2tnmExt} $( X,X^{[t]}, N,$&&\\
			& $Y, Y^{[t]}, M )$& & \\
			& & & $d(Z_0 \vert \tilde{B}) \leq \mathcal{O}(\eps)$\\
		 \\
				Alice: $(X,X^{[t]},Z_0,$&&Bob: $(Y,Y^{[t]},$ & \\
		
				 $Z_0^{[t]},G,G^{[t]},N)$&&$G,G^{[t]},M)$\\
			\hline
					& Protocol~\ref{prot:block2t}'~$( X,X^{[t]},Z_0,Z_0^{[t]},$&\\
        	&$G,G^{[t]},N,Y,Y^{[t]},G,G^{[t]},M)$ & \\
        	
        		&& \\
				Alice: $(X,X^{[t]},Z,N)$ &  & Bob: $(Y,Y^{[t]},Z^{[t]},M)$ & $\quad$\\
			\hline
			&& \\
			$L^1 =\Ext_6(X,S^1)$&$S^1 \longleftarrow S^1$ &  $S^1 =\Ext_4(Y^1,Z^1)$ &$d( Z \vert \tilde{B} ) \leq \cO(\eps)$ \\ 
			
			\quad \quad \quad \vdots&  \vdots& \quad \quad \vdots&  $\vdots$ \hspace{1cm} \\
				$L^t =\Ext_6(X,S^t)$&$S^t \longleftarrow S^t$ &  $S^t =\Ext_4(Y^t,Z^t)$ &$d( Z \vert \tilde{B} ) \leq \cO(\eps)$ \\ \\
		   
		    & $ Z \longrightarrow Z$ &  $S =\Ext_4(Y,Z)$ &$d( S \vert \tilde{A} ) \leq \cO(\eps)$ \\ \\
		    
		     & $ L^1 \longrightarrow L^1$ & &$d( S \vert \tilde{A} ) \leq \cO(\eps)$  \\
		    &   \vdots& &\vdots \hspace{1cm} \\ 
		     & $ L^t \longrightarrow L^t$ & &$d( S \vert \tilde{A} ) \leq \cO(\eps)$  \\ \\
		    $L =\Ext_6(X,S)$	&$S \longleftarrow S$ &  &$d( L \vert \tilde{B} ) \leq \cO(\eps)$ \\
		\hline \\
			Alice($L, N$)                & & Bob($L^{[t]}, Y, Y^{[t]}, M$) &

		\end{tabular}
		{\small {\caption{ \label{prot:2tnmExt_full}
					 $( X, \hat{X}, X^{[t]}, \hat{X}^{[t]}, N, Y, \hat{Y}, Y^{[t]}, \hat{Y}^{[t]},  M )$.}}
		}}
	\end{center}
	
\end{Protocol}

\restoregeometry

\newpage
\pagenumbering{arabic}
\setcounter{page}{55}
\section{Privacy amplification against an active adversary} 
\label{sec:PA}
\subsection*{Preliminaries}
We begin with some useful definitions, facts and claims. Let $n, m, d, z$ be positive integers and $k, \eps>0$.

\begin{definition}\label{def:mac1}
	A function $\mac:\{0,1\}^{2m} \times\{0,1\}^m \to \{0,1\}^m$ is an \emph{$\eps$-secure one-time message authentication code} if for all $\mathcal{A}:\{0,1\}^m \times \{0,1\}^m \to \{0,1\}^m\times \{0,1\}^m$ and $b' \in \{0,1\}^m$,
$$\Pr_{s\leftarrow U_{2m}} [ \mathsf{P}(s,\mathcal{A}(b',\mac(s,b')),b')=1] \leq \eps,$$
	where predicate
	$\mathsf{P}:\{0,1\}^{2m} \times \{0,1\}^m \times \{0,1\}^m \times \{0,1\}^m \rightarrow\{0,1\}$ is defined as 
	$$\mathsf{P}(s,b,t,b') \defeq (\mac(s,b) = t ) \, \wedge \, (b' \neq b).$$
\end{definition}
Efficient constructions of $\mac$ satisfying the conditions of Definition~\ref{def:mac1} are known. 
\begin{fact}[Proposition 1 in~\cite{KR09}]\label{fact:mac1}
	For any integer $m > 0$, there exists an efficient family of  $2^{-m}$-secure one-time message authentication code $\mac:\{0,1\}^{2m} \times\{0,1\}^m \to \{0,1\}^m.$
\end{fact}
\begin{definition}\label{markov}
We say joint random variables $ABC$, form a Markov-chain, denoted $A \leftrightarrow B \leftrightarrow C$, iff 
$$\forall b\in \supp(B): (AC|B=b) = (A|B=b) \otimes (C|B=b).$$
\end{definition}
We have the following corollaries of Theorem~\ref{intro:thm:nmext}.
\begin{corollary}
\label{cor:nmextext}
Let  $d = \cO \left(\log^7 \left( \frac{n}{\eps} \right) \right)$ and $k \geq 5d$. Let $\sigma_{X\hat{X}NY\hat{Y}M}$ be a $(k) \mhyphen \qpas$ with $\vert X \vert=n$ and $\vert Y \vert=d$. Let $\nmext:\{0,1\}^n \times\{0,1\}^{d} \to \lbrace 0,1\rbrace^{k/4}$ be an efficient  $(k, \eps)$-quantum secure non-malleable extractor from Theorem~\ref{intro:thm:nmext}. Let $S=\nmext(X,Y)$. Then,
$$ \| \sigma_{SYM} - U_{k/4} \otimes \sigma_{YM} \|_1 \leq \eps.$$
\end{corollary}
\begin{proof}
Let $V: \cH_Y \rightarrow \cH_Y \otimes \cH_{Y'} \otimes  \cH_{\hat{Y}'}$ be a (safe) isometry such that for $\rho = V \sigma V^\dagger$, we have $Y'$ classical (with copy $\hat{Y}'$) and $\Pr(Y \ne Y')_\rho =1.$~\footnote{It is easily seen that such an isometry exists.} Notice the state $\rho$ is a $(k)\mhyphen\nmas$. Since $\nmext$ is a  $(k, \eps)$-quantum secure non-malleable extractor (see Definition~\ref{nme}), we have $$ \| \rho_{SS'YY'M} - U_{k/4} \otimes \rho_{S'YY'M} \|_1 \leq \eps
.$$ 
Using Fact~\ref{fact:data}, we get $$ \| \rho_{SYM} - U_{k/4} \otimes \rho_{YM} \|_1 \leq \eps.$$ 
The desired now follows by noting $\sigma_{XNMY} =  \rho_{XNMY}.$
\end{proof}
\begin{corollary}[$\nmext$ is a quantum secure extractor]
 \label{corr:extractornmext}
Let $\nmext:\{0,1\}^n \times\{0,1\}^{d} \to \lbrace 0,1\rbrace^{k/4}$ be an efficient  $(k, \eps)$-quantum secure non-malleable extractor from Theorem~\ref{intro:thm:nmext}. $\nmext$ is a $(k,\eps)$-quantum secure  $(n,d,k/4)$-extractor for parameters $d = \cO( \log^7(n/\eps))$ and $k\geq 5d$.
\end{corollary}
\begin{proof}
    Let $\rho_{XEY} =\rho_{XE} \otimes U_d$ be a c-q state ($XY$ classical) such that $\hmin{X}{E}_\rho \geq k$. Consider the following purification $\rho_{X\hat{X}\hat{E}EY\hat{Y}} $ of $\rho_{XEY}$, 
    \[ \rho_{X\hat{X}\hat{E}EY\hat{Y}}= \rho_{X\hat{X}\hat{E}E} \otimes \rho_{Y\hat{Y}},\]
    where $\rho_{X\hat{X}\hat{E}E}$ is a purification of $\rho_{XEY}$ ($\hat{X}$ a copy of $X$) and $\rho_{Y\hat{Y}}$ is the canonical purification of $\rho_Y$.
    Note $\rho$ is a $(k) \mhyphen \qpas$. Let $S = \nmext(X,Y)$. Using Corollary~\ref{cor:nmextext} (by setting  $\sigma_{X\hat{X}NY\hat{Y}M}\leftarrow \rho_{X\hat{X}\hat{E}Y\hat{Y}E})$, we get \[\| \rho_{SYE} - U_{k/4} \otimes \rho_{YE} \|_1 \leq \eps. \qedhere\] 
\end{proof}
\suppress{
\begin{corollary}
 \label{corr:extractornmextqma}
 
 Let $\sigma_{X\hat{X}NY\hat{Y}M}$ be a $(k) \mhyphen \qmas$ with $\vert X \vert =n$,  $\vert Y \vert =d$.  Let $\nmext$ be the $(k,\eps)$-quantum secure strong $(n,d,k/4)$-extractor from Corollary~\ref{corr:extractornmext}. Let $S=\nmext(X,Y)$. Then,
 \[\Delta_B(\sigma_{SYM} , U_{k/4} \otimes \sigma_{YM} ) \leq \sqrt{\eps} \quad  \textnormal{and}\quad \Vert \sigma_{SYM} - U_{k/4} \otimes \sigma_{YM} \Vert_1 \leq \sqrt{8\eps}. \]
\end{corollary}
  \begin{proof}
    We use Lemma~\ref{lem:2},  with the following assignment of registers (below the registers on the left are from Lemma~\ref{lem:2} and the registers on the right are the registers in this proof)
 $$(\theta_X, \theta_A, \theta_{S}, \theta_{B}) \leftarrow (\sigma_{X},  \sigma_{\hat{X}N} , \sigma_{Y}, \sigma_{\hat{Y}M}),$$
 and $\Ext$ from Lemma~\ref{lem:2} as $\nmext$ (here), we get
  $$\Delta_B(\sigma_{S\hat{Y}M} , U_{k/4} \otimes \sigma_{\hat{Y}M} ) \leq \sqrt{\eps}.$$
  Noting $\hat{Y}$ is a copy of $Y$ in $\sigma$, and using Fact~\ref{fidelty_trace} we get the desired.
  \end{proof}
   }

\begin{claim}\label{claim:macprop1c}
 Let $\mac$ be an $\eps$-secure one-time message authentication code from Definition~\ref{def:mac1}. Let $SB'T'$ be such that $SB'= U_{2m} \otimes U_m$ and $T'= \mac(S,B')$. Let $BT$ be such that 
$S \leftrightarrow B'T' \leftrightarrow BT$ and $\vert B\vert =\vert T \vert =m$. Then, 
	$$\Pr\big[\mathsf{P}(S,B,T,B')=1 \big] \,\leq\,\eps.$$ 
\end{claim}
\begin{proof}
For each $(b',t')$, define  $g_{(b't')}:\{0,1 \}^m \times \{0,1 \}^m \to [0,1]$ as 
$$g_{(b't')}(b,t) \defeq \expect{s\leftarrow S^{b't'}}{\mathsf{P}(s,b,t,b')=1}.$$
Define, $\mathcal{A}: \{0,1 \}^m \times \{0,1 \}^m \to \{0,1 \}^m \times \{0,1 \}^m$ as
$$\mathcal{A}(b',t') \defeq \argmax \{ g_{(b't')}(b,t) \}.~\footnote{If there are more than one achieving maximum, pick one of them arbitrarily.}$$
Consider,
\begin{align}
& \Pr\big[\mathsf{P}(S,B,T,B')=1 \big] &\nonumber \\ 
& =  \expect{(b't')\leftarrow B'T'}{\expect{(bt)\leftarrow BT^{b't'}} {\expect{s\leftarrow S^{b't'}}{\mathsf{P}(s,b,t,b')=1 }}} &\mbox{(since $S \leftrightarrow B'T' \leftrightarrow BT$)}\nonumber  \\ 
& \leq  \expect{(b't')\leftarrow B'T'}{\expect{s\leftarrow S^{b't'}}{\mathsf{P}(s,\mathcal{A}(b',t'),b')=1 }} & \mbox{(Definition of $\mathcal{A}$)}\nonumber \\ 
& = \expect{(sb') \leftarrow SB'}{  \mathsf{P}(s,\mathcal{A}(b',\mac(s,b')),b')=1} & \mbox{(Definition of $T'$)}\nonumber \\
&\leq \eps \enspace.& \mbox{(Definition~\ref{def:mac1})} \nonumber & \qedhere
\end{align}
\end{proof}
\begin{claim}\label{claim:macprop1q}
Let $\mac$ be an $\eps$-secure one-time message authentication code from Definition~\ref{def:mac1}. Let $\rho_{SB'T'E'}$ be a c-q state ($SB'T'$ classical) such that 
\[\rho_{SB'T'E'} = \rho_{SB'T'} \otimes \rho_{E'} \quad ; \quad \rho_{SB'} = U_{2m} \otimes U_m \quad ; \quad T'=\mac(S,B') .\]  Let classical registers $BT$ be generated by a quantum adversary using (safe on classical registers) isometry $V: \mathcal{H}_{E^\prime} \otimes \mathcal{H}_{B^\prime} \otimes \mathcal{H}_{T^\prime} \rightarrow \mathcal{H}_{E^{\prime \prime}}  \otimes \mathcal{H}_{B'} \otimes \mathcal{H}_{T'} \otimes \mathcal{H}_{B} \otimes \mathcal{H}_{T}$~\footnote{We included the copies of classical registers $B,T$ in register $E''$.}. Let $\sigma_{SB'T'BTE''} = V \rho_{SB'T'E'} V^\dagger$. Then, 
	$$\Pr\big[\mathsf{P}(S,B,T,B')=1 \big]_\sigma \,\leq\,\eps.$$ 
\end{claim}
\begin{proof}
Note in state $\rho$, for every $b't' \in \supp(B'T')$, we have $\rho^{b't'}_{SE'} =\rho^{b't'}_{S} \otimes \rho_{E'}.$ Since $V$ is safe on registers $B'T'$, we have 
$$\sigma^{b't'}_{SE''BT} =\sigma^{b't'}_{S} \otimes \sigma^{b't'}_{E''BT},$$ 
where $\sigma^{b't'}_{S}=\rho^{b't'}_{S}.$ Using Fact~\ref{fact:data}, we get $\sigma^{b't'}_{SBT} =\sigma^{b't'}_{S} \otimes \sigma^{b't'}_{BT}$. Using Definition~\ref{markov} we have (in state $\sigma$), $S \leftrightarrow B'T' \leftrightarrow BT$. Using Claim~\ref{claim:macprop1c} (for state $\sigma_{SB'T'BT}$) while noting $\sigma_{SB'T'}=\rho_{SB'T'}$, we get the desired.\suppress{Consider,
\begin{align}
 \mutinf{S}{B'T'}_\rho & = \mutinf{S}{B'T'E'}_\rho - \condmutinf{S}{E'}{B'T'}_\rho &  \mbox{(Definition~\ref{condmutinfo})}  \nonumber\\
 & = \mutinf{S}{B'T'E'}_\rho  &   \mbox{($\rho_{SB'T'E'} = \rho_{SB'T'} \otimes \rho_{E'}$)} \nonumber\\
     &\geq    \mutinf{S}{B'T'BTE''}_\sigma  & \mbox{(Fact~\ref{monotonicitymap})}\nonumber \\
      &\geq    \mutinf{S}{B'T'BT}_\sigma  & \mbox{(Fact~\ref{monotonicitymap})}\label{eq9876}.
\end{align}Now, \begin{align}
 \condmutinf{S}{BT}{B'T'}_\sigma & = \mutinf{S}{BTB'T'}_\sigma - \mutinf{S}{B'T'}_\sigma & \mbox{(Definition~\ref{condmutinfo})}  \nonumber\\
 & =\mutinf{S}{BTB'T'}_\sigma - \mutinf{S}{B'T'}_\rho &   \mbox{($\sigma_{SB'T'} = \rho_{SB'T'}$)}  \nonumber\\
     &\leq  0 & \mbox{(Eq.~\eqref{eq9876})}\nonumber .
\end{align}Thus, using Fact~\ref{nonneg}, we get $\condmutinf{S}{BT}{B'T'}_\sigma=0$. Now, using Claim~\ref{claim:macprop1c} (for state $\sigma_{SB'T'BT}$) and noting $\sigma_{SB'T'}=\rho_{SB'T'}$, we get the desired.  }
\end{proof}
\begin{claim}\label{claim:macprop2q}
Let $\mac$ be an $\eps$-secure one-time message authentication code from Definition~\ref{def:mac1}. 
Let $STBB'$ be random variables such that,   
$STBB' = U_{2m} \otimes TBB'$ and $B' =U_m$. Then, 
$$\Pr[\mathsf{P}(S,B,T,B') =1] \leq \eps .$$
\end{claim}
\begin{proof}
Note that $S \leftrightarrow B' \leftrightarrow BT$. This implies $S \leftrightarrow B'T' \leftrightarrow BT$, where $T'= \mac(S,B')$. Using Claim~\ref{claim:macprop1c}, the desired follows.  
\end{proof}

We start with the definition of a quantum secure privacy amplification (PA) protocol against active adversaries. The following description is from~\cite{ACLV18}. 
A PA protocol $(P_A, P_B)$ is defined as follows.
The protocol is executed by two parties (Alice and Bob) sharing a secret $X \in \lbrace 0,1 \rbrace^n$.
Their actions are described by $P_A$ and $P_B$ respectively. In addition there is an active, computationally
unbounded adversary Eve, who might have some quantum side information $E$
correlated with $X$, where $\rho$ denotes the initial state of the protocol.
Note that, for the definition, it is not necessary  to specify exactly how the protocols are formulated; informally, each player's action is described by a sequence of efficient algorithms that compute the player's next message, given the past interaction.

The protocol should have the following property:  in case protocol does not terminate with a rejection, output keys $R_A,R_B$ should be random and statistically independent of  Eve's view.
Moreover, they must output the
same keys $R_A=R_B$ with overwhelming probability. 
We assume that Eve is in full control of the
communication channel between Alice and Bob, and can arbitrarily
insert, delete, reorder or modify messages sent by Alice and Bob.
At the end of protocol, Alice outputs $R_A\in
\{0,1\}^z \cup \{\perp\}$, where $\perp$ is a special symbol indicating
rejection. Similarly, Bob outputs $R_B \in \{0,1\}^z \cup
\{\perp\}$. For a random variable $R\in
\{0,1\}^z \cup \{\perp\}$, let $\mathsf{purify}(R)$ be a random variable on $z$-bit strings that is deterministically equal to $\perp$ if $R=\perp$, and is otherwise uniformly distributed over $\lbrace 0, 1\rbrace ^z$. The following definition generalizes the classical definition in~\cite{DLWZ14}.

\begin{definition}[\cite{ACLV18}]\label{privamp}
	Let  $\Theta$ be the joint state of Alice, Bob and Eve at the end of the protocol given by $(P_A,P_B)$ including $\mathsf{purify}(R_A)$ and $\mathsf{purify}(R_B)$. We say that a PA protocol $(P_A,P_B)$ is $(k,z,\epsilon)$-secure against quantum adversaries if for any initial state $\rho_{XE}$ such that $\Hmin(X|E)_\rho \geq k$ it satisfies the following three properties. 

	\begin{enumerate}
		\item \emph{Correctness.} If the adversary does not interfere with the protocol, then $\Pr\left(R_A=R_B \neq \perp \right)_{\Theta}=1$. 
		\item \emph{Robustness.} In the presence of an active adversary, $\Pr\left(Q(R_A,R_B)=1\right)_{\Theta}\leq \eps$,
		where $Q(R_A,R_B)$ is the predicate $(R_A \neq R_B \land~ R_A \neq \perp \land~ R_B \neq \perp)$.
		
		\item \emph{Extraction.} Let $\Theta_{\tilde{E}}$ be the final quantum state possessed by Eve (including the transcript of the protocol). The following should hold: 
		\[\Vert \Theta_{R_A  \tilde{E}}- \Theta_{\mathsf{purify}(R_A) \tilde{E}} \Vert_1 \leq \eps		~~~~\mbox{and}~~~~
	\Vert \Theta_{R_B \tilde{E}}- \Theta_{\mathsf{purify}(R_B) \tilde{E}} \Vert_1 \leq \eps\;.\]
		In other words, whenever a party does not reject, the party's key is (approximately) indistinguishable from a fresh random string to the adversary.
	\end{enumerate}
\end{definition}
\subsection*{PA Protocol}
Let $\delta>0$ be a small constant.
\begin{itemize}
	\item Let $\nmext:\{0,1\}^n \times\{0,1\}^{d} \to \lbrace 0,1\rbrace^{2m}$ be a $(k,\eps)$-quantum secure $(n,d,2m)$-non-malleable extractor from Theorem~\ref{intro:thm:nmext} with following choice of parameters,
	\[ d = \cO \left(\log^7 \left( \frac{n}{\eps} \right) \right) \quad ; \quad k \geq 5d \quad ; \quad k \geq 8m. \]
	\suppress{
	We choose $\eps_1=2^{-n^{\delta}}$.}
	\item Let $\mac:\{0,1\}^{2m} \times \{0,1\}^{m} \to \{0,1\}^{m}$ be an $2^{-m}$-secure one-time message authentication code from Fact~\ref{fact:mac1} for $m= \cO \left(\log^3(\frac{n}{\eps}) \right)$. Note $2^{-m} \le \eps$. 
	\item Let $\mathsf{Ext}:\{0,1\}^n\times\{0,1\}^{m} \to \{0,1\}^{z}$ be a $(2z,\eps)$-quantum secure strong extractor from Fact~\ref{fact:extractor}. Taking $2z= (1-2\delta)k$ suffices for the PA application.
\end{itemize}

\begin{Protocol}[htb]
	\begin{center}
		\begin{tabular}{l c r}
			Alice:  $X$ & Eve: $E$ & ~~~~~~~~~~~~Bob: $X$ \\
			
			\hline\\
			Generate $Y \leftarrow U_{d}$\\
			& $Y \xrightarrow{\mathsf{T}_1} Y^\prime$ & \\
					  $S = \nmext(X,Y)$ &&  $S' = \nmext(X,Y^\prime)$\\
		  && Generate $B^\prime \leftarrow U_{m}$ \\
		  	& & $R_B = \Ext(X,B')$\\ 
		  & & $T^\prime= \mac_{}(S^\prime,B^\prime)$\\
			& $B,T \xleftarrow{\mathsf{T}_2} B^\prime,T^\prime$ & \\
		
			If $T \neq \mac_{}(S,B)$: reject ($R_A = \perp$) &&\\
			Otherwise: && \\
			$R_A = \Ext(X,B)$&& \\
			\hline
		\end{tabular}
		{\small {\caption{\label{prot:priv-amp}
					  PA protocol.}}
		}
	\end{center}
\end{Protocol}

\begin{remark}
	In Protocol~\ref{prot:priv-amp}, registers $Y$, $B'$ are generated uniformly and independently of the state of the protocol at that point.   
\end{remark}

\begin{definition}[Active attack]
An active attack against PA protocol is described by 3 parameters.
\begin{itemize}
\item A c-q state $\rho_{XE}$ (of adversary choice) such that $\hmin{X}{E}_{\rho} \geq k$.
\item A CPTP map $\mathsf{T}_1: \mathcal{H}_E \otimes \mathcal{H}_{Y} \rightarrow \mathcal{H}_{E^\prime} \otimes \mathcal{H}_Y \otimes \mathcal{H}_{Y^\prime}$.
\item A CPTP map $\mathsf{T}_2: \mathcal{H}_{E^\prime} \otimes \mathcal{H}_{B^\prime} \otimes \mathcal{H}_{T^\prime} \rightarrow \mathcal{H}_{E^{\prime \prime}}  \otimes \mathcal{H}_{B'} \otimes \mathcal{H}_{T'} \otimes \mathcal{H}_{B} \otimes \mathcal{H}_{T}$.
\end{itemize} 
\end{definition}

\subsection*{Result}
\begin{theorem}
\label{thm:PA_added1}
For any active attack $(\rho_{XE},T_1,T_2)$, Protocol~\ref{prot:priv-amp} is $\left(k, \left(\frac{1}{2} -\delta \right)k, \cO(\sqrt{\eps}) \right)$-secure as defined in Definition~\ref{privamp} with communication $\cO \left(d\right)$. 
\end{theorem}
\begin{proof}
For the purpose of this proof, without any loss of generality, we can consider $\mathsf{T}_1$ and $\mathsf{T}_2$ to be isometries because tracing out registers after applying an isometry (which amounts to applying a CPTP map) will only weaken the adversary holding the side information. We can assume isometries to be safe and  registers $(YY'B'T'BT)$ to be classical, since both Alice and Bob when executing the protocol, keep a copy of the registers they send and also make a copy of the registers they receive~\footnote{We do not mention it in the protocol or in the analysis, instead we assume and proceed for simplifying the analysis.}. We keep the transcript of the protocol in adversary side information at different stages of the protocol. 

Correctness of the protocol follows by observation. Let the adversary choose state $\rho_{XE}$. Let $\rho_{X\hat{X} E \hat{E}Y\hat{Y}} = \rho_{X\hat{X} E \hat{E}} \otimes \rho_{Y\hat{Y}}$ be a pure state such that,
\[\rho_{X\hat{X}  \hat{E}EY\hat{Y}} = \rho_{X\hat{X}  \hat{E}E} \otimes \rho_{Y\hat{Y}} \quad ; \quad  \hmin{X}{E}_\rho \geq k \quad ; \quad \rho_Y = U_{d},\]
where registers $XY$ are classical with copies $\hat{X}\hat{Y}$. Note in Protocol~\ref{prot:priv-amp}, register $X$ (is held by both Alice and Bob), register $E$ is the quantum side information with Eve, register $Y$ is generated by Alice in the first step and $\hat{E}$ is the purification inaccessible to any of Alice, Bob or Eve throughout the protocol.

 Let $\sigma_{} = \mathsf{T}_1(\rho)$. Note $\sigma$ is $(k) \mhyphen \qpas$ since $\hmin{X}{E'YY'\hat{Y}}_\sigma \geq k$ and $\sigma_Y =U_{\vert Y \vert}$. Using Corollary~\ref{cor:nmextext}, we get
\begin{equation}\label{eq:priv100}
    \Vert \sigma_{SYY'E'} - U_{2m} \otimes \sigma_{YY'E'}  \Vert_1 \leq \eps.
\end{equation}
 Let $\tau_{}$ be a pure state after Alice generates $S$ (with copy $\hat{S}$), Bob generates classical registers $S'B'R_BT'$ (with copies $\hat{S'}\hat{B'}\hat{R_B}\hat{T'}$). Note, 
\begin{equation}\label{eq:priv1}
    \tau_{XSS'YY'E'B'}= \sigma_{XSS'YY'E'} \otimes U_m.
\end{equation}Let $\Theta_{} $ be the final pure state at the end of protocol including $\mathsf{purify}(R_A)\mathsf{purify}(R_B)$ along with their copies $\hat{\mathsf{purify}(R_A)} \hat{\mathsf{purify}(R_B)}$. \\

\noindent {\bf Robustness property:} We now show the robustness property of Protocol~\ref{prot:priv-amp}. Let $p= \Pr(Y^\prime \neq Y)_\sigma$. Based on the value of $p$, we divide our analysis into three parts. \\
 
\noindent \textbf{Case 1 ($p =0$):} In this case we have $\Pr(S=S')_\tau=1$. From Eq.~\eqref{eq:priv100} and~\eqref{eq:priv1} we have 
\begin{equation}\label{eq:priv767}\nonumber
    \Vert  {\tau}_{SB'YY'E' } - U_{2m} \otimes U_{m} \otimes  {\sigma}_{YY'E' }   \Vert_1 \leq \eps.
\end{equation}Let $\hat{\tau}_{SB'YY'E'}  = U_{2m} \otimes U_{m} \otimes  {\sigma}_{YY'E' }.$ Using Fact~\ref{fact:data}, we get 
\begin{equation}\label{eq:priv468}
    \Vert  {\tau}_{SB'T'YY'E' } - \hat{\tau}_{SB'T'} \otimes  \hat{\tau}_{YY'E'}  \Vert_1 \leq \eps.
\end{equation}
 Let $\hat{\Theta}$ be the final state if protocol is run on $\hat{\tau}$ instead of $\tau$. Using Fact~\ref{fact:data}, we get 
 \begin{equation}\nonumber
    \Vert  {\Theta}_{SBTB'T'YY'E'' } - \hat{\Theta}_{SBTB'T'YY'E'' }  \Vert_1 \leq \eps.
\end{equation}
Using Fact~\ref{fact:data} again, we get 
 \begin{equation}\label{eq:priv768}
    \Vert  {\Theta}_{SBB'T'} - \hat{\Theta}_{SBB'T'}  \Vert_1 \leq \eps.
\end{equation}From Claim~\ref{claim:macprop1q},  with the following assignment of registers and isometry, 
 $$( \rho_{SB'T'}, \rho_{E'}) \leftarrow (\hat{\tau}_{SB'T'} ,\hat{\tau}_{YY'E'}), \quad V \leftarrow \mathsf{T}_2,$$ 
 we get \[\Pr\left(\mathsf{P}(S,B,T,B')=1   \right)_{\hat{\Theta}}\leq \eps.\]Using Eq.~\eqref{eq:priv768} and Fact~\ref{fact:close} we get,
 \[\Pr\left(\mathsf{P}(S,B,T,B')=1   \right)_{{\Theta}}\leq 2\eps.\]Noting, 
\[ \Pr\left( Q(R_A,R_B)=1    \right)_{{\Theta}}\leq \Pr\left(\mathsf{P}(S,B,T,B')=1\right)_{{\Theta}}\]
we get 
\begin{equation}\label{eq:priv54}
    \Pr\left( Q(R_A,R_B)=1   \right)_{{\Theta}} \leq 2\eps.
\end{equation}
This establishes the robustness property in Case~$1$. \\

\noindent \textbf{Case 2 ($p =1$):} In this case, $\sigma$ is a $(k) \mhyphen \nmas$. Since $\nmext$ is $(k,\eps)$-quantum secure non-malleable extractor (Definition~\ref{nme}), we get \begin{equation}\label{eq:priv2004}\nonumber
     \Vert  {\sigma}_{SS'YY'E' } - U_{2m} \otimes  {\sigma}_{S'YY'E' }   \Vert_1 \leq \eps.
\end{equation}
Since $\tau_{SS'YY'E'B'} = \sigma_{SS'YY'E'} \otimes U_{m}  $, we have $$\Vert  {\tau}_{SB'S'YY'E' } - U_{2m} \otimes U_{m} \otimes  {\sigma}_{S'YY'E' }   \Vert_1 \leq \eps.$$Let $\hat{\tau}_{SB'S'YY'E'}  = U_{2m} \otimes U_{m} \otimes  {\sigma}_{S'YY'E' }.$  Using Fact~\ref{fact:data}, we get 
\begin{equation}\label{eq:priv701}\nonumber
    \Vert  {\tau}_{SB'T'S'YY'E' } - U_{2m} \otimes  \hat{\tau}_{B'T'S'YY'E'}  \Vert_1 \leq \eps.
\end{equation}Also note $\hat{\tau}_{B'} =U_{m}$. Let $\hat{\Theta}$ be the final state if protocol is run on $\hat{\tau}$ instead of $\tau$. Using Fact~\ref{fact:data}, we get 
\begin{equation}\nonumber
    \Vert  {\Theta}_{SB'T'BTS'YY'E'' } - U_{2m} \otimes  \hat{\Theta}_{B'T'BTS'YY'E''}  \Vert_1 \leq \eps.
\end{equation}Using Fact~\ref{fact:data} again, we get 
\begin{equation}\label{eq:priv702}
    \Vert  {\Theta}_{SB'BT} - U_{2m} \otimes  \hat{\Theta}_{B'BT}  \Vert_1 \leq \eps.
\end{equation}Since $\mathsf{T}_2$ is safe on register $B'$, we also have $\hat{\Theta}_{B'} =U_{m}$. From Claim~\ref{claim:macprop2q}, with the following assignment of registers (below the registers on the left are from Claim~\ref{claim:macprop2q} and the registers on the right are the registers in this proof)
 $$( SB'TB) \leftarrow (\hat{\Theta}_{SB'TB}),$$
we get \[\Pr\left(\mathsf{P}(S,B,T,B')=1   \right)_{ \hat{\Theta}}\leq \eps.\]Using Eq.~\eqref{eq:priv702}  and Fact~\ref{fact:close} we get, 
\[\Pr\left(\mathsf{P}(S,B,T,B')=1   \right)_{{\Theta}}\leq 2\eps.\]
Noting, 
\[ \Pr\left( Q(R_A,R_B)=1 \right)_{{\Theta}}\leq \Pr\left(\mathsf{P}(S,B,T,B')=1\right)_{{\Theta}}\]
we get \begin{equation}\label{eq:priv55}
     \Pr\left( Q(R_A,R_B)=1 \right)_{{\Theta}} \leq 2\eps.
\end{equation}This establishes the robustness property in Case~$2$. 
 
 \vspace{0.2cm}
 
\noindent \textbf{Case 3: $0<p <1$.} 
In the analysis, we consider a pure state $\tilde{\sigma}$ which is generated from $\rho$, in the following way: 
\begin{itemize}
    \item Generate $\sigma =\mathsf{T_1}(\rho)$.
    \item Generate one bit classical register $C$ (with copy $\hat{C}$) such that $C=1$ indicates $Y \ne Y'$ in state $\sigma$. 
    \item Conditioned on $C=0$, generate classical register $Y''$ (with copy $\hat{Y''}$)  such that $Y'' \ne Y$. 
    \item Conditioned on $C=1$, generate classical register $Y''$ (with copy $\hat{Y''}$) such that $Y'' = Y'$. 
\end{itemize}
Note $\Pr(Y \ne Y'')_{\tilde{\sigma}} =1$ and $\tilde{\sigma}$ is a $(k) \mhyphen \nmas$ (by Fact~\ref{fact102}).  Since $\nmext$ is $(k,\eps)$-quantum secure non-malleable extractor (Definition~\ref{nme}), we get
\begin{equation}\label{eq:priv3}
     \Vert  \tilde{\sigma}_{SS'YY'Y''CE'} - U_{2m} \otimes  \tilde{\sigma}_{S'YY'Y'' C E'}   \Vert_1 \leq \eps,
\end{equation}
where $S= \nmext(X,Y)$ and $S'= \nmext(X,Y'')$. Note by construction of state $\tilde{\sigma}$, we have $\Pr(C=1)_{\tilde{\sigma}} = p$. Let $\tilde{\sigma}^1 = \tilde{\sigma} \vert (C=1)$ and $\sigma^1 = \sigma \vert (Y \ne Y')$. Thus, from Eq.~\eqref{eq:priv3} and Fact~\ref{traceavg}, we get 
\begin{equation}\label{eq:priv4}
    \Pr(C=1)_{\tilde{\sigma}}  \Vert  \tilde{\sigma}^1_{SS'YY'Y'' E' } - U_{2m} \otimes  \tilde{\sigma}^1_{S'YY'Y'' E' }   \Vert_1 \leq \eps,
\end{equation}
where $S= \nmext(X,Y)$, $S'= \nmext(X,Y'')$ and $Y''$ is a copy of $Y'$ in $\tilde{\sigma}^1$.
Noting $\Pr(C=1)_{\tilde{\sigma}} = p$ and 
 $\tilde{\sigma}^1_{XYY'Y''E'}$ is the same state as $\sigma^1_{XYY'Y''E'}$ (with additional copy of $Y'$ in $Y''$), from Eq.~\eqref{eq:priv4} and using Fact~\ref{fact:data}, we further get 
 \begin{equation}\label{eq:priv101}
     \Vert  \sigma^1_{SS'YY'E' } - U_{2m} \otimes  \sigma^1_{S'YY'E' } \Vert_1 \leq \frac{\eps}{p}.
 \end{equation}We further consider a pure state $\hat{\sigma}$ which is generated from $\rho$, in the following way: 
\begin{itemize}
    \item Generate $\sigma =\mathsf{T_1}(\rho)$.
    \item Generate one bit classical register $C$ (with copy $\hat{C}$) such that $C=1$ indicates $Y \ne Y'$ in state $\sigma$. 
\end{itemize}Note by construction of state $\hat{\sigma}$, we have $\Pr(C=0)_{\hat{\sigma}} = 1-p$ and  $\hat{\sigma}$ is a $(k) \mhyphen \qpas$ (by Fact~\ref{fact102}). Let $\hat{\sigma}^0 = \hat{\sigma} \vert (C=0)$ and $\sigma^0 = \sigma \vert (Y=Y').$ Using Corollary~\ref{cor:nmextext}, we get \begin{equation}\label{eq:priv5}\nonumber
     \Vert  \hat{\sigma}_{SYY'CE' } - U_{2m} \otimes  \hat{\sigma}_{YY'CE' }   \Vert_1 \leq  \eps.
\end{equation}Using Fact~\ref{traceavg}, we get 
\begin{equation}\label{eq:priv6}\nonumber
     \Vert  \hat{\sigma}^0_{SYY'E' } - U_{2m} \otimes  \hat{\sigma}^0_{YY'E'}   \Vert_1 \leq \frac{\eps}{1-p}.
\end{equation}Noting  
 $\hat{\sigma}^0_{XYY'E'}$ is the same state as $\sigma^0_{XYY'E'}$, we get 
\begin{equation}\label{eq:priv106}
     \Vert  {\sigma}^0_{SYY'E' } - U_{2m} \otimes  {\sigma}^0_{YY'E'}   \Vert_1 \leq \frac{\eps}{1-p}.
\end{equation}
Note, 
\begin{equation}
    \sigma_{SS'YY'E'} = (1-p) \cdot  {\sigma}^0_{SS'YY'E'} +  p \cdot  {\sigma}^1_{SS'YY'E'}. \label{eq:convex}
\end{equation} 
Let $\Theta^0$ and $\Theta^1$ be the final states if we proceed PA protocol with states $\sigma^0$ and $\sigma^1$ after the first round. Using using Eq.~\eqref{eq:priv106} and arguments similar to case~$1$, we get 
\begin{equation}\label{eq:priv7}
    \Pr\left( Q(R_A,R_B)=1  \right)_{{\Theta}^0} \leq  \frac{\eps}{1-p}+ \eps.
\end{equation}Similarly, using Eq.~\eqref{eq:priv101} and similar arguments of case~$2$, we get 
\begin{equation}\label{eq:priv8}
    \Pr\left( Q(R_A,R_B)=1    \right)_{{\Theta}^1} \leq  \frac{\eps}{p}+ \eps.
\end{equation}
Thus, from Eq.~\eqref{eq:convex},~\eqref{eq:priv7}~and~\eqref{eq:priv8}, we have 
\begin{equation}\label{eq:priv9}
    \Pr\left( Q(R_A,R_B)=1   \right)_{{\Theta}} \leq (1-p)\left(\frac{\eps}{1-p}+ \eps\right) + p \left(\frac{\eps}{p}+ \eps\right)  =  3\eps ,
\end{equation}
i.e. the robustness property in Case~$3$. 

\vspace{0.2cm}

\noindent \textbf{Extraction property:} We now show the extraction property of PA protocol. From Eq.~\eqref{eq:priv1}, we have 
$\tau_{XE'YY'SS'B'} = \sigma_{XE'YY'SS'}  \otimes U_{m}$. Consider,\begin{align}
 \hmin{X}{E'YY'SS'}_\tau& = \hmin{X}{E'YY'SS'}_\sigma&   \nonumber\\
     &\geq   \hmin{X}{E'YY'}_\sigma - \vert SS' \vert & \mbox{(Fact~\ref{fact2})}\nonumber \\
    &\geq \hmin{X}{E}_\rho - \vert SS' \vert & \mbox{(Fact~\ref{fact102} and $\rho_{XEY} = \rho_{XE} \otimes U_d$) } \nonumber \\
    &\geq k-2m & \nonumber \label{eq:priv2} \\
     &\geq (1-2\delta)k. & \nonumber  
\end{align}Thus, noting $2z =(1-2\delta)k$ and  $\Ext$ is a quantum secure extractor (see Definition~\ref{qseeded}), we get $$\Vert \tau_{R_BE'YY'SS'B'} - U_z \otimes  \tau_{E'YY'SS'B'} \Vert_1 \leq \eps.$$ Using Fact~\ref{fact:data}, we get 
$$\Vert \tau_{R_BE'YY'T'SS'B'} - U_z \otimes  \tau_{E'YY'T'SS'B'} \Vert_1 \leq \eps,$$
and further, 
\begin{equation}\label{eq:privbob}
    \Vert {\Theta}_{R_B \tilde{E}SS'} - U_z \otimes {\Theta}_{ \tilde{E}SS'}  \Vert_1 \leq   \eps,
\end{equation}where $\tilde{E}$ denotes the registers held by Eve and transcript of the protocol. Using Fact~\ref{fact:data} again, we have
\begin{equation}\label{eq:privbob1}
    \Vert {\Theta}_{R_B \tilde{E}} - U_z \otimes {\Theta}_{ \tilde{E}}  \Vert_1 \leq   \eps,
\end{equation}
the extraction property for Bob. From the robustness property of the  protocol, i.e. Eq.~\eqref{eq:priv54},~\eqref{eq:priv55}~and~\eqref{eq:priv9}, we have $$\Pr\left(Q(R_A,R_B)=1\right)_{\Theta}\leq \cO( \eps ).$$Thus, $\Pr( (R_A=R_B) \vee  (R_A = \perp) )_\Theta \geq 1-\cO( \eps )$. Using Fact~\ref{fact:gentle_measurement} and Fact~\ref{fidelty_trace},
\begin{equation} \label{eq:last_tilde_to_normal}
    \Vert \Theta - \tilde{\Theta} \Vert_1 \leq    \cO\left(\sqrt{ \eps}\right), \text{where $\tilde{\Theta}= \Theta \vert ((R_A=R_B) \vee  (R_A = \perp))$}.
\end{equation} Let $C$ be the predicate to indicate $(\mac(S,B)=T)$. Let  $\tilde{\Theta}^{=\perp} = \tilde{\Theta} \vert (C=0)$ and  $\tilde{\Theta}^{\ne \perp} = \tilde{\Theta} \vert (C=1)$. Note $\Pr(R_A=R_B)_{\tilde{\Theta}^{\ne \perp}}=1$.  
Consider,\begin{align}
 &\Vert \tilde{\Theta}_{R_A \tilde{E}} -\tilde{\Theta}_{  \mathsf{purify}(R_A) \tilde{E}}  \Vert_1 \nonumber\\
 & \leq \Vert \tilde{\Theta}_{R_A \tilde{E}SS'} -\tilde{\Theta}_{  \mathsf{purify}(R_A) \tilde{E}SS'}  \Vert_1& \mbox{(Fact~\ref{fact:data})}  \nonumber\\
  &=   ( \Pr(C=1)_{\tilde{\Theta}})\Vert \tilde{\Theta}^{\ne \perp}_{R_A \tilde{E}SS'} - U_z \otimes \tilde{\Theta}^{\ne \perp}_{   \tilde{E}SS'}  \Vert_1 & \nonumber \\
     &=   (\Pr(C=1)_{\tilde{\Theta}}) \Vert \tilde{\Theta}^{\ne \perp}_{R_B \tilde{E}SS'} - U_z \otimes \tilde{\Theta}^{\ne \perp}_{   \tilde{E}SS'}  \Vert_1  & \mbox{($\Pr(R_A=R_B)_{\tilde{\Theta}^{\ne \perp}}=1$)}\nonumber \\
     &\leq    \Vert \tilde{\Theta}_{R_B \tilde{E}CSS'} - U_z \otimes \tilde{\Theta}_{   \tilde{E}CSS'}  \Vert_1 & \mbox{(Fact~\ref{traceavg})} \nonumber \\
    &=    \Vert \tilde{\Theta}_{R_B \tilde{E}SS'} - U_z \otimes \tilde{\Theta}_{   \tilde{E}SS'}  \Vert_1 & \mbox{($C$ can be generated from $\tilde{E}SS'$)} \nonumber \\
    &\leq  \cO(  \sqrt{ \eps}). \label{eq:tilde_purify_bound}&  \mbox{(Claim~\ref{claim:traingle_rho_rho_prime} and  Eq.~\eqref{eq:privbob})} 
\end{align}

\begin{align}
   \nonumber &\Vert {\Theta}_{R_A \tilde{E}} -{\Theta}_{  \mathsf{purify}(R_A) \tilde{E}}  \Vert_1 \\
   &\leq \Vert {\Theta}_{R_A \tilde{E}} -  \tilde{{\Theta}}_{R_A \tilde{E}}\Vert_1 + \Vert  \tilde{{\Theta}}_{R_A \tilde{E}}-{\Theta}_{  \mathsf{purify}(R_A) \tilde{E}}  \Vert_1& \mbox{(Triangle inequality)}\\ \nonumber
    &\leq \cO\left(\sqrt{ \eps } \right)+ \Vert  \tilde{{\Theta}}_{R_A \tilde{E}}-{\Theta}_{  \mathsf{purify}(R_A) \tilde{E}}  \Vert_1 & \mbox{(Eq.~\eqref{eq:last_tilde_to_normal})} 
    \\ \nonumber &\leq  \cO\left(\sqrt{ \eps } \right)+ \Vert  \tilde{{\Theta}}_{R_A \tilde{E}}- \tilde{{\Theta}}_{  \mathsf{purify}(R_A) \tilde{E}}\Vert_1+ \Vert \tilde{{\Theta}}_{  \mathsf{purify}(R_A) \tilde{E}} -{\Theta}_{  \mathsf{purify}(R_A) \tilde{E}}  \Vert_1 & \mbox{(Triangle inequality)}
    \\&\leq \cO\left( \sqrt{ \eps }\right) & \mbox{(Eq.\eqref{eq:tilde_purify_bound} and Eq.\eqref{eq:last_tilde_to_normal})}. \label{eq:priv2}
\end{align}
By our choice of parameters, using Eq.~\eqref{eq:priv54},~\eqref{eq:priv55},~\eqref{eq:priv9},~\eqref{eq:privbob1}~and~\eqref{eq:priv2}, the theorem follows. 

\end{proof}

\begin{corollary}
 For any active attack $(\rho_{XE},T_1,T_2)$, Protocol~\ref{prot:priv-amp} is $\left(k, \left(\frac{1}{2} -\delta \right)k,  \cO(2^{-n^{\delta}/2})\right)$-secure as defined in Definition~\ref{privamp} with communication $\cO(n^{7\delta})$  as long as $k \geq \Omega(n^{7\delta})$. 
\end{corollary}
\begin{proof}
Choosing $\eps = 2^{-n^{\delta}}$, the corollary follows from Theorem~\ref{thm:PA_added1}.
\end{proof}
\begin{corollary}
 For any active attack $(\rho_{XE},T_1,T_2)$, Protocol~\ref{prot:priv-amp} is $\left(k, \left(\frac{1}{2} -\delta \right)k,   \frac{1}{\mathsf{poly}(n)}\right)$-secure as defined in Definition~\ref{privamp} with communication $\cO(\log^7(n))$ as long as $k \geq \Omega(\log^7(n))$. 
\end{corollary}
\begin{proof}
Choosing $\eps = \frac{1}{\mathsf{poly}(n)}$, the corollary follows from Theorem~\ref{thm:PA_added1}.
\end{proof}

\newpage

\bibliography{References}
\bibliographystyle{alpha}

\end{document}